\documentclass[11pt]{article}
    
\usepackage{setspace}
\usepackage[a4paper, margin=1in]{geometry}

\usepackage{newtxtext}     
\usepackage{newtxmath}     

\usepackage{authblk}
\title{Empirical Models of the Time Evolution of SPX Option Prices}
\author[1]{Alessio Brini}
\author[2]{David A. Hsieh}
\author[1]{Patrick Kuiper}
\author[1]{Sean Moushegian}
\author[3]{David Ye}
\affil[1]{Duke University Pratt School of Engineering}
\affil[2]{Duke University Fuqua School of Business}
\affil[3]{Duke University Department of Mathematics and Statistical Science}

\usepackage{natbib}
\usepackage{amsmath,mathtools,bm}
\usepackage{graphicx,subcaption,booktabs,multirow,float}
\usepackage{microtype,nicefrac,enumitem,xcolor}
\usepackage[colorlinks=true,linkcolor=blue,citecolor=blue,urlcolor=blue]{hyperref}
\usepackage{cleveref}
\usepackage{tikz}
\usepackage{algorithm}
\usepackage{algpseudocode}
\usepackage{array}
\usepackage{placeins}
\usepackage{endnotes}

\newcount\Comments
\newcommand{\kibitz}[2]{\ifnum\Comments=1\textcolor{#1}{#2}\fi}

\DeclareCaptionLabelFormat{andtable}{#1~#2  \&  \tablename~\thetable}
\date{} 

\begin{document}
\maketitle

\begin{abstract}
The key objective of this paper is to develop an empirical model for pricing SPX options that can be simulated over future paths of the SPX. To accomplish this, we formulate and rigorously evaluate several statistical models, including neural network, random forest, and linear regression. These models use the observed characteristics of the options as inputs -- their price, moneyness and time-to-maturity, as well as a small set of external inputs, such as the SPX and its past history, dividend yield, and the risk-free rate. Model evaluation is performed on historical options data, spanning 30 years of daily observations. Significant effort is given to understanding the data and ensuring explainability for the neural network. A neural network model with two hidden layers and four neurons per layer, trained with minimal hyperparameter tuning, performs well against the theoretical Black-Scholes-Merton model for European options, as well as two other empirical models based on the random forest and the linear regression. It delivers arbitrage-free option prices without requiring these conditions to be imposed.
\end{abstract}

\section{Introduction}\label{Sec:intro}

Accurately modeling the joint evolution of asset prices and option valuations remains a central challenge in finance \citep{cont2002dynamics,gatheral2011volatility}. This task is especially critical in contexts such as risk management \citep{bollerslev2011dynamic}, tail-risk hedging \citep{ramos2020liquidity}, and portfolio stress-testing \citep{banerjee2007implied}, where one must jointly consider the trajectory of the underlying asset price and the option price. The typical approach is to model the joint evolution of the asset price and the implied volatility surface of options across different strikes and maturities. As market conditions change, the entire implied volatility surface shifts in ways that influence option pricing, trading strategies, and the valuation of complex derivatives portfolios \citep{mixon2002factors,fanelli2019seasonality,cao2020neural}. Consequently, developing tools capable of simulating this joint evolution over time holds significant practical importance.

The implied volatility surfaces of options are typically generated from the foundational \cite{black1973pricing} model, which introduced a closed-form solution for European options under the assumptions of constant volatility and log-normal asset returns. However, its assumption of constant volatility does not capture well-documented empirical regularities, such as the volatility smile and term structure of implied volatilities \citep{derman1994riding,dupire1997pricing}. This shortcoming has motivated a rich literature of model extensions. The \cite{heston1993closed} model introduces stochastic volatility through a mean-reverting square-root process, offering analytical tractability while addressing some of the deficiencies of the Black-Scholes framework. Other models aim to capture observed option pricing features by introducing additional complexities. The jump-diffusion model in \cite{merton1976option} accounts for discontinuities in asset returns, while the \cite{bates1996jumps} model combines stochastic volatility with jumps to more accurately match the dynamics of both asset returns and implied volatility surfaces. Local volatility models, such as the one in \cite{dupire1994pricing}, relax the assumption of constant or stochastic volatility by allowing the volatility to be a deterministic function of the asset price and time. These approaches improve fit. However, they typically require frequent recalibration and may still struggle to model the full joint evolution of the volatility surface through time.

In practice, the implied volatility surface is often reconstructed daily from market data and interpolated across strikes and maturities. While this approach ensures consistency with observed option prices on each trading day, it treats the surface as a static snapshot rather than a dynamic object. As a result, the temporal evolution of the surface must be approximated indirectly, commonly through heuristics such as sticky strike or sticky delta rules in \cite{daglish2007volatility}. While these methods offer simplicity and tractability, they may not fully capture the dynamic nature of the market, especially during periods of significant price movement, as noted in \cite{marabel2010dynamics}.

To address the limitations of static implied volatility surfaces and heuristic-based dynamic modeling, we propose a hybrid framework that integrates econometric volatility modeling with data-driven option pricing via a non-parametric model. Specifically, we employ a Generalized Autoregressive Conditional Heteroskedasticity (GARCH) model of \cite{bollerslev1986generalized} to capture the time-varying volatility inherent in asset returns, effectively modeling features such as volatility clustering and persistence. We use these volatility estimates as inputs to a supervised learning model, in which we train a neural network to approximate the option pricing function under realistic, evolving market conditions. To benchmark the performance of our neural network approach, we also implement a linear regression model and a classical nonlinear machine learning model, namely, a random forest.

Our framework enables the simulation of joint paths for asset prices and corresponding option prices, assuming only that the asset price follows a GARCH process. We directly model the price of the option rather than the implied volatility surface, since the Black-Scholes model cannot generate implied volatilities for some options. Our machine learning models provide a flexible and non-parametric mapping by learning the functional relationship between market features and option prices from historical data. 

We train and test our models on a comprehensive dataset of S\&P 500 index put options spanning from 1996 to 2022. Our empirical strategy relies on both expanding and rolling window training methodologies, allowing us to evaluate how model performance evolves over time. The results show that neural networks consistently outperform linear and tree-based benchmarks, particularly in pricing deep out-of-the-money put options. They also outperform during periods of heightened volatility. We further investigate the models’ behavior through interpretability analyses and tests of economic consistency, verifying that the predicted prices respect key no-arbitrage conditions such as monotonicity and convexity. Our contribution is to provide an empirical evaluation of machine learning models in approximating the option pricing function conditional on time-varying features. This yields a simulation framework to generate internally consistent paths of asset prices and option prices, which is especially useful in risk management settings involving tail events in derivatives portfolios.

The remainder of the paper is organized as follows: Section~\ref{Sec:literature} reviews the relevant literature, Section~\ref{Sec:data} describes the dataset and features, Section~\ref{Sec:method} outlines the modeling approach, Section~\ref{Sec:empres} presents the empirical pricing results, and Section~\ref{Sec:interpret} discusses model interpretability via SHAP. Section~\ref{Sec:no_arbitrage_check} examines the extent to which the trained models violate no-arbitrage conditions. Section~\ref{Sec:takeaways} summarizes the main takeaways from the empirical analysis. Section~\ref{Sec:conclusion} concludes.

\section{Related Literature}\label{Sec:literature}

Traditional parametric models, such as Black-Scholes and Heston, remain widely used in practice and serve as classical benchmarks for option pricing. A comprehensive discussion of these traditional approaches can be found in \cite{bakshi1997empirical,bates2003empirical,guyon2013nonlinear,bates2022empirical}. Building on these foundational models, the application of machine learning models to option pricing has received increasing attention in recent years. Broadly, the literature can be categorized into two main lines of research: studies using neural networks, commonly referred to as deep learning, for option pricing and volatility modeling, and research exploring alternative machine learning models, such as tree-based methods, support vector regressions, and Gaussian processes. Below, we summarize key contributions in each of these lines of research.

A substantial body of work has focused on the use of neural networks for option pricing, originating from foundational work by \cite{RePEc:nbr:nberwo:4718}.  \cite{bloch2019option} describe the integration of neural network-based models into the option pricing framework.  \cite{ivașcu2021option} examine the performance of various machine learning models, including feedforward neural networks, and evaluate their effectiveness on European call options. \cite{almeida2023can} employ a neural network to correct pricing errors from classical parametric models, while \cite{andreou2023stock} extend this idea by incorporating firm characteristics as additional inputs. Other studies aim to ensure consistency with financial theory when applying neural networks to pricing derivatives. As a few examples, \cite{itkin2019deep} introduce a soft constraint in the loss function to enforce no-arbitrage conditions. \cite{funahashi2021artificial} propose a hybrid approach combining asymptotic expansions and neural networks to improve the stability, accuracy, and computational efficiency of option pricing under complex stochastic volatility models. \cite{pagnottoni2019neural} explore the application of neural networks to less liquid markets, particularly cryptocurrency options.

Beyond standalone models, several studies have explored the integration of neural networks into existing option pricing frameworks. \cite{cao2020neural} and \cite{jang2021deepoption} propose models where deep learning architectures are first trained on synthetic data generated from classical pricing models such as Black-Scholes before being fine-tuned on real market data. Similarly, \cite{huge2020differential} introduce differential machine learning, where models are trained on price inputs and on the differentials of labels with respect to the inputs, enhancing pricing accuracy. \cite{cao2021deep} leverages a neural network to improve the pricing of exotic options. Recurrent architectures have also been explored. \cite{yang2017gated} develop a gated recurrent neural network that incorporates prior knowledge of no-arbitrage conditions, while \cite{liang2022time} applies convolutional and recurrent networks to capture the time series nature of option pricing. A related approach is taken by \cite{horvath2021deep}, which employ convolutional neural networks for volatility surface calibration. \cite{van2023machine} analyzes the effect of different neural network architectures on the task of option pricing. \cite{gan2024option} employ a residual neural network for the option pricing problem. \cite{fan2024machine} develops a hybrid framework in which neural networks are embedded into a stochastic differential equation (SDE)-based pricing model. \cite{buehler2019deep} explores reinforcement learning techniques to optimize hedging strategies under non-linear reward structures.

Beyond neural networks, alternative machine learning models have also been explored. \cite{bali2023option} evaluates various linear and nonlinear models, including Lasso, Ridge, Elastic Net, tree-based methods, and ensemble models, using a dataset of approximately 12 million transactions. \cite{sood2023black} compares the Black-Scholes model to several machine learning models, finding that recurrent networks outperform others. Gaussian process regression has been investigated as an alternative to deep learning, as seen in \cite{park2014parametric,de2018machine,goudenege2020machine}, where Gaussian processes are applied to price European and American options while also fitting the implied volatility surface. A continuous-time approach combining neural networks with parametric PDE solutions is proposed by \cite{glau2022deep} to enhance pricing accuracy.

Tree-based models, particularly random forests and gradient boosting methods, have been proposed as alternatives to neural networks. \cite{arin2022deep} implements a classifier that determines whether to use Black-Scholes or a neural network for pricing, based on predefined error thresholds. Hybrid approaches have also been explored, where machine learning models are combined with classical pricing techniques. \cite{wei2021intelligent} presents ensemble methods that integrate convolutional neural networks with traditional option pricing models. \cite{brini2024pricing} also employ tree-based ensemble methods to price cryptocurrency options and demonstrate their effectiveness in capturing pricing patterns in less liquid markets.

Several papers have compared different machine learning models for option pricing. \cite{li2022application} provides a broad literature review of existing methods, distinguishing between direct price prediction models and indirect approaches that forecast volatility before applying a classical pricing model. \cite{chen2023teaching} proposes a hybrid method that first pretrains a neural network on theoretical prices generated from the Black-Scholes-Merton framework before incorporating real market data. This study also explores the impact of including prior volatility information (e.g., VIX) and evaluates the performance of neural networks against random forests and regression-based models.

Overall, the literature highlights the potential of machine learning techniques in option pricing, with deep learning models offering high flexibility and alternative methods providing robust, interpretable benchmarks. Our study contributes to this growing body of research by systematically evaluating neural networks, random forests, and linear regression models in a unified framework, and analyze their relative performance for option pricing.


\section{S\&P 500 Put Options Dataset}\label{Sec:data}
 
The dataset comprises European-style put options on the S\&P 500 index, with daily data sourced from OptionMetrics, covering the period from January 1996 to December 2022. A similar data preprocessing approach can be applied to European-style call options written on the same underlying index. We classify put options as out-of-the-money (OTM) or in-the-money (ITM) based on the ratio $\frac{S}{K}$, where options with $\frac{S}{K} > 1$ are OTM, while those with $\frac{S}{K} \leq 1$ are ITM. From the gathered option data, we apply the following filters:

\begin{itemize}
    \item To avoid very illiquid contracts, we restrict the sample to put options with moneyness between 50\% OTM and 50\% ITM.
    \item We retain only contracts with at least one month until expiry to ensure they have some remaining time to expiration. We also exclude options with more than 18 months to expiration.
    \item Options must have a positive bid price. We do not impose additional liquidity filters, so some illiquid options with wide bid-ask spreads remain in the dataset.
    \item For each expiration, we include AM-settled options and PM-settled options with strikes unique from AM-settled options. This ensures consistency in pricing, as including both AM- and PM-settled options with the same strikes would yield duplicate entries for the same contract, complicating model training.
    
\end{itemize}

After this filtering, the dataset contains approximately 4.57 million OTM put options and 2.84 million ITM put options. Each row corresponds to a specific option on a given trading day. As a result, the same option may appear multiple times within the training window, each time with updated characteristics such as time to maturity, the underlying price, or volatility.

We enrich the OptionMetrics data with additional variables obtained from external sources. Table \ref{Tab:dataset} lists both direct observations and derived variables used in our analysis. Each day, we interpolate the spot rate linearly using zero-coupon rates. For durations up to one year, we use Treasury bill prices from CRSP. For periods between one and five years, we rely on data from \cite{gurkaynak2007us}. We compute the dividend yield from month-end data and adjust it within each month based on changes in the SPX index.

\begin{table}[t]
    \centering
    \begin{tabular}{ll}
        \toprule
        \textbf{Variable} & \textbf{Description} \\
        \midrule
        Time to maturity & Time to maturity of the option in a fraction of years. \\
        S\&P 500 index & Value of the S\&P 500 index at market close. \\
        Strike price & Strike price of the quoted option. \\
        Spot rate & Continuously compounded zero-coupon rate. \\
        Dividend yield & Dividend yield of the S\&P 500 index. \\
        GARCH volatility & GARCH volatility extracted from the GARCH-normal model. \\
        Black-Scholes price & Black-Scholes option price using inputs from this table. \\
        Option price & Midpoint between the best bid and ask price for the option. \\
        \bottomrule
    \end{tabular}
    \caption{Overview of the variables included in the dataset, representing option characteristics, market conditions, and model-implied values.}

    \label{Tab:dataset}
\end{table}

We obtain the S\&P 500 index from the CRSP database through October 31, 2022. Starting November 1, 2022, we use the CBOE index settlement for SPX options, as daily SPX expirations began at that time. This ensures consistency with the options market, as SPX options settle on the CBOE and its prices are observed on expiration days. 

We estimate the GARCH volatility\footnote{We estimate the model using the rugarch package in R \citep{rugarch}.} for the S\&P 500 index using a rolling 252-day window, which provides a daily measure of market volatility.  The model follows the GARCH(1,1) specification \citep{bollerslev1986generalized}, applied to the log returns of the S\&P 500 index:

\begin{align}
    r_t &= \mu + \sigma_t e_t, \\
    \sigma_t^2 &= a_0 + a_1 \sigma_{t-1}^2 + b_1 \sigma_{t-1}^2e_{t-1}^2,
\end{align}
where $ r_t $ is the log return of SPX from $ t-1 $ to $ t $ and $e_t \sim \text{IID } N(0,1)$. We re-estimate the GARCH model daily using the most recent 252 trading days. We then use the estimated parameters in the variance function to forecast the variance of the underlying asset from $t$ to $t + d$, represented as $\sum_{s=t}^{t+d} \sigma_s^2$. We compute each $\sigma_s^2$ recursively using the estimated variance equation and the assumption that the expected variance of $e_s$ equals one for all $s > t - 1$.

We compute the Black-Scholes option price for a continuous dividend-paying asset using the S\&P 500 index, spot rate, dividend yield, time to maturity, and the GARCH-estimated volatility. The Black-Scholes model price plays two roles in our analysis. First, we include it as an additional input feature in a subset of models, using it as a reference value to inform the learning process. Second, we use it as a baseline model to assess whether the trained models outperform a traditional closed-form pricing approach.

\begin{figure}[t]
    \centering
    \begin{subfigure}[t]{0.32\textwidth}
        \centering
        \includegraphics[width=\textwidth]{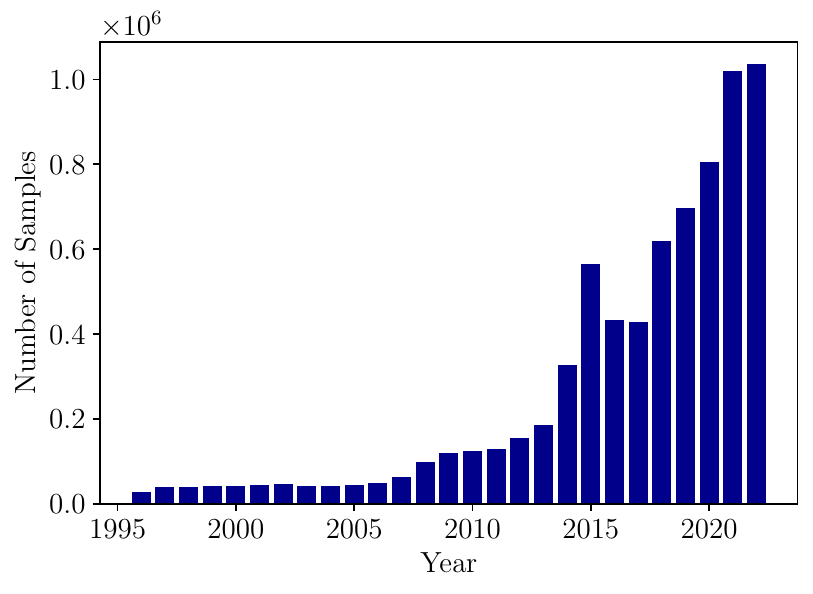}
        \label{fig:date_hist}
    \end{subfigure}
    \hfill
    \begin{subfigure}[t]{0.32\textwidth}
        \centering
        \includegraphics[width=\textwidth]{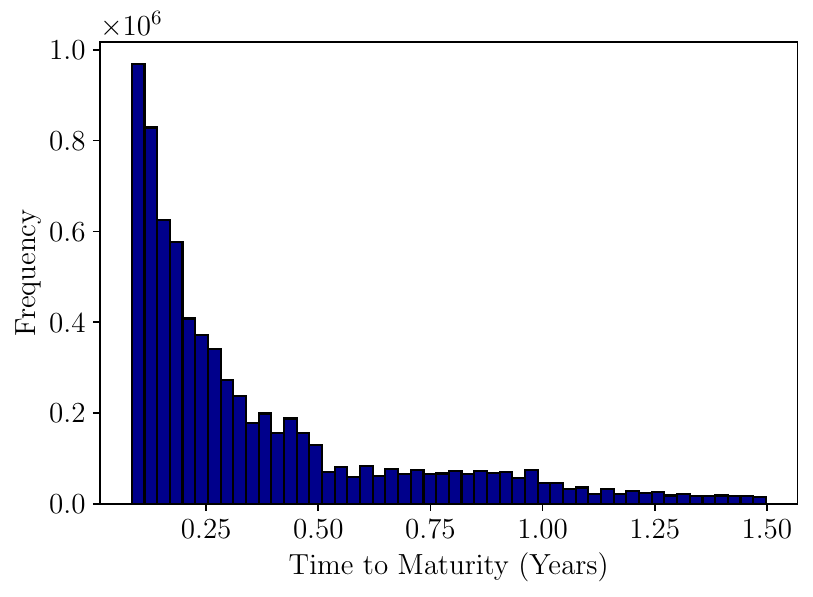}
        \label{fig:TTM_hist}
    \end{subfigure}
    \hfill
    \begin{subfigure}[t]{0.32\textwidth}
        \centering
        \includegraphics[width=\textwidth]{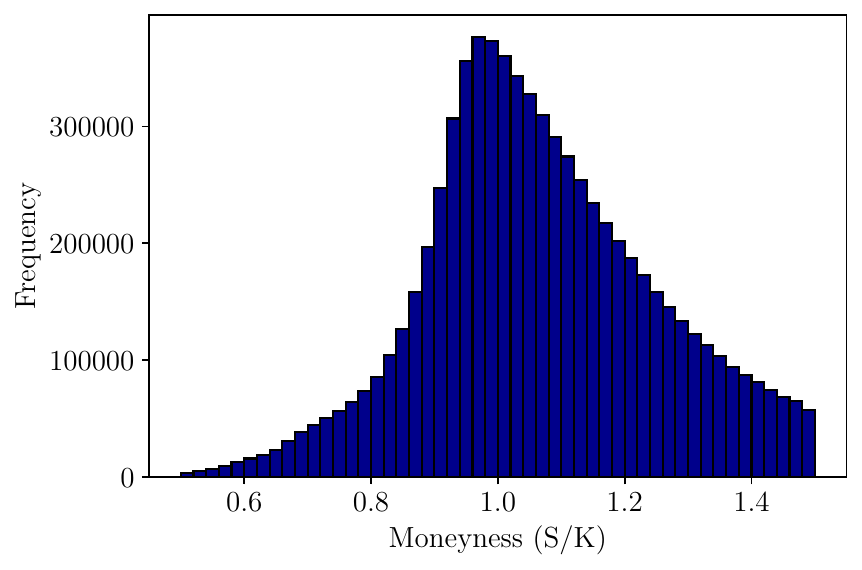}
        \label{fig:OTM_hist}
    \end{subfigure}
    \caption{Empirical distributions of options data observations across different dimensions. 
    The left panel shows the frequency of observations by year over the analyzed time period. 
    The middle panel presents the distribution of observations by time to maturity. 
    The right panel displays the frequency of observations by out-of-moneyness level, where out-of-moneyness is calculated as $\frac{S}{K}$.}
    \label{fig:options_data}
\end{figure}

Fig. \ref{fig:options_data} shows the empirical distributions of the options data across three key dimensions: the date of the option price quote, the time to maturity, and the moneyness level, calculated as $\frac{S}{K}$. The left panel shows that the number of options increases significantly over the years, both by strike and by maturity. The middle panel highlights the predominance of short-maturity options over long-maturity options. The right panel shows that OTM options are more numerous than ITM options, with the distribution slightly skewed to the right.

\section{Option Pricing Models}\label{Sec:method}

In this section, we describe the pricing models considered in this study and establish consistent notation for clarity. We train a feedforward neural network (NN) model for option pricing. For comparison, we also train a linear regression (LR) model and a random forest (RF) model.

We denote $ X \in \mathbb{R}^{n \times p} $ as the matrix of input features, where each row corresponds to an option transaction and each column represents a continuous variable from Tab.~\ref{Tab:dataset}. The target vector, $y \in \mathbb{R}^{n}$, contains the option prices. Each model predicts an option price, represented as $\hat{y} \in \mathbb{R}^{n}$, with a superscript indicating the model used. We follow this notation in the corresponding subsections for each model.

Rather than conducting extensive hyperparameter tuning, which is common in machine learning applications for regression problems, we adopt an off-the-shelf training approach with informed hyperparameter selection. We deliberately avoid conducting an exhaustive hyperparameter search, which could potentially improve option pricing accuracy. However, it would also significantly increase computational costs and compromise the fairness of our comparisons. Minimizing hyperparameter tuning allows for a more balanced evaluation between machine learning-based pricing models and classical option pricing methods, where model calibration does not involve comparable computational effort. While training a NN or RF on a large dataset inherently demands more computation than traditional closed-form models, our approach shows that strong performance can still be achieved with minimal tuning, reinforcing the practical viability of these models.

This modeling choice aligns with prior work in machine learning-based financial modeling, where researchers have trained models with minimal tuning to highlight their baseline performance relative to traditional approaches \citep{christensen2023machine}. When deploying one of these models in practice, a more refined hyperparameter tuning process could further improve predictive accuracy, but we defer such optimization to future work focused on the model deployment phase.

\subsection{Feedforward Neural Network (NN)}
A NN approximates the function mapping $ X $ to $ y $ using a fully connected feedforward architecture. The prediction is given by:
\begin{equation}
    \hat{y}^{\text{NN}} = f_{\theta}(X),
\end{equation}
where $f_{\theta}$ denotes the neural network with parameters $\theta = \{W^{(l)}, b^{(l)}\}_{l=1}^{L}$, consisting of weights $W^{(l)}$ and biases $b^{(l)}$ across $L$ hidden layers. Each hidden layer applies a nonlinear transformation:
\begin{equation}
    h^{(l)} = \sigma(W^{(l)} h^{(l-1)} + b^{(l)}),
\end{equation}
where $ \sigma $ is a nonlinear activation function. We train the model using stochastic gradient descent by minimizing the Huber loss, which is less sensitive to outliers than the mean squared error \cite{mnih2015human,jadon2024comprehensive}:
\begin{equation}
    \min_{\theta} \sum_{i=1}^{n} L_{\delta}(y_i, \hat{y}_{i}^{\text{NN}}),
\end{equation}
where $y_i$ is the true price of the options for the $i$-th observation and the Huber loss function is defined as:
\begin{equation}
    L_{\delta}(y, \hat{y}^{\text{NN}}) =
    \begin{cases}
        \frac{1}{2} (y - \hat{y}^{\text{NN}})^2, & \hspace{-3.5mm}\text{if } |y - \hat{y}^{\text{NN}}| \leq \delta, \\
        \delta (|y - \hat{y}^{\text{NN}}| - \frac{1}{2} \delta), & \hspace{-3.5mm}\text{otherwise}.
    \end{cases}
\end{equation}

Each NN consists of two hidden layers with four neurons each, using ReLU as the activation function. We optimize the model using the Adam optimizer with a fixed learning rate of $0.0001$. To prevent overfitting, we apply weight decay with a coefficient of $10^{-3}$, which penalizes large weights and encourages simpler models that generalize better. We train with a batch size of $512$ and apply early stopping with a patience of $20$ training epochs to halt training when no significant improvements are observed. Before training the NN, we also standardize all input features to have zero mean and unit variance. This is a common practice when training these models, as it helps improve convergence, prevents numerical instability, and ensures that each feature contributes proportionally to the learning process \citep{lecun2002efficient, goodfellow2016deep}. We set the Huber loss parameter $\delta$ to $1$ throughout all experiments.

\subsection{Random Forest}
We test the RF \cite{breiman2001random} as an alternative nonlinear model for option pricing. RF is an ensemble of $M$ decision trees, where each tree $ T_m(X) $, for $m = 1, \dots, M$, is trained on a bootstrapped sample of the training data. At each split, the tree considers a randomized subset of features to reduce overfitting. This approach follows bagging (bootstrap aggregating) technique, which improves predictive stability by averaging multiple low-bias models \citep{hastie2009elements}. The model makes predictions by averaging the outputs of all trees, which enhances stability and reduces variance:
\begin{equation}
    \hat{y}^{\text{RF}} = \frac{1}{M} \sum_{m=1}^{M} T_m(X).
\end{equation}
Each tree recursively partitions the feature space, learning decision rules that minimize prediction error while maintaining diversity across trees.

We train each RF model with $100$ decision trees and a maximum tree depth of $10$ to balance predictive power and computational efficiency. We enable bootstrapping, which means that each tree is trained on a randomly drawn sample of size equal to the total number of training observations in $X$. Since sampling is performed with replacement,each tree is trained on a bootstrapped sample that contains approximately $63.2\%$ of unique observations from the dataset.

\subsection{Linear Regression}
We include LR as a baseline to assess whether nonlinear models improve pricing performance. The LR model assumes a linear relationship between the input variables and the option price:
\begin{equation}
    \hat{y}^{\text{LR}} = X \beta + \epsilon,
\end{equation}
where $ \beta $ is the vector of regression coefficients, and $ \epsilon \sim \mathcal{N}(0, \sigma^2 I)$ is an IID error term. The coefficients $ \beta $ are estimated by minimizing the sum of squared errors through OLS:
\begin{equation}
    \min_{\beta} \sum_{i=1}^{n} (y_i - X_i \beta)^2.
\end{equation}

The set of regressors includes SPX, moneyness, time to maturity, dividend yield, the risk-free rate, the predicted GARCH volatility, and squared terms and interaction terms among these variables. We optionally include the Black-Scholes price computed from these variables as a regressor.

\section{Empirical Pricing Results}\label{Sec:empres}
In this section, we present the performance of the models trained and tested on the option dataset described in Section~\ref{Sec:data}. The evaluation considers multiple versions of the models, each trained under a different data-splitting strategy and on distinct subsets of the dataset.

Because the dataset spans several years, we adopt two training methodologies. The first is an \textit{expanding window approach}, where models are trained on an initial three-year window of data and tested on the subsequent six months. With each iteration, the test set is added to the training set, progressively expanding the training window. By the final iteration, the models are trained on all available data points and tested on the last six months. The second methodology is a \textit{rolling window approach}, where we fix the training window at three years and slide it forward in time. In this case, the models are always trained on the most recent three years of data and tested on the following six months, ensuring consistency in the size of the training and test sets.

To explore model performance across different scenarios, we trained models on subsets of the dataset based on moneyness. Specifically, we consider options classified as OTM and ITM, as defined in Section~\ref{Sec:data}. Since we train models exclusively on put options, it is meaningful to distinguish between ITM and OTM contracts, as their pricing errors may follow different patterns.

For each combination of data-splitting strategy and moneyness subset, we train the models in two configurations: one that includes the Black-Scholes (BS) price in the input space and one that excludes it. This design allows us to evaluate the impact of including the BS price. It also facilitates a direct comparison with the BS model itself. We denote models that include the BS price with a "+" (plus) symbol, and those that exclude it with a "-" (minus) symbol. For baseline comparison, we also report the performance of the Black-Scholes model itself in pricing put options. Following this setup, we train two nonlinear models, NN and RF, and one linear model, LR.

The evaluation metric used in this ex-post evaluation is the Mean Absolute Percentage Error (MAPE). We motivate this choice by the fact that option prices can vary by several orders of magnitude depending on moneyness and time to maturity, making absolute error metrics less informative for assessing model performance across different pricing regimes. In particular, within our filtered sample—limited to options no more than 50\% OTM or ITM, prices still vary considerably. Deep OTM and near-expiry options tend to have low prices, whereas at-the-money and longer-maturity options are significantly more expensive. Standard absolute error measures would disproportionately emphasize mispricing in high-priced options while neglecting relative errors in cheaper options, potentially distorting model evaluation. MAPE focuses on percentage errors rather than absolute errors and provides a more meaningful comparison across models and pricing conditions. It also normalizes performance across different periods of the dataset, making models trained on distinct timeframes more comparable. Furthermore, since option prices strongly depend on the level of the underlying index, which varies significantly over time, using a relative error measure like MAPE accounts for these fluctuations and provides a more robust assessment of model accuracy. While MAPE is known to be more sensitive to small values, this characteristic aligns with the main objective of our study, which emphasizes accuracy in pricing deep OTM put options. These options are particularly relevant for tail-risk hedging strategies, and their effective pricing is crucial despite their lower market prices.

\subsection{Out-of-the-Money Put Options}

\begin{figure}[t]
    \centering
    \includegraphics[width=0.78\textwidth]{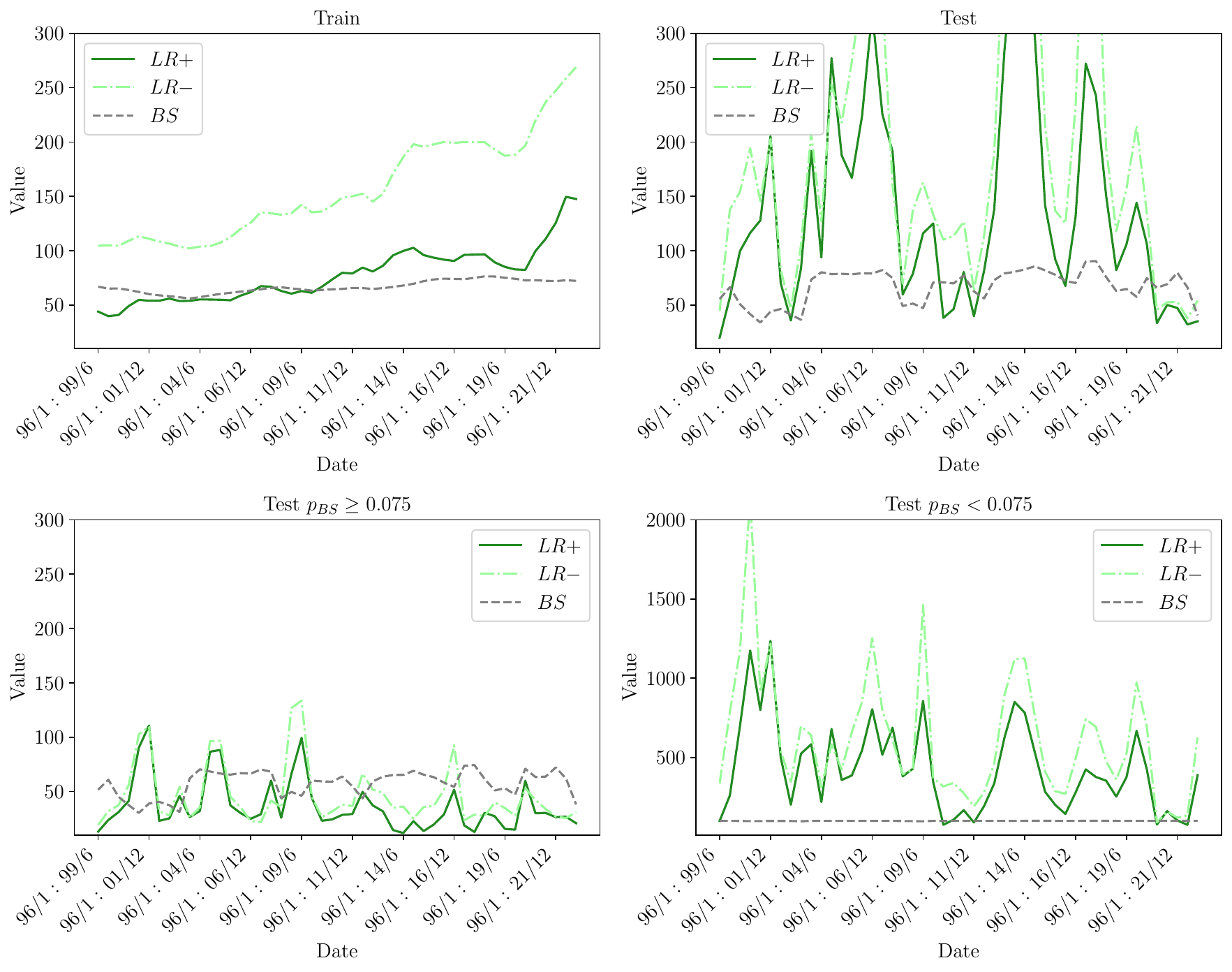}
    \caption{Evolution over time of the MAPE values for OTM options trained using an expanding window schema with the LR model. The values correspond to Tab.~\ref{Tab:OTM_expanding_LR}. }
    \label{Fig:OTM_series_expanding_LR}
\end{figure}

\begin{figure}[t]
    \centering
    \includegraphics[width=0.78\textwidth]{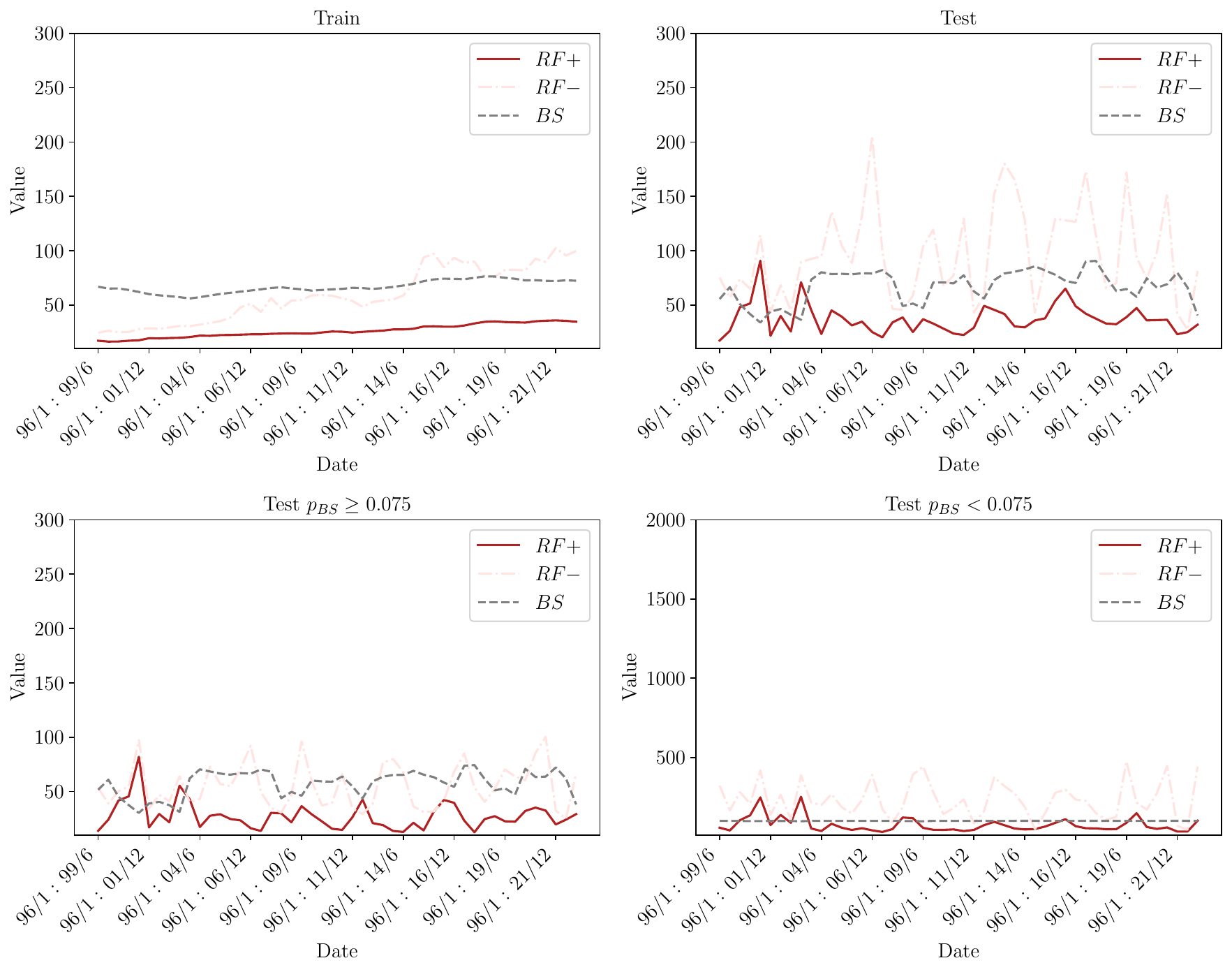}
    \caption{Evolution over time of the MAPE values for OTM options trained using an expanding window schema with the RF model. The values correspond to Tab.~\ref{Tab:OTM_expanding_RF}. }
    \label{Fig:OTM_series_expanding_RF}
\end{figure}

\begin{figure}[t]
    \centering
    \includegraphics[width=0.78\textwidth]{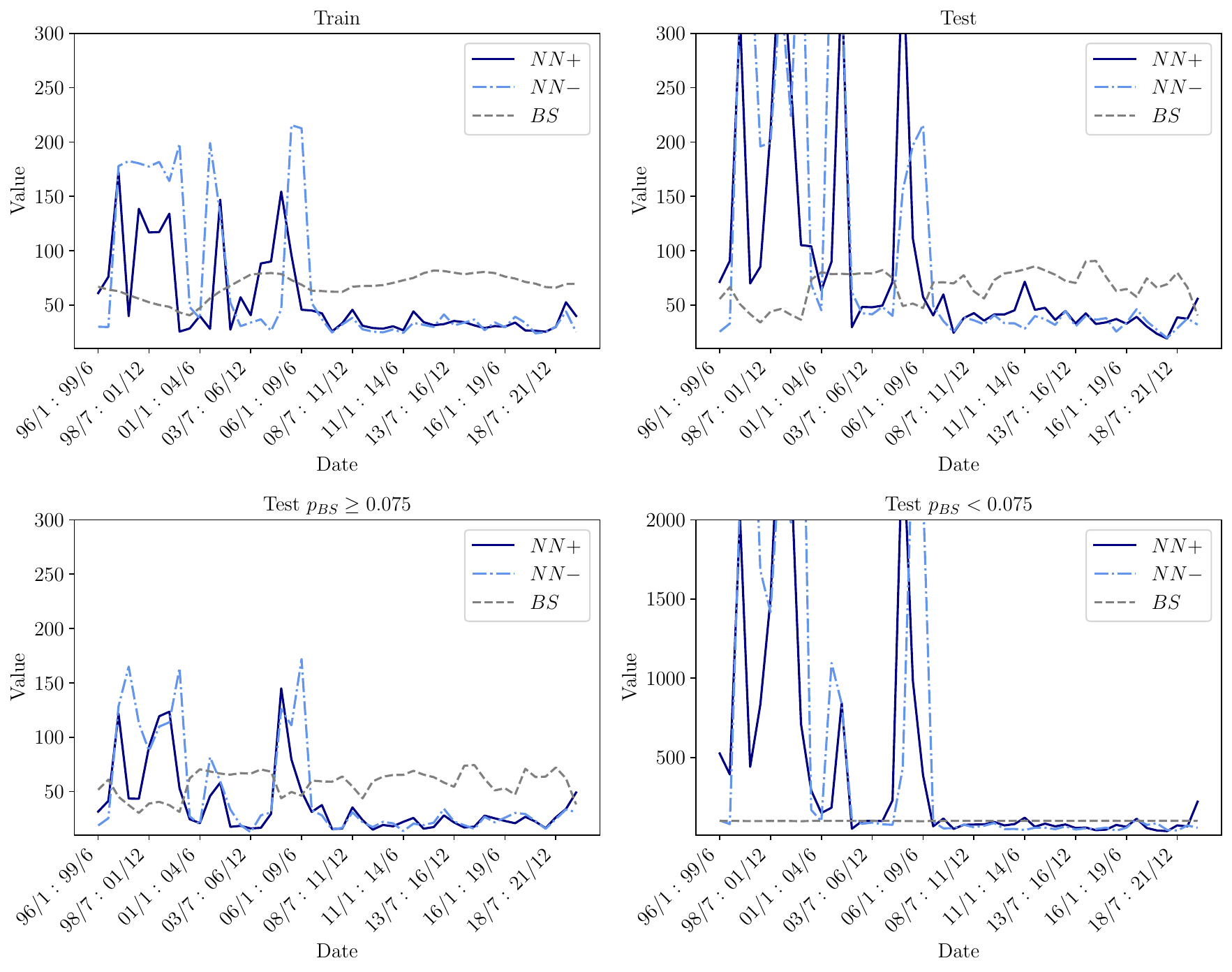}
    \caption{Evolution over time of the MAPE values for OTM options trained using a rolling window schema with the NN model. The values correspond to Tab.~\ref{Tab:OTM_rolling_NN}.}
    \label{Fig:OTM_series_rolling_NN}
\end{figure}

\begin{figure}[t]
    \centering
    \includegraphics[width=0.78\textwidth]{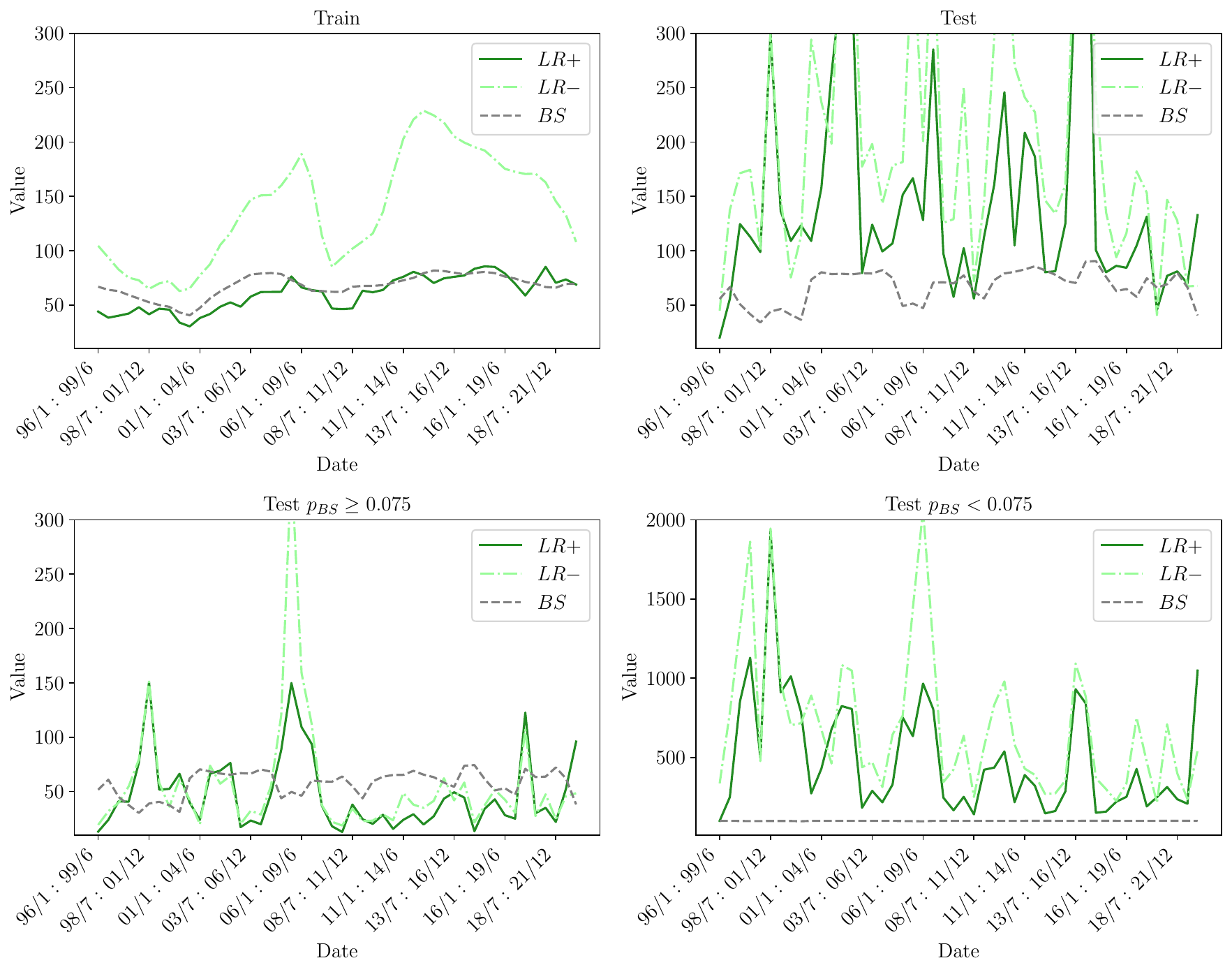}
    \caption{Evolution over time of the MAPE values for OTM options trained using a rolling window schema with the LR model. The values correspond to Tab.~\ref{Tab:OTM_rolling_LR}.}
    \label{Fig:OTM_series_rolling_LR}
\end{figure}

\begin{figure}[t]
    \centering
    \includegraphics[width=0.78\textwidth]{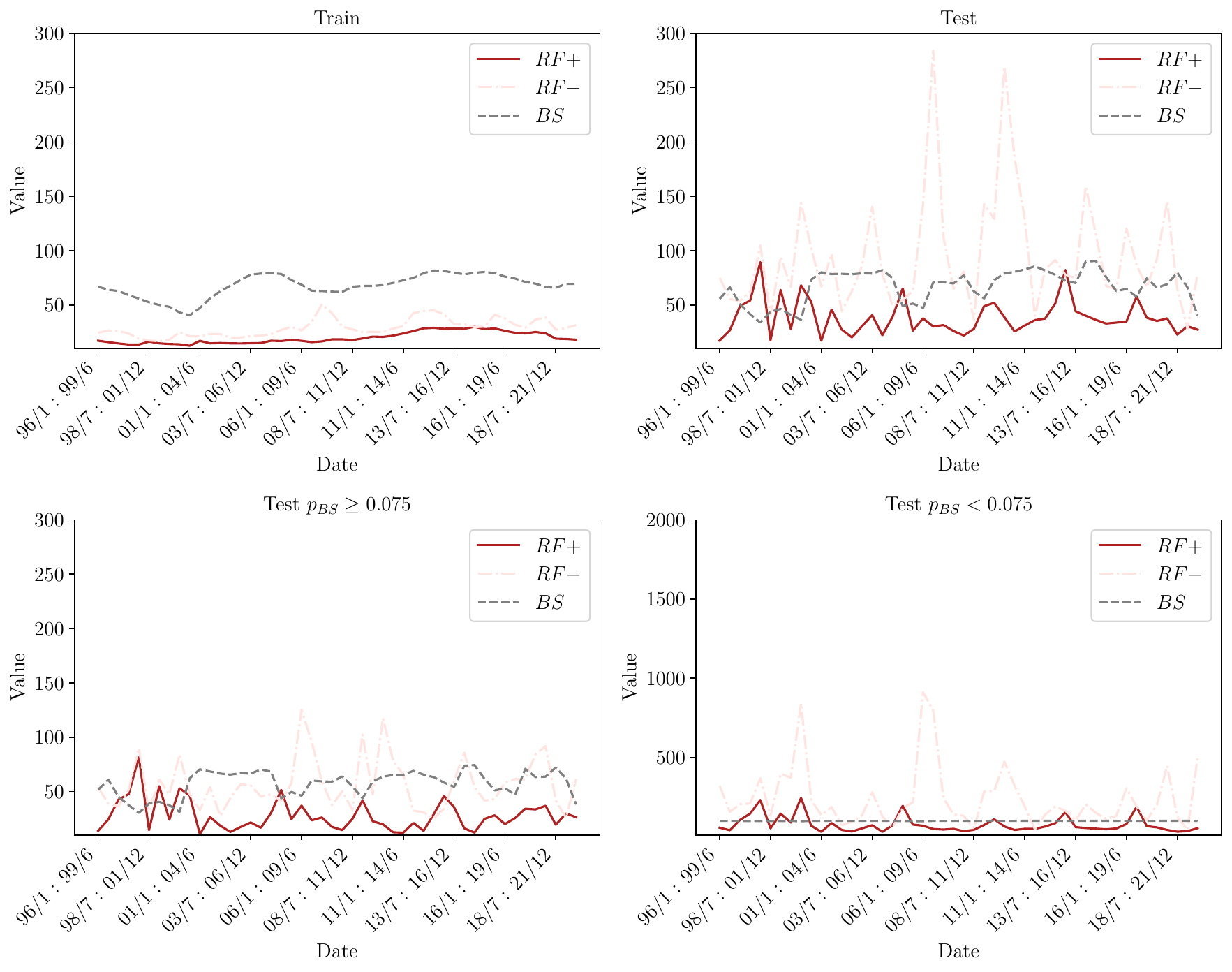}
    \caption{Evolution over time of the MAPE values for OTM options trained using a rolling window schema with the RF model. The values correspond to Tab.~\ref{Tab:OTM_rolling_RF}.}
    \label{Fig:OTM_series_rolling_RF}
\end{figure}

In this subsection, we evaluate the performance of models trained and tested on OTM put options\footnote{We report the numerical results corresponding to all the figures of this and the next subsections in Appendix~\ref{App:tables}.}. Figures~\ref{Fig:OTM_series_expanding_NN}–\ref{Fig:OTM_series_rolling_RF} show the evolution of the MAPE over time for each model (NN, LR, and RF) using both expanding and rolling window approaches. In these figures, we present the dynamics of the MAPE over different periods into subsets of four subplots, one for each model, identified by different colors. Each subplot uses x-axis labels in the format YY/MM - YY/MM, where the first date marks the start of the training set and the second the end of the six-month test set. The subplots in the second row of each subset focus on segmenting the test set based on BS price. Specifically, we evaluate the trained models under both methodologies on two test subsets. This segmentation isolates very low-priced options in the dataset, which the Black-Scholes model struggles to price accurately. By isolating these options, we assess how the competing models perform under these conditions. The results show that very low-priced options contribute significantly to the overall MAPE. Consequently, all models and configurations show lower errors on the subset where $p_{\text{BS}} > 0.075$. Additional insights emerge when OTM options are divided into different moneyness levels, as discussed in the next section.

Including the BS price as a feature improves model performance across both training methodologies, namely the expanding and rolling window approaches. NN and RF consistently outperform LR. The BS feature contributes substantially to LR and RF, leading to large performance gains when included. For RF in particular, its predictive accuracy benefits considerably from the BS input, and the model remains more stable than NN, with MAPE values exhibiting fewer large spikes across the dataset's time span. When BS is removed, RF loses predictive power and often performs worse than the BS baseline. In contrast, NN retains stronger performance even when BS is excluded. While including BS generally improves NN performance, the effect is less pronounced and not always consistent across model runs or training periods.

The choice of training methodology significantly impacts model behavior. Under the expanding window approach, NN and RF consistently outperform LR, with RF maintaining greater stability when BS is included. The rolling window approach reveals additional insights. Before the 2008 financial crisis, NN exhibits instability, which is not observed in the expanding window setup. After the crisis, NN outperforms BS consistently, with similar results whether we include BS price as a feature or not. RF maintains stability across both training approaches. Including BS continues to benefit RF, reinforcing the trend observed in the expanding window setup. Under the rolling window approach, NN remains the best-performing model on the test set, consistently outperforming BS.

\begin{figure}[t]
    \centering
    \includegraphics[width=0.78\textwidth]{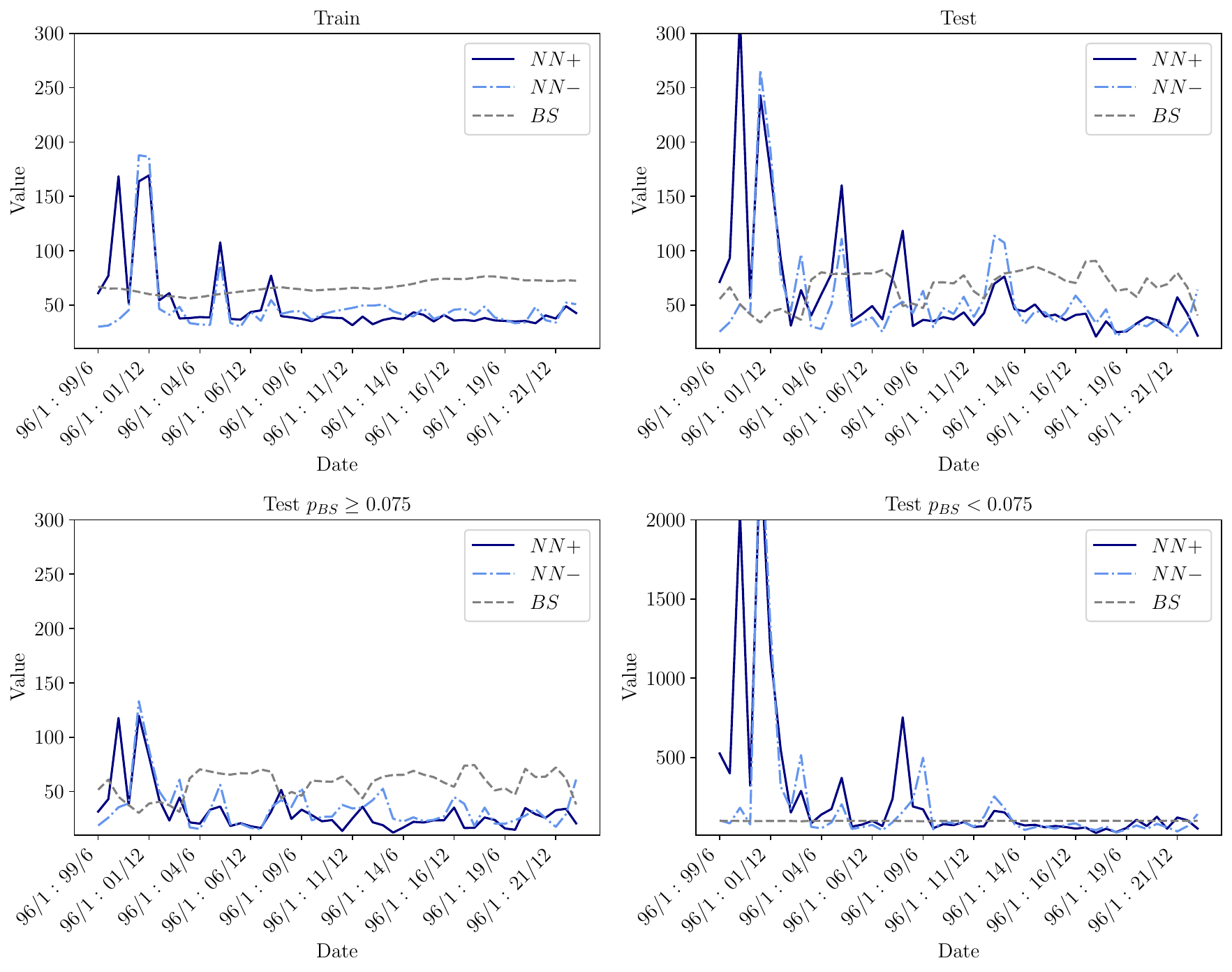}
    \caption{Evolution over time of the MAPE values for OTM options trained using an expanding window schema with the NN model. The values correspond to Tab.~\ref{Tab:OTM_expanding_NN}.}
    \label{Fig:OTM_series_expanding_NN}
\end{figure}
\FloatBarrier

\FloatBarrier
\FloatBarrier

\begin{figure}[t]
    \centering
    \includegraphics[width=\linewidth]{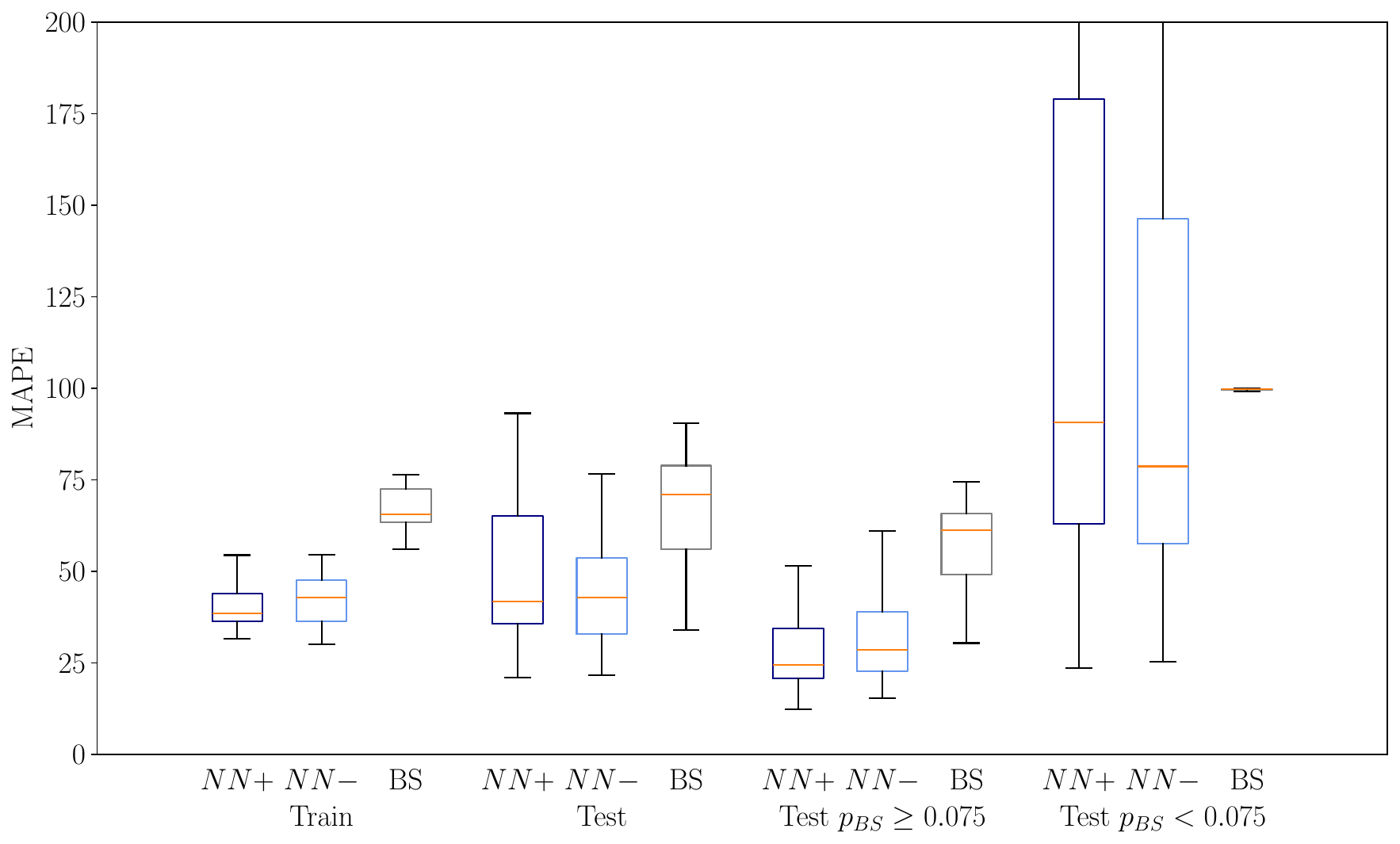}
    \includegraphics[width=\linewidth]{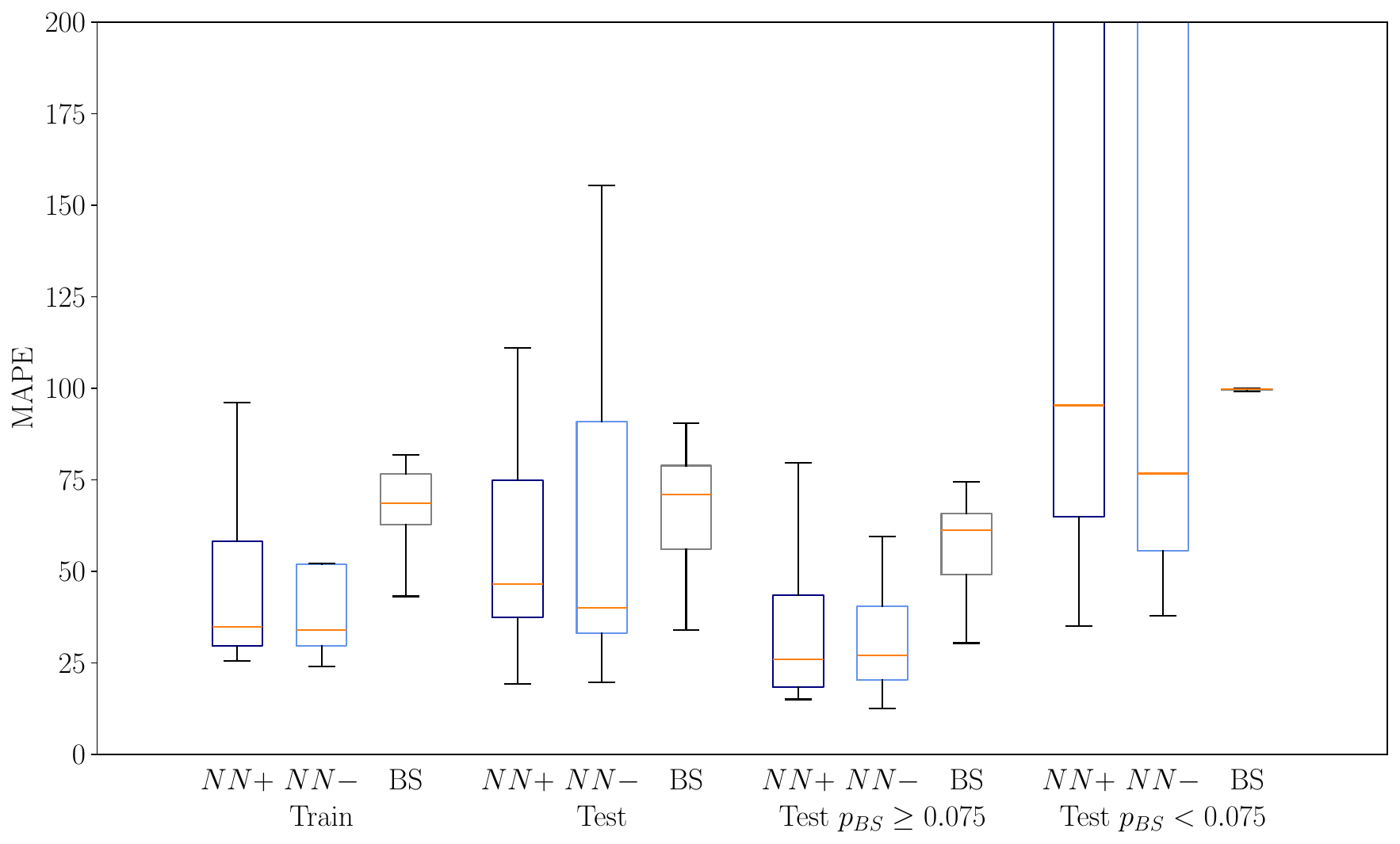}
    \caption{Boxplots of the mean absolute percentage error (MAPE) for OTM options under two data-splitting schemas. The top panel corresponds to the expanding window schema, while the bottom panel corresponds to the rolling window schema. In both panels, the data is grouped by subsets: the training set, the test set, the test set filtered by Black-Scholes price ($p_{\text{BS}} \geq 0.075$), and the test set filtered by $p_{\text{BS}} < 0.075$. Each group contains three boxplots, corresponding to the NN trained with Black-Scholes information (NN+), the NN trained without Black-Scholes information (NN-), and the Black-Scholes model (BS) itself. The exact numerical values corresponding to the expanding and rolling window results are presented in Tables~\ref{Tab:OTM_expanding_NN} and \ref{Tab:OTM_rolling_NN}, respectively.}
    \label{Fig:OTM_combined_NN}
\end{figure}

\begin{figure}[t]
    \centering
    \includegraphics[width=\linewidth]{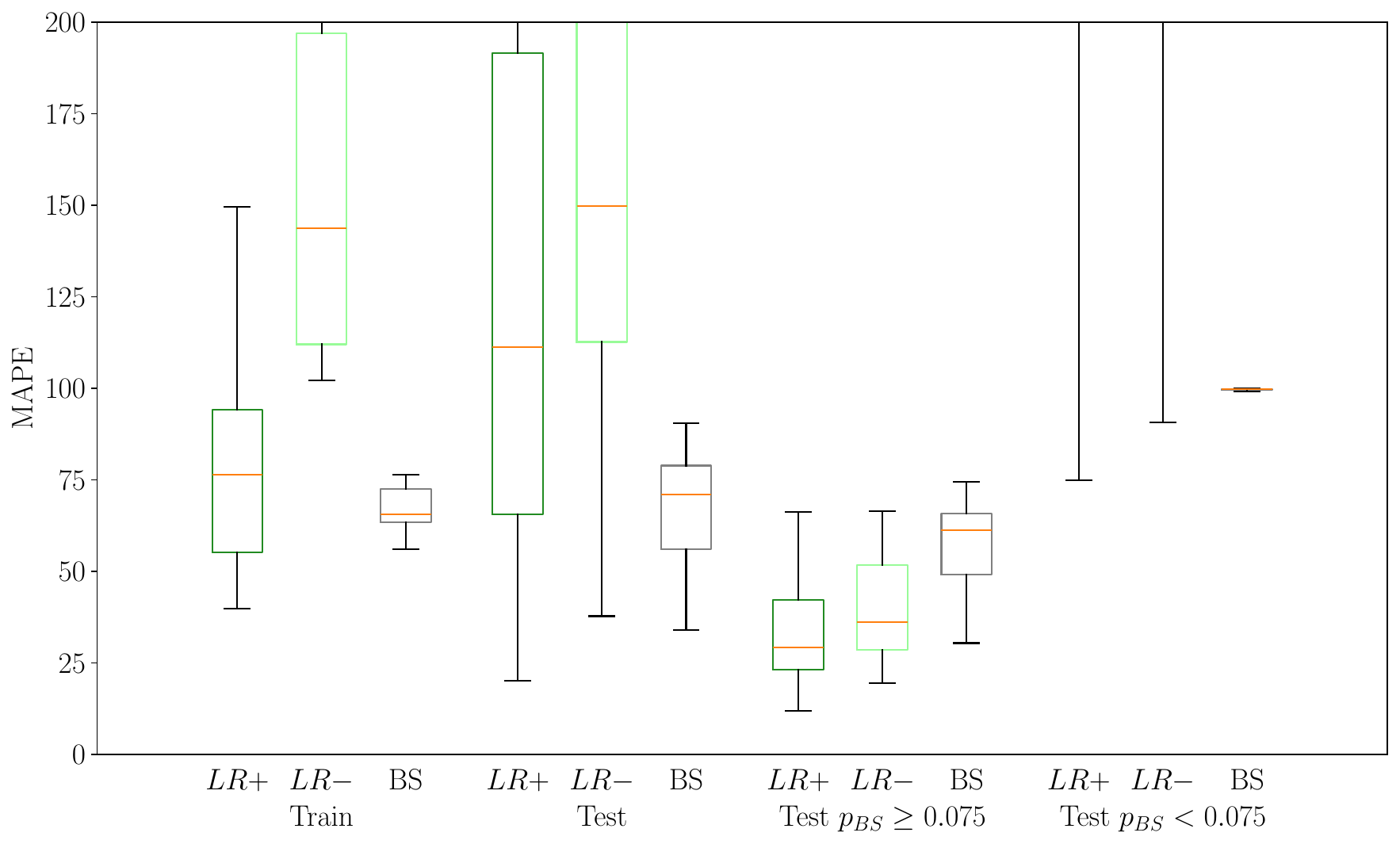}
    \includegraphics[width=\linewidth]{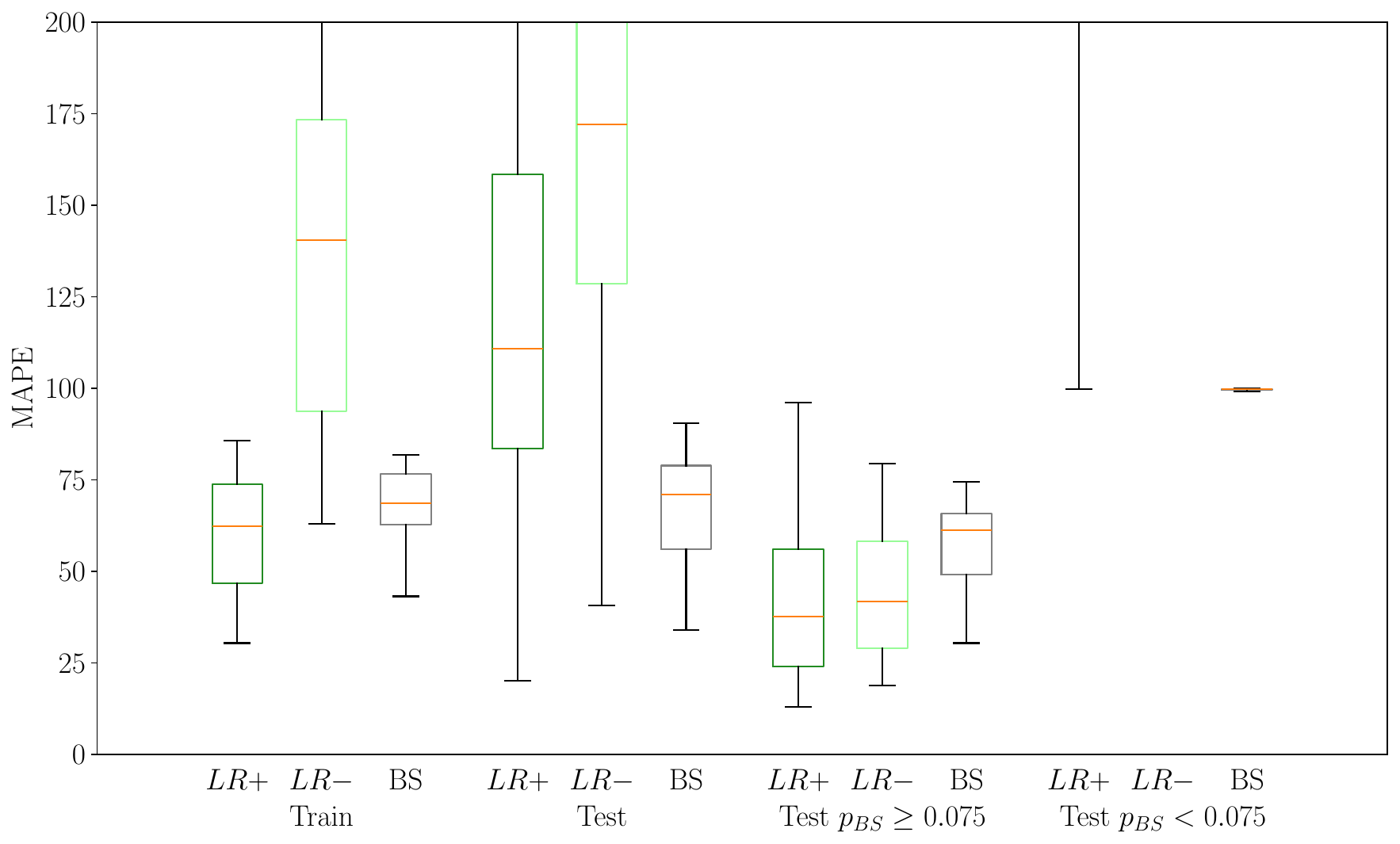}
    \caption{Boxplots of the mean absolute percentage error (MAPE) for OTM options under two data-splitting schemas. The top panel corresponds to the expanding window schema, while the bottom panel corresponds to the rolling window schema. In both panels, the data is grouped by subsets: the training set, the test set, the test set filtered by Black-Scholes price ($p_{\text{BS}} \geq 0.075$), and the test set filtered by $p_{\text{BS}} < 0.075$. Each group contains three boxplots, corresponding to the LR trained with Black-Scholes information (LR+), the LR model trained without Black-Scholes information (LR), and the Black-Scholes model (BS) itself. The exact numerical values corresponding to the expanding and rolling window results are presented in Tables~\ref{Tab:OTM_expanding_LR} and \ref{Tab:OTM_rolling_LR}, respectively.}
    \label{Fig:OTM_combined_LR}
\end{figure}

\begin{figure}[t]
    \centering
    \includegraphics[width=\linewidth]{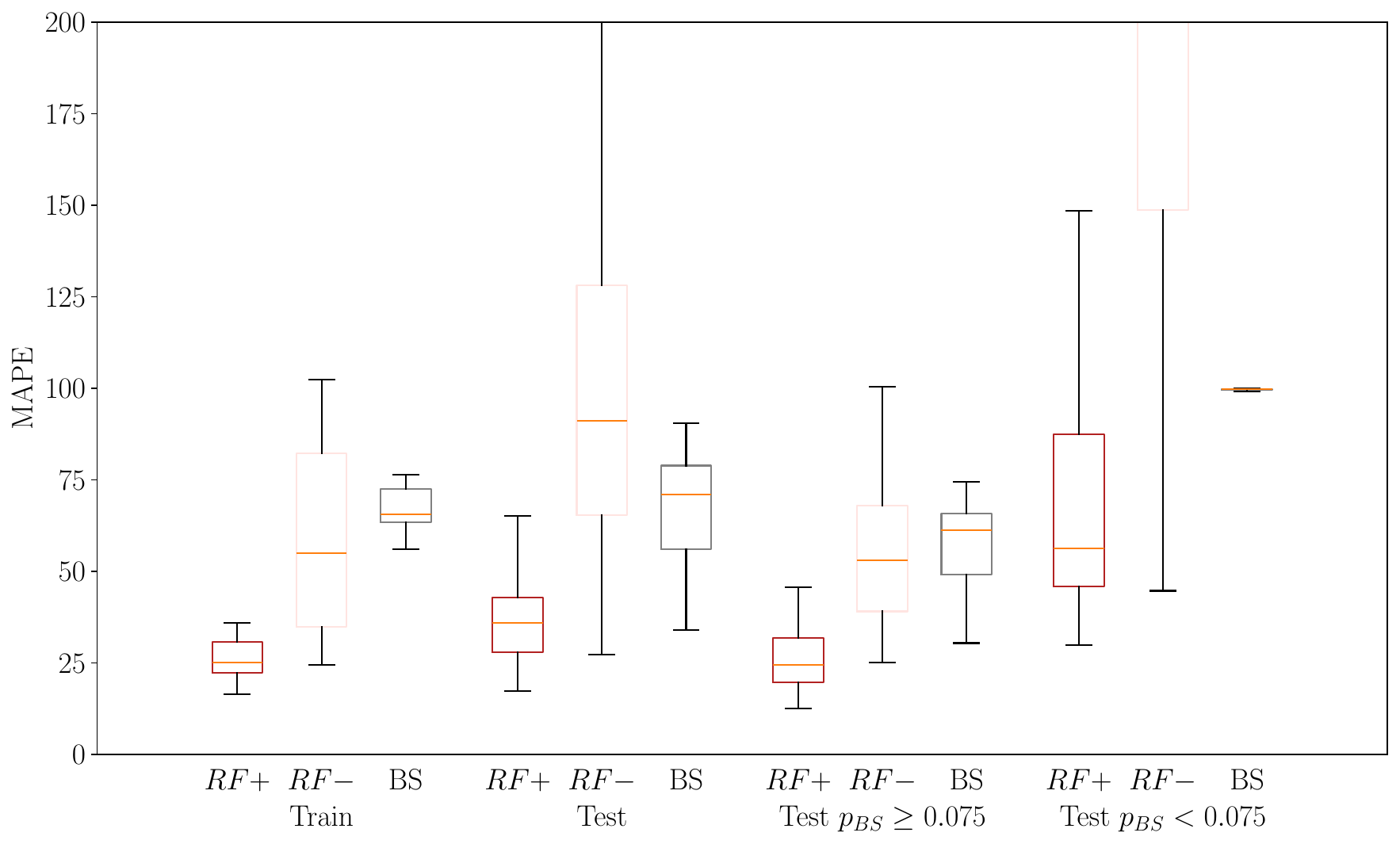}
    \includegraphics[width=\linewidth]{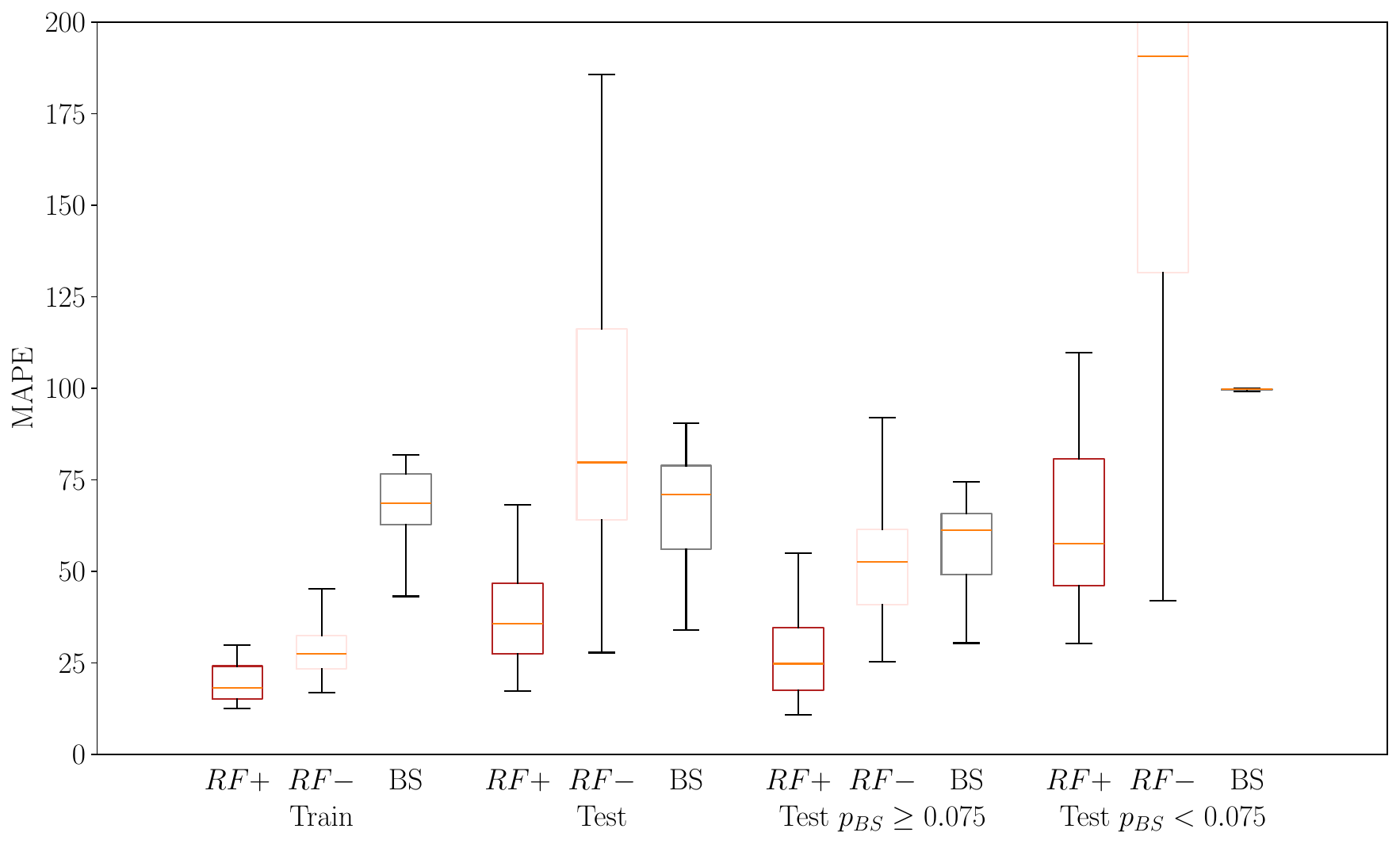}
    \caption{Boxplots of the mean absolute percentage error (MAPE) for OTM options under two data-splitting schemas. The top panel corresponds to the expanding window schema, while the bottom panel corresponds to the rolling window schema. In both panels, the data is grouped by subsets: the training set, the test set, the test set filtered by Black-Scholes price ($p_{\text{BS}} \geq 0.075$), and the test set filtered by $p_{\text{BS}} < 0.075$. Each group contains three boxplots, corresponding to the NN trained with Black-Scholes information (RF+), the NN trained without Black-Scholes information (RF-), and the Black-Scholes model (BS) itself. The exact numerical values corresponding to the expanding and rolling window results are presented in Tables.~\ref{Tab:OTM_expanding_RF} and \ref{Tab:OTM_rolling_RF}, respectively.}
    \label{Fig:OTM_combined_RF}
\end{figure}

Figures~\ref{Fig:OTM_combined_NN}–\ref{Fig:OTM_combined_RF} present the distribution of MAPE values across all training periods and data-splitting strategies for each model using boxplots, allowing us to analyze the variability in out-of-sample performance. Each figure pertains to a single model—NN, LR, or RF—and contains four groups of boxplots: one for the full train-test setup and three for the segmented test sets, based on different thresholds of the BS price. Within each group, we show results for models trained with and without the BS feature, alongside the BS benchmark, which remains constant across all figures.

The key takeaway from these figures is that including BS in the input space generally reduces the average MAPE compared to models trained without it. The only exception occurs in the NN case, where adding BS improves average performance but increases standard deviation, indicating greater variability in the results across training windows. 

These plots confirm the earlier finding gained from the time-series evolution of MAPE: the largest contribution to the overall error originates from low-priced put options, as evident from the boxplots showing high variance. Finally, when analyzing the segmented test set where option prices exceed the selected threshold, RF and NN achieve comparable performance when BS is included as a feature. However, NN demonstrates greater flexibility, achieving similar results even without BS, a flexibility not matched by RF.

\FloatBarrier

\FloatBarrier

\FloatBarrier

\subsection{Segmentation by Moneyness}

The insights from the previous section highlight that a significant portion of the MAPE originates from low-priced OTM options across all trained models, including the baseline BS model. To further investigate this effect, at each period, we segment the test dataset into four moneyness intervals, decreasing as the options move further from 1. Specifically, we define the segments as follows: $ \frac{S}{K} \in (1, 1.1] $, $ \frac{S}{K} \in (1.1, 1.2] $, $ \frac{S}{K} \in (1.2, 1.3] $, and $ \frac{S}{K} \in (1.3, 1.5] $.

\begin{figure}[t]
    \centering
    \includegraphics[width=0.78\textwidth]{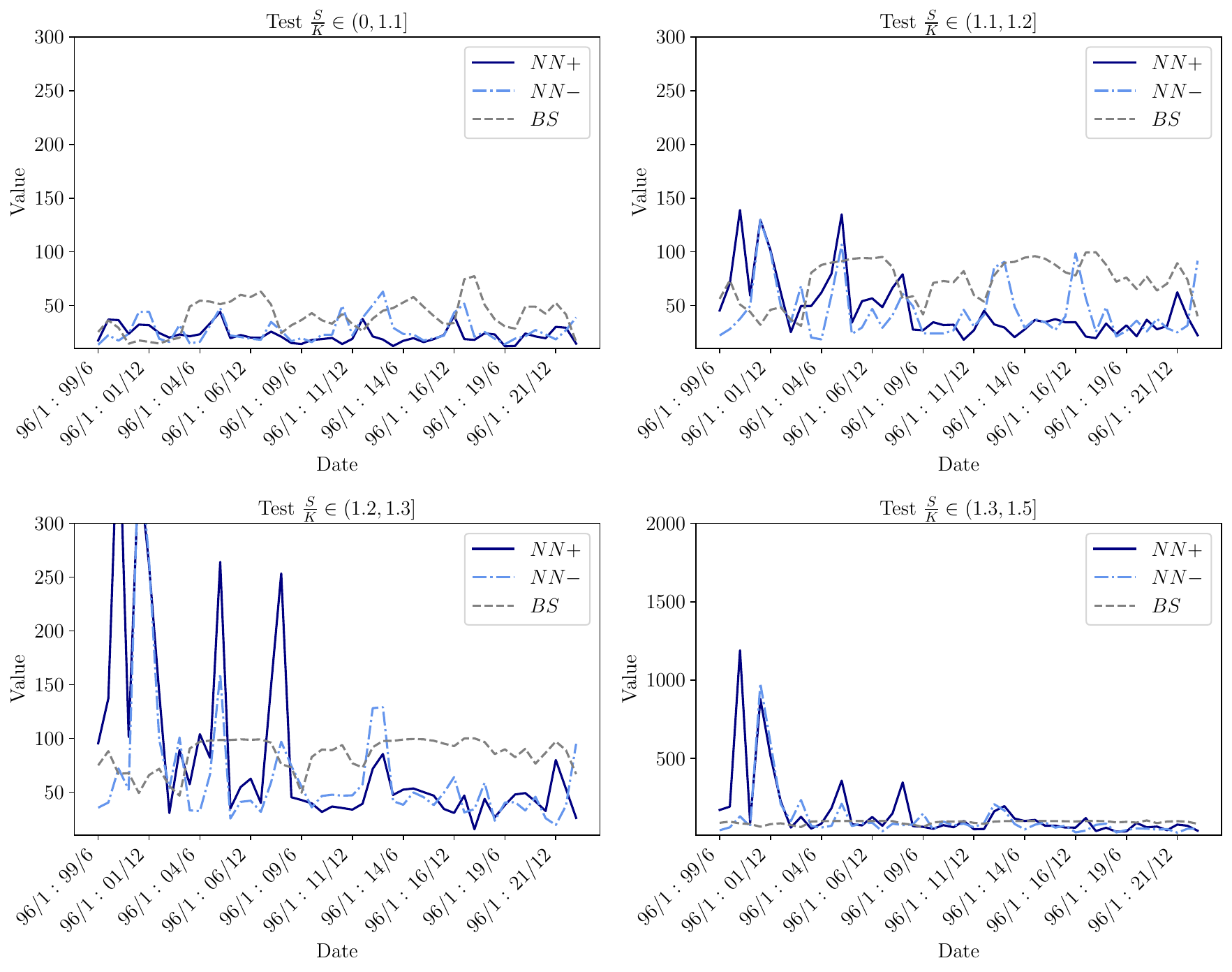}
    \caption{Evolution over time of the MAPE values for OTM options grouped by moneyness ranges trained using an expanding window schema with the NN model. The values correspond to Tab.~\ref{Tab:OTM_moneyness_expanding_NN}.}
    \label{Fig:OTM_series_moneyness_expanding_NN}
\end{figure}

\begin{figure}[t]
    \centering
    \includegraphics[width=0.78\textwidth]{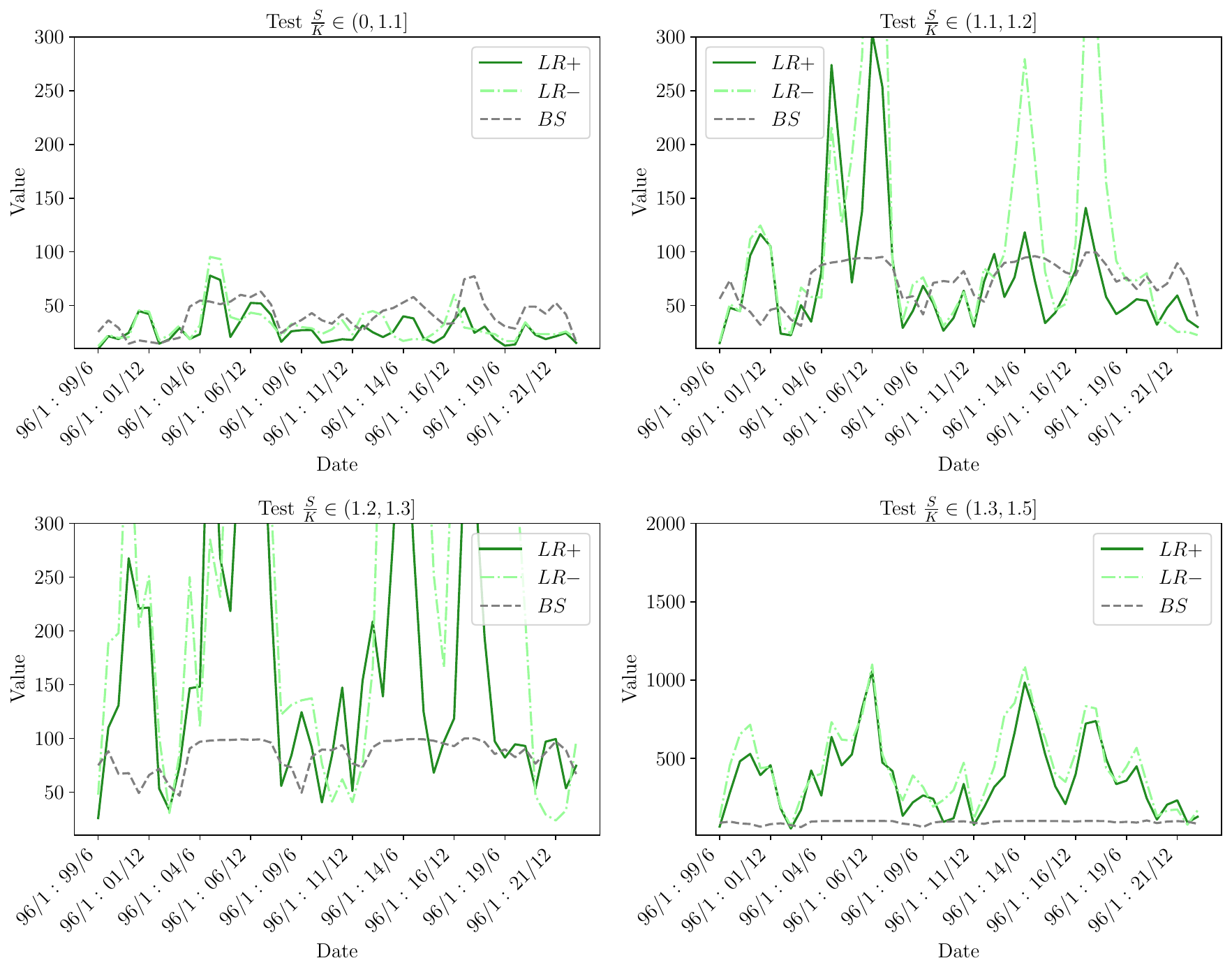}
    \caption{Evolution over time of the MAPE values for OTM options grouped by moneyness ranges trained using an expanding window schema with the LR model. The values correspond to Tab.~\ref{Tab:OTM_moneyness_expanding_LR}.}
    \label{Fig:OTM_series_moneyness_expanding_LR}
\end{figure}

\begin{figure}[t]
    \centering
    \includegraphics[width=0.78\textwidth]{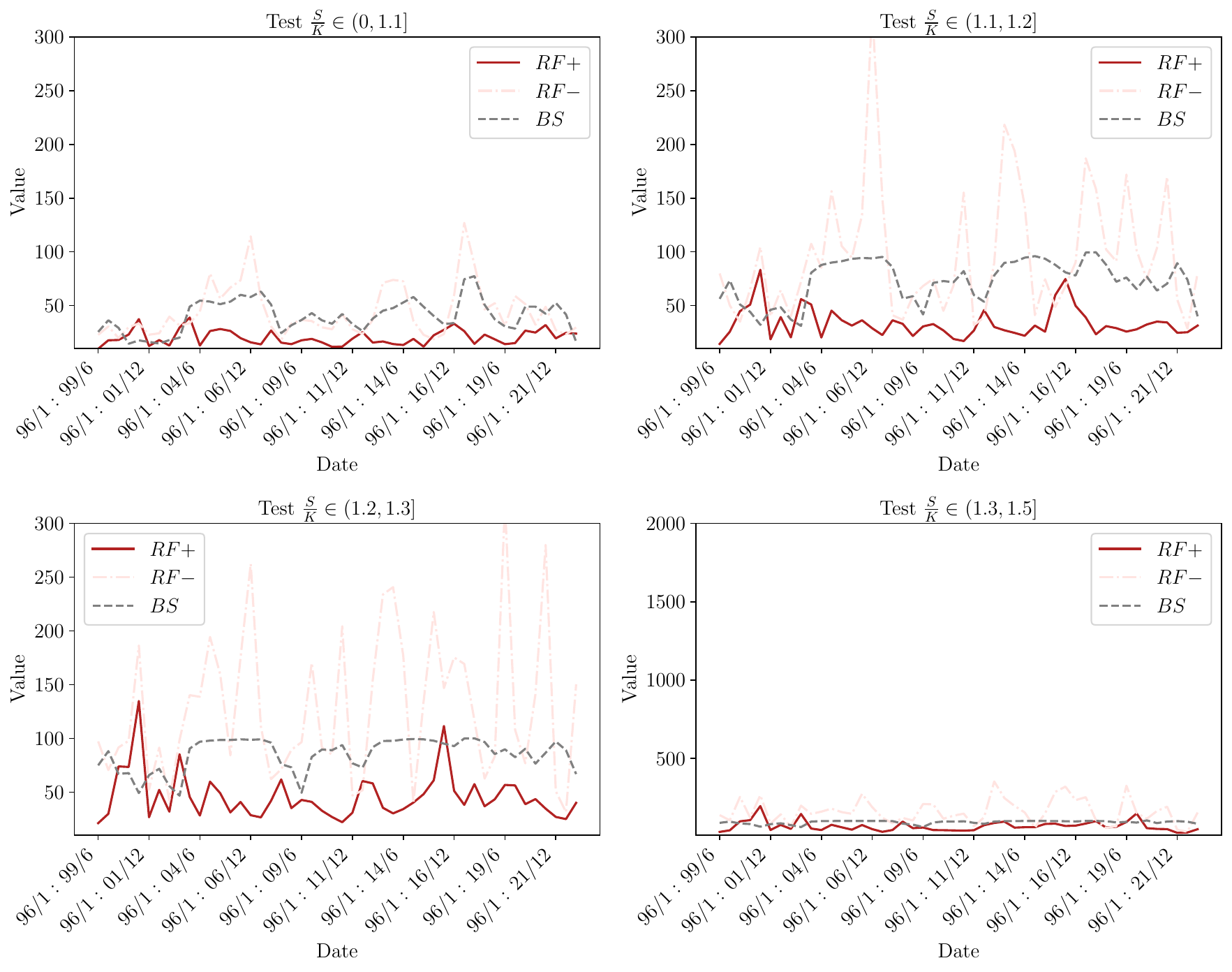}
    \caption{Evolution over time of the MAPE values for OTM options grouped by moneyness ranges trained using an expanding window schema with the RF model. The values correspond to Tab.~\ref{Tab:OTM_moneyness_expanding_RF}.}
    \label{Fig:OTM_series_moneyness_expanding_RF}
\end{figure}

\begin{figure}[t]
    \centering
    \includegraphics[width=0.78\textwidth]{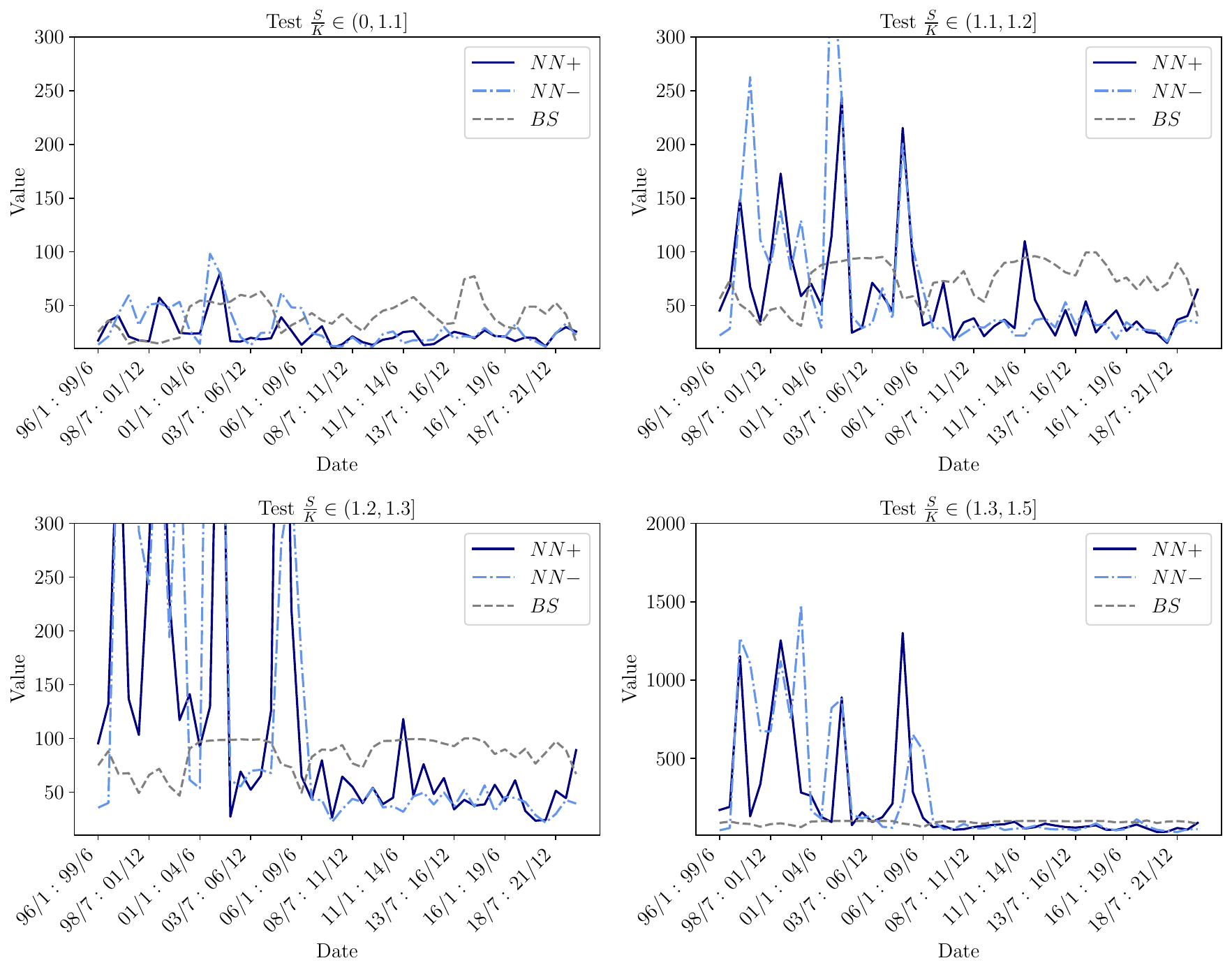}
    \caption{Evolution over time of the MAPE values for OTM options grouped by moneyness ranges trained using a rolling window schema with the NN model. The values correspond to Tab.~\ref{Tab:OTM_moneyness_rolling_NN}.}
    \label{Fig:OTM_series_moneyness_rolling_NN}
\end{figure}

\begin{figure}[t]
    \centering
    \includegraphics[width=0.78\textwidth]{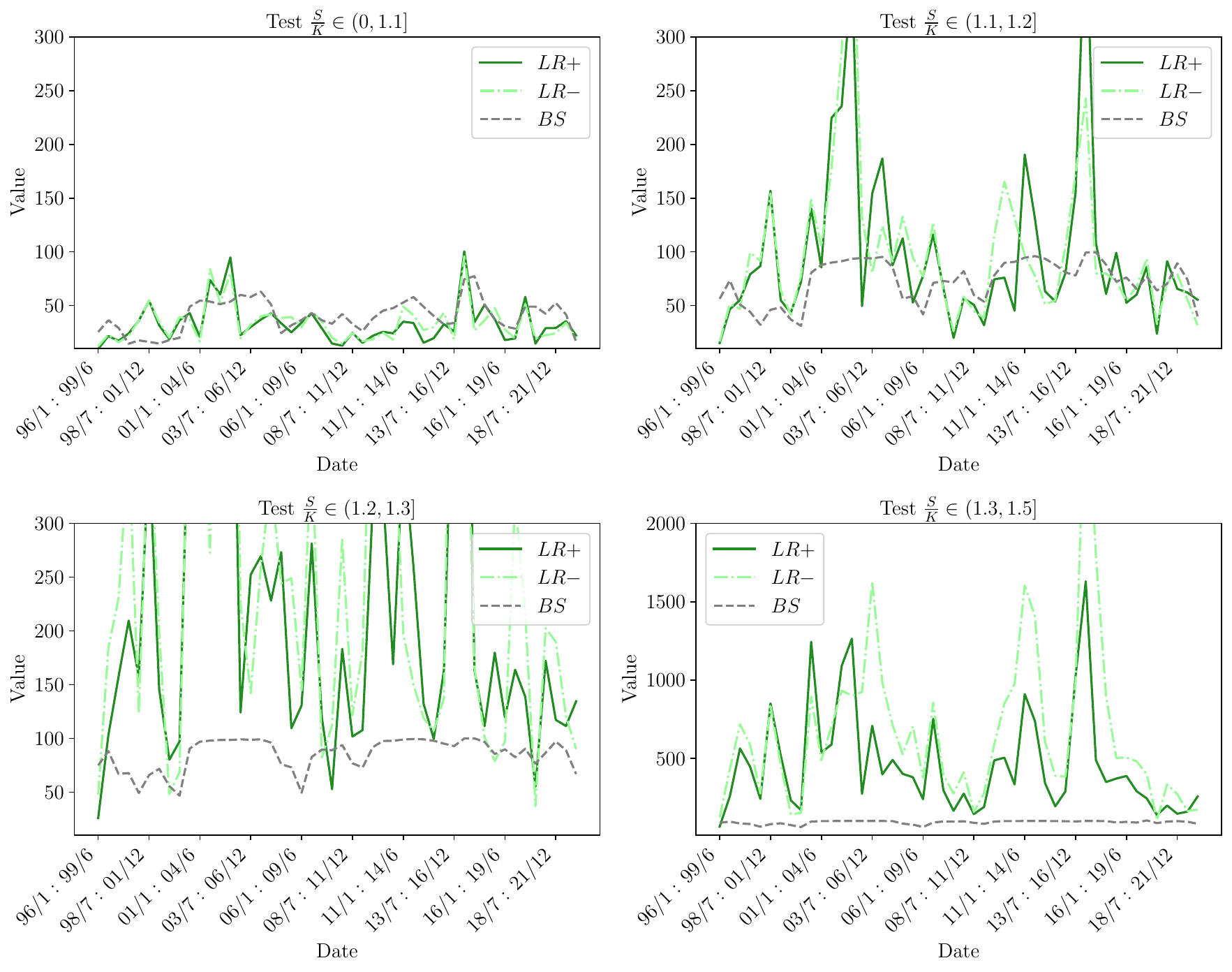}
    \caption{Evolution over time of the MAPE values for OTM options grouped by moneyness ranges trained using a rolling window schema with the LR model. The values correspond to Tab.~\ref{Tab:OTM_moneyness_rolling_LR}.}
    \label{Fig:OTM_series_moneyness_rolling_LR}
\end{figure}

\begin{figure}[t]
    \centering
    \includegraphics[width=0.78\textwidth]{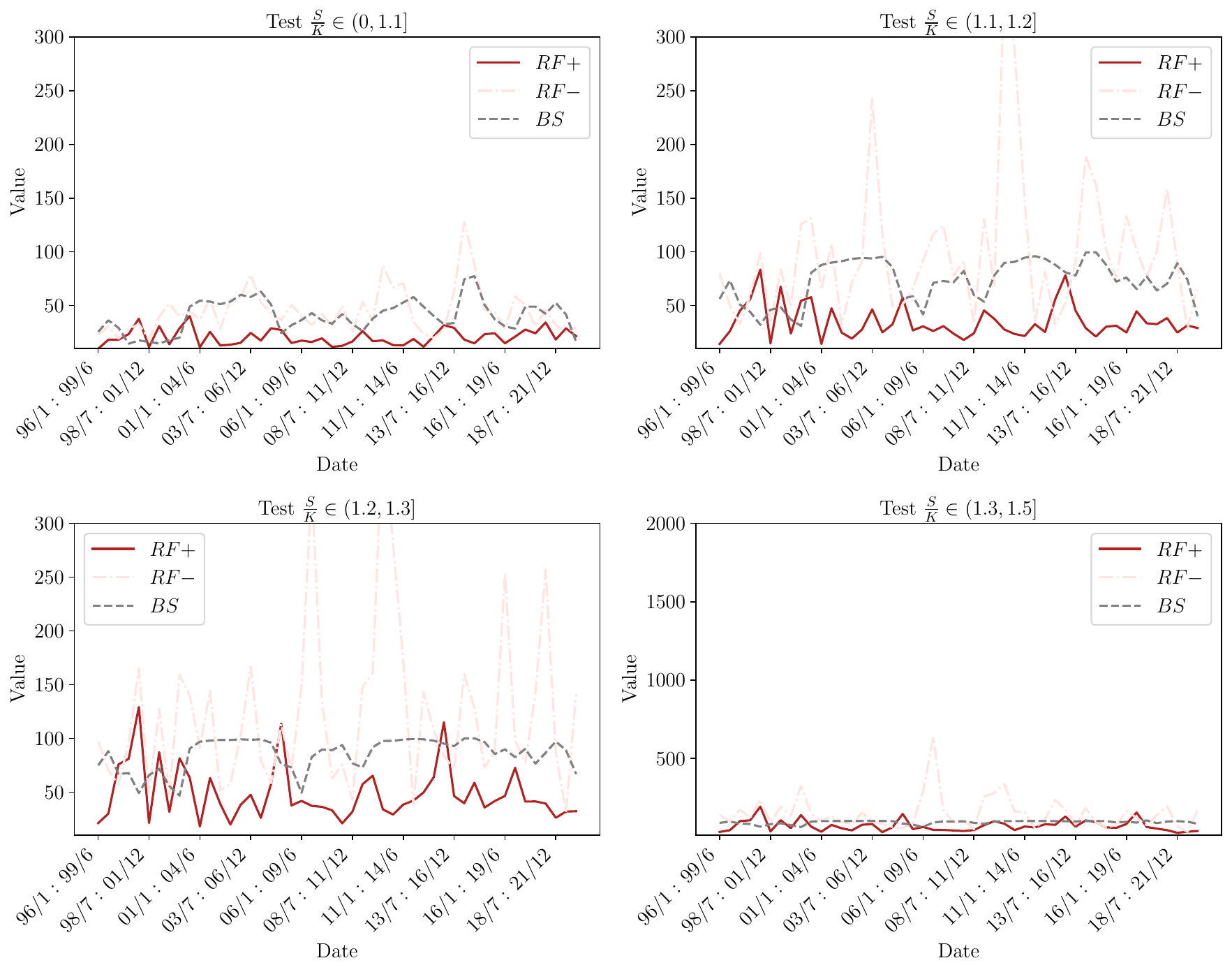}
    \caption{Evolution over time of the MAPE values for OTM options grouped by moneyness ranges trained using a rolling window schema with the RF model. The values correspond to Tab.~\ref{Tab:OTM_moneyness_rolling_RF}.}
    \label{Fig:OTM_series_moneyness_rolling_RF}
\end{figure}

Figures~\ref{Fig:OTM_series_moneyness_expanding_NN}–\ref{Fig:OTM_series_moneyness_expanding_RF} and Figures~\ref{Fig:OTM_series_moneyness_rolling_NN}–\ref{Fig:OTM_series_moneyness_rolling_RF} illustrate the MAPE evolution under the expanding and rolling window approaches, respectively. As expected, the MAPE increases as options become deeper OTM. This pattern holds consistently across all models. NN exhibits slightly higher instability across different test periods when trained using a rolling window but remains the top-performing model for near-the-money contracts ($S/K \in (1, 1.1]$).

\begin{figure}[t]
    \centering
    \includegraphics[width=\linewidth]{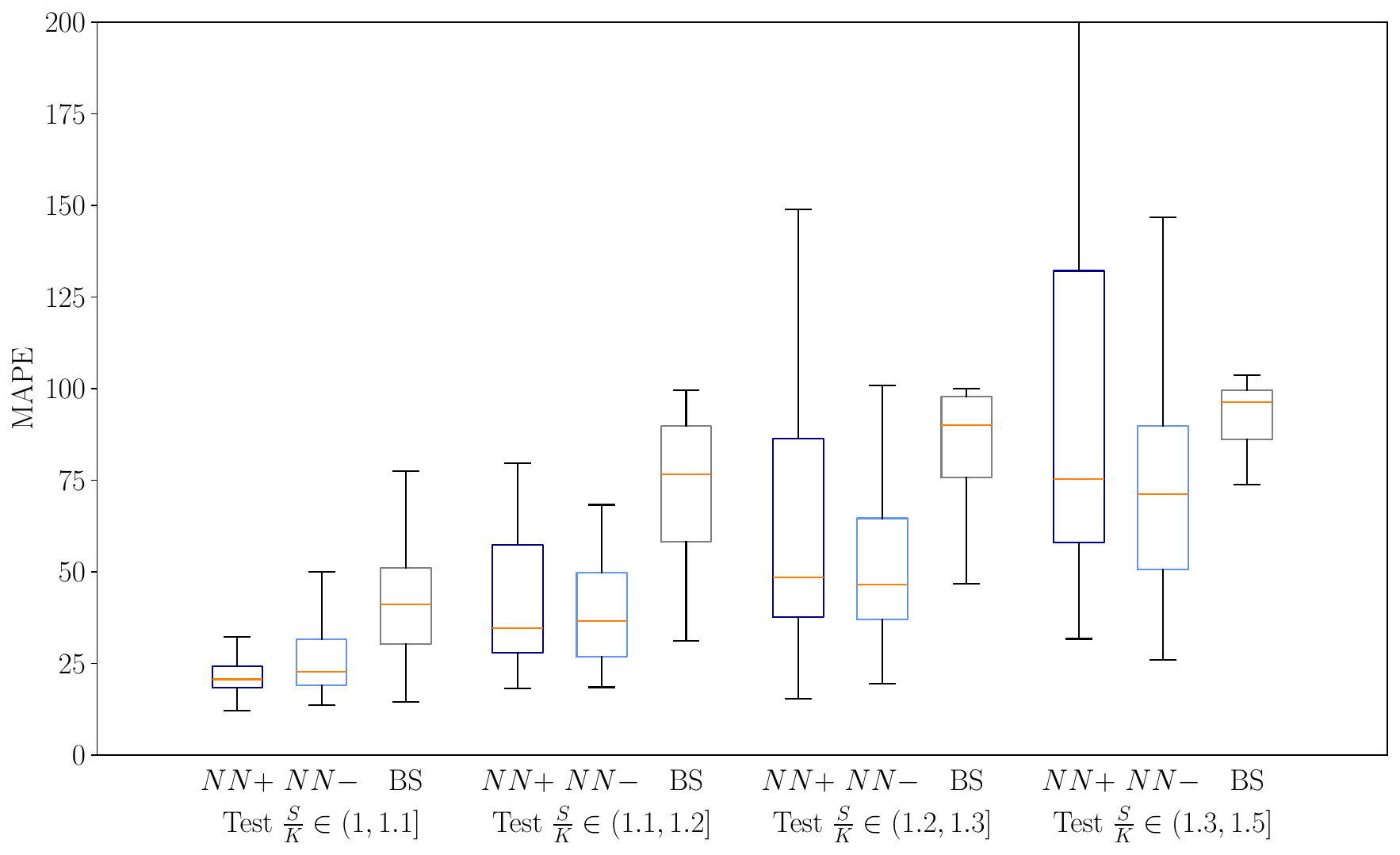}
    \includegraphics[width=\linewidth]{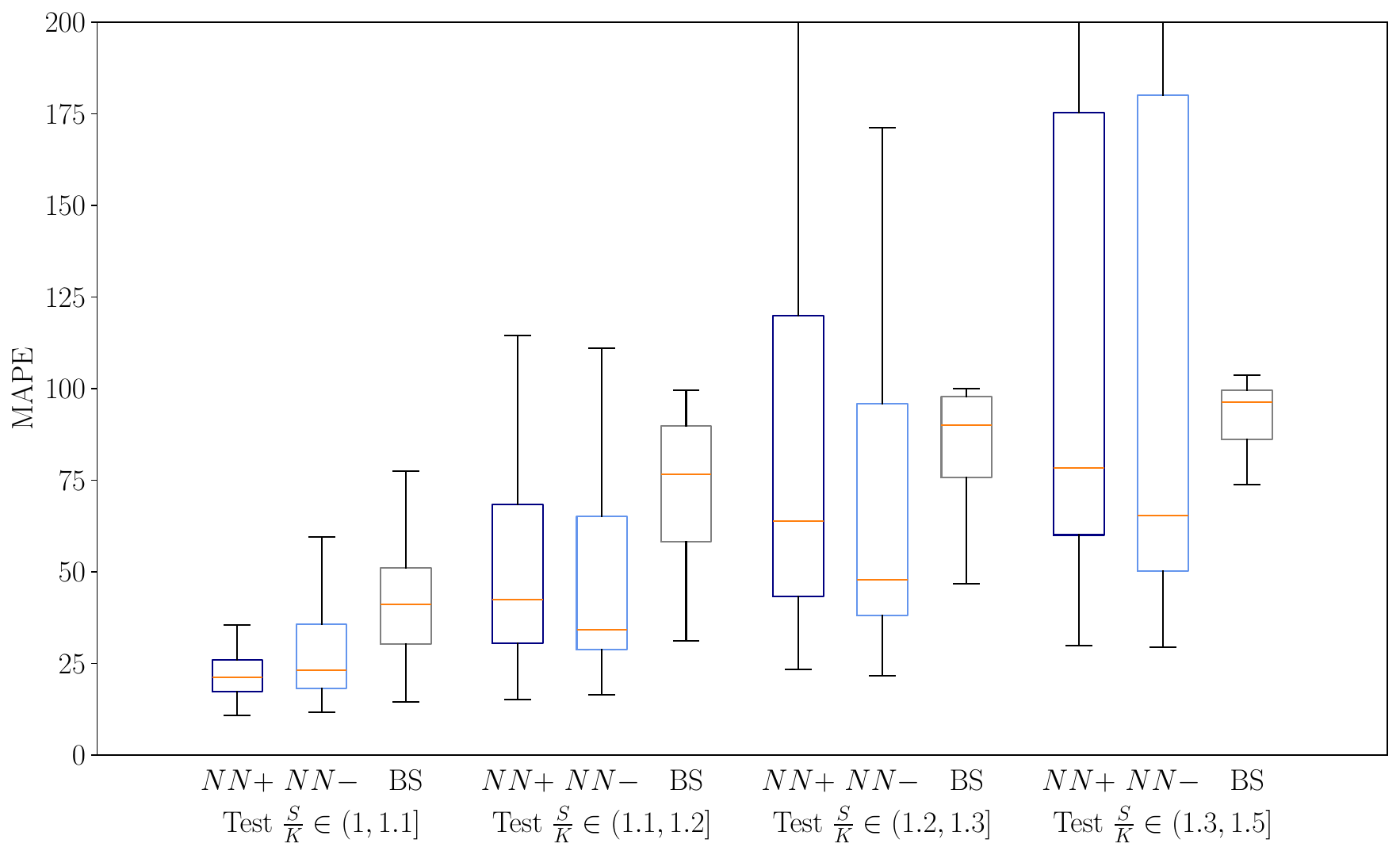}
    \caption{Boxplots of the mean absolute percentage error (MAPE) for OTM options grouped by moneyness ranges under two data-splitting schemas. The top panel corresponds to the expanding window schema, while the bottom panel corresponds to the rolling window schema. The test set is divided into subsets based on moneyness ranges. Each group contains three boxplots, corresponding to the NN trained with Black-Scholes information (NN+), the NN trained without Black-Scholes information (NN-), and the Black-Scholes model (BS) itself. The exact numerical values corresponding to the expanding and rolling window results are presented in Tables.~\ref{Tab:OTM_moneyness_expanding_NN} and \ref{Tab:OTM_moneyness_rolling_NN}, respectively.}
    \label{Fig:OTM_moneyness_combined_NN}
\end{figure}

\begin{figure}[t]
    \centering
    \includegraphics[width=\linewidth]{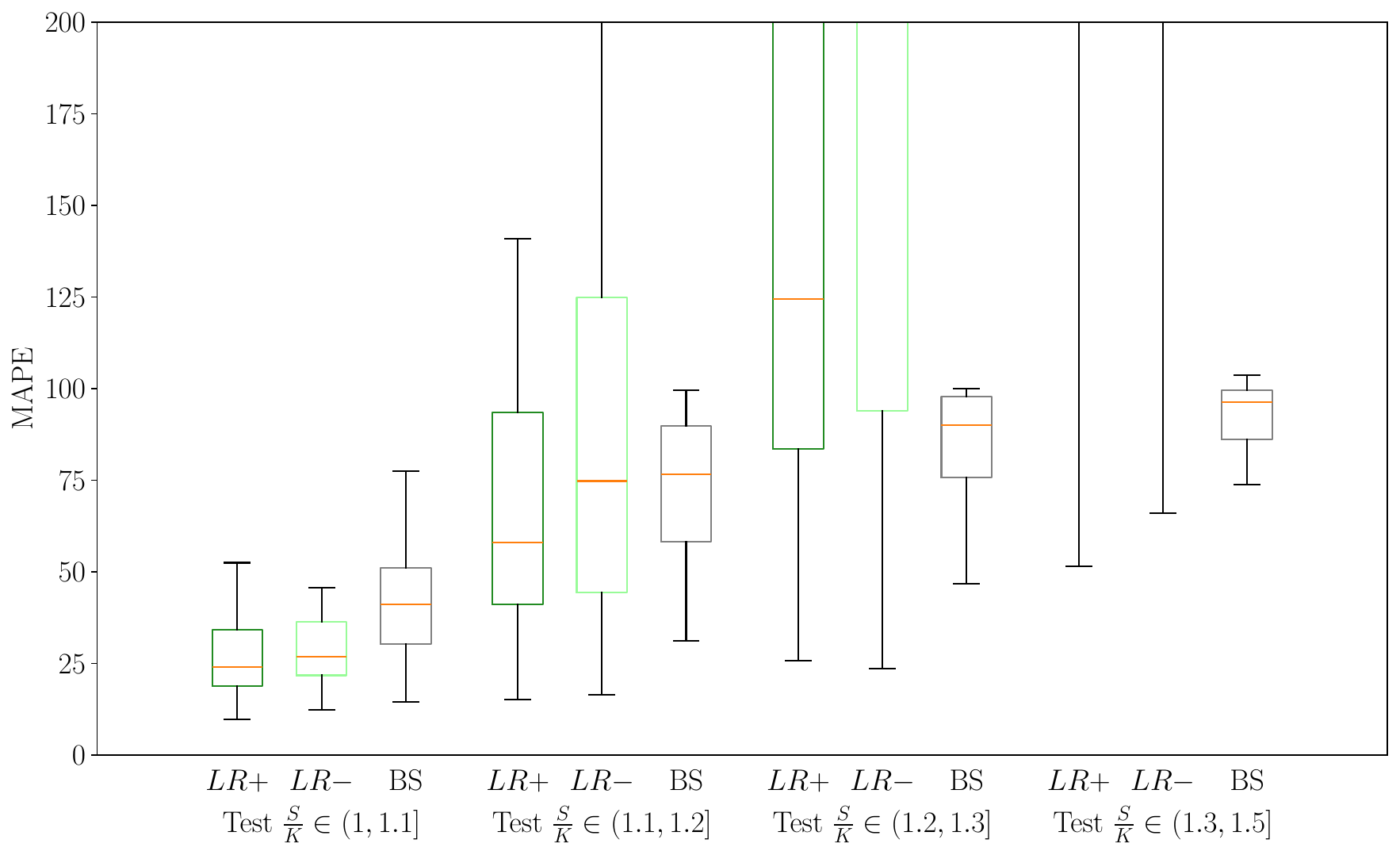}
    \includegraphics[width=\linewidth]{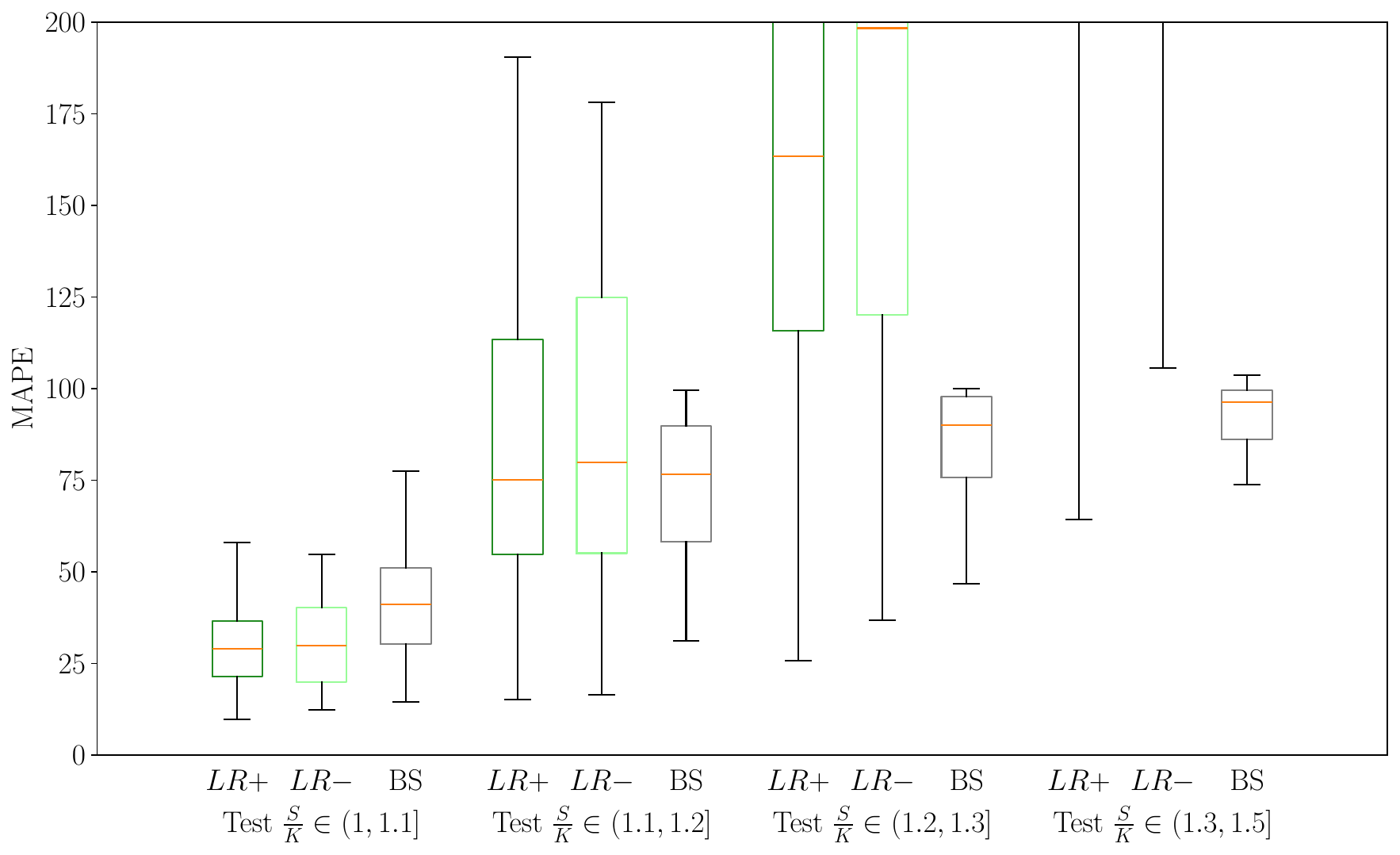}
    \caption{Boxplots of the mean absolute percentage error (MAPE) for OTM options grouped by moneyness ranges under two data-splitting schemas. The top panel corresponds to the expanding window schema, while the bottom panel corresponds to the rolling window schema. The test set is divided into subsets based on moneyness ranges. Each group contains three boxplots, corresponding to the LR model trained with Black-Scholes information (LR+), the LR model trained without Black-Scholes information (LR-), and the Black-Scholes model (BS) itself. The exact numerical values corresponding to the expanding and rolling window results are presented in Tables.~\ref{Tab:OTM_moneyness_expanding_LR} and \ref{Tab:OTM_moneyness_rolling_LR}, respectively.}
    \label{Fig:OTM_moneyness_combined_LR}
\end{figure}

\begin{figure}[t]
    \centering
    \includegraphics[width=\linewidth]{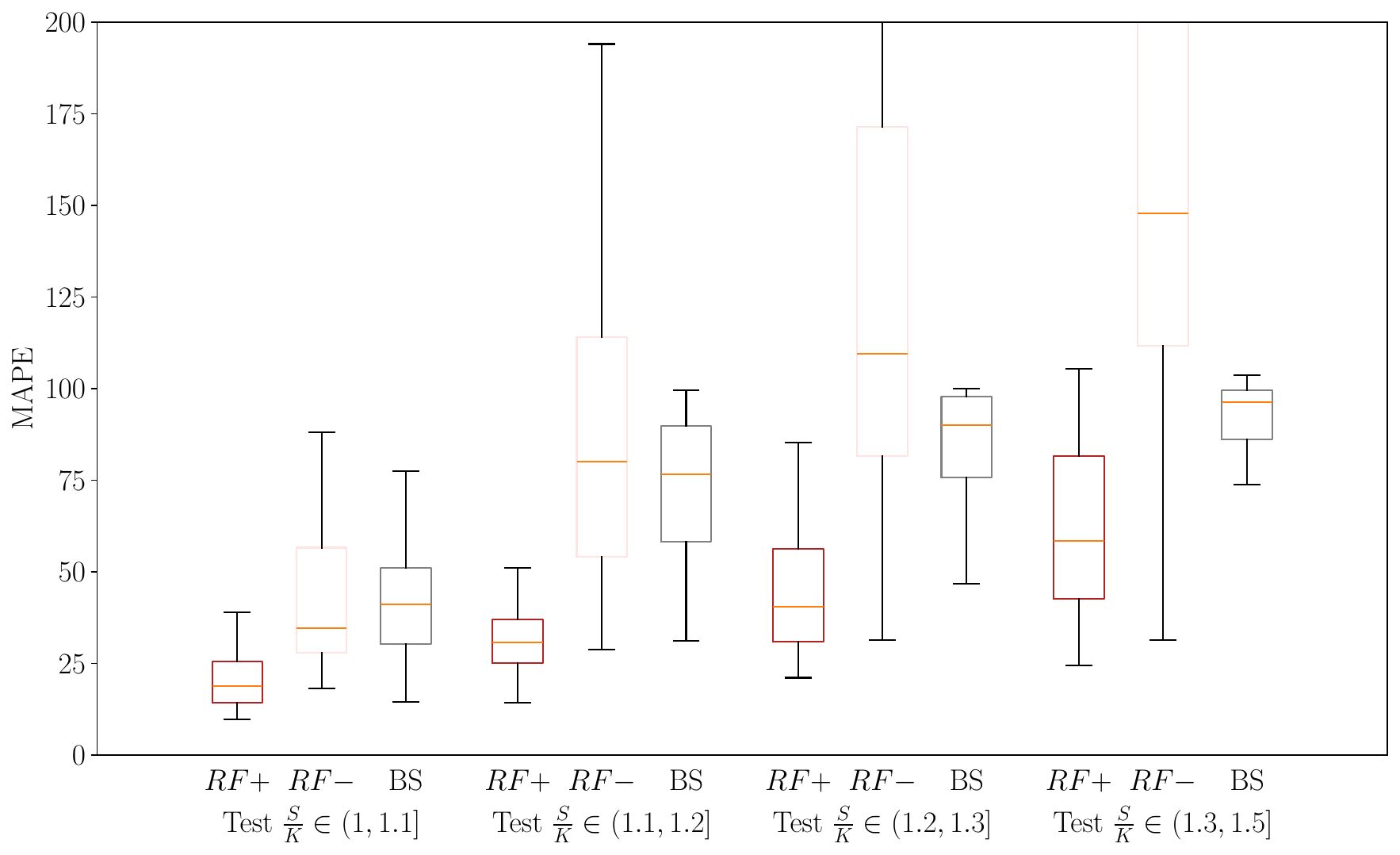}
    \includegraphics[width=\linewidth]{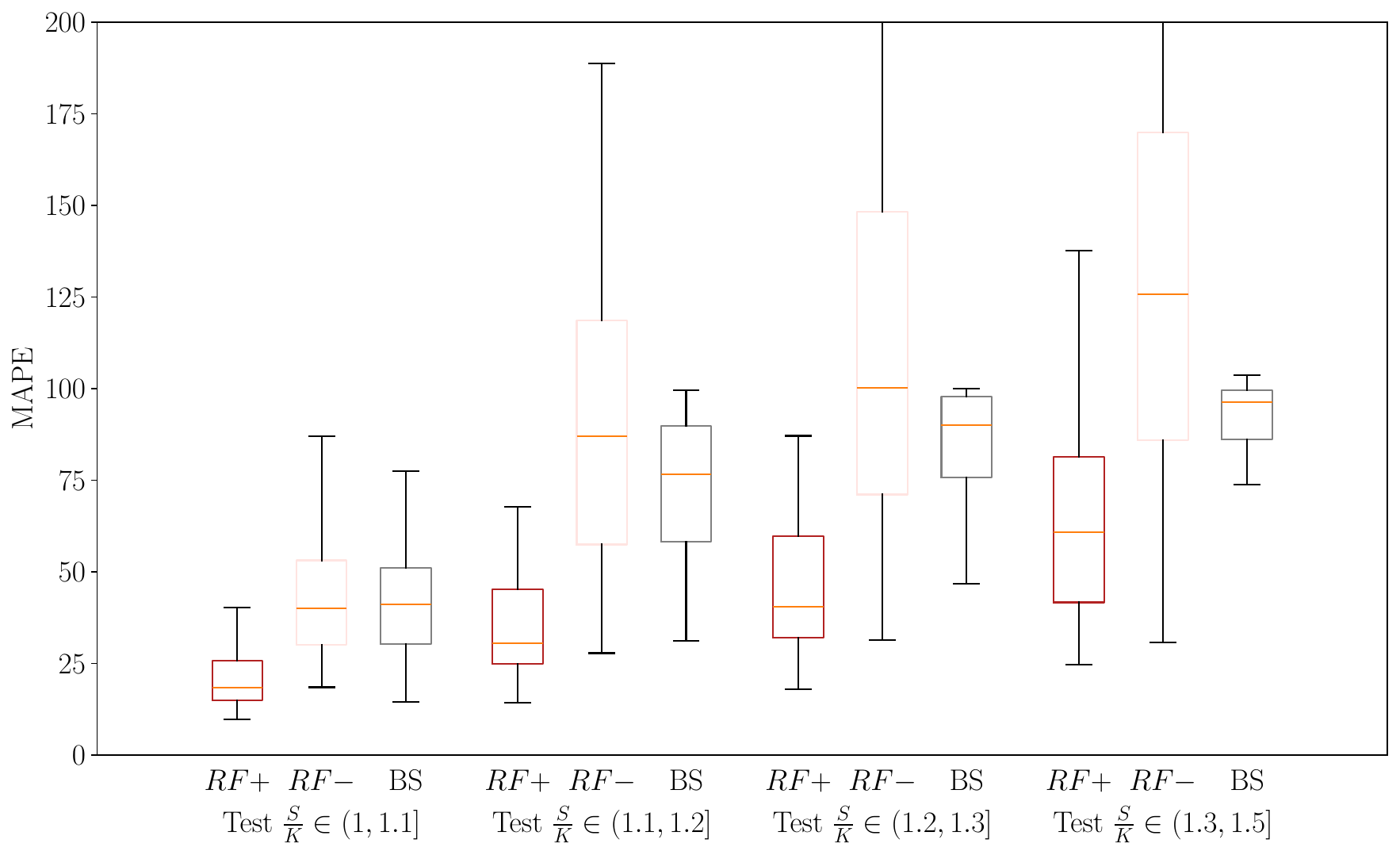}
    \caption{Boxplots of the mean absolute percentage error (MAPE) for OTM options grouped by moneyness ranges under two data-splitting schemas. The top panel corresponds to the expanding window schema, while the bottom panel corresponds to the rolling window schema. The test set is divided into subsets based on moneyness ranges. Each group contains three boxplots, corresponding to the RF model trained with Black-Scholes information (RF), the RF model trained without Black-Scholes information (RF-), and the Black-Scholes model (BS) itself. The exact numerical values corresponding to the expanding and rolling window results are presented in Tables.~\ref{Tab:OTM_moneyness_expanding_RF} and \ref{Tab:OTM_moneyness_rolling_RF}, respectively.}
    \label{Fig:OTM_moneyness_combined_RF}
\end{figure}

The discrepancy between RF models trained with and without BS persists. When BS is excluded, RF consistently underperforms, showing significantly higher MAPE compared to when BS is included. This effect is also evident in Figures~\ref{Fig:OTM_moneyness_combined_NN}, \ref{Fig:OTM_moneyness_combined_LR}, and \ref{Fig:OTM_moneyness_combined_RF}, which present the boxplots of the MAPE distribution for each moneyness segment for NN, LR, and RF, respectively.


\FloatBarrier

\FloatBarrier
\FloatBarrier

\FloatBarrier
\FloatBarrier
\FloatBarrier

\FloatBarrier

\FloatBarrier

\FloatBarrier

\subsection{In-the-Money Put Options}

\begin{figure}[t]
    \centering
    \begin{subfigure}[b]{0.78\textwidth}
        \includegraphics[width=\textwidth]{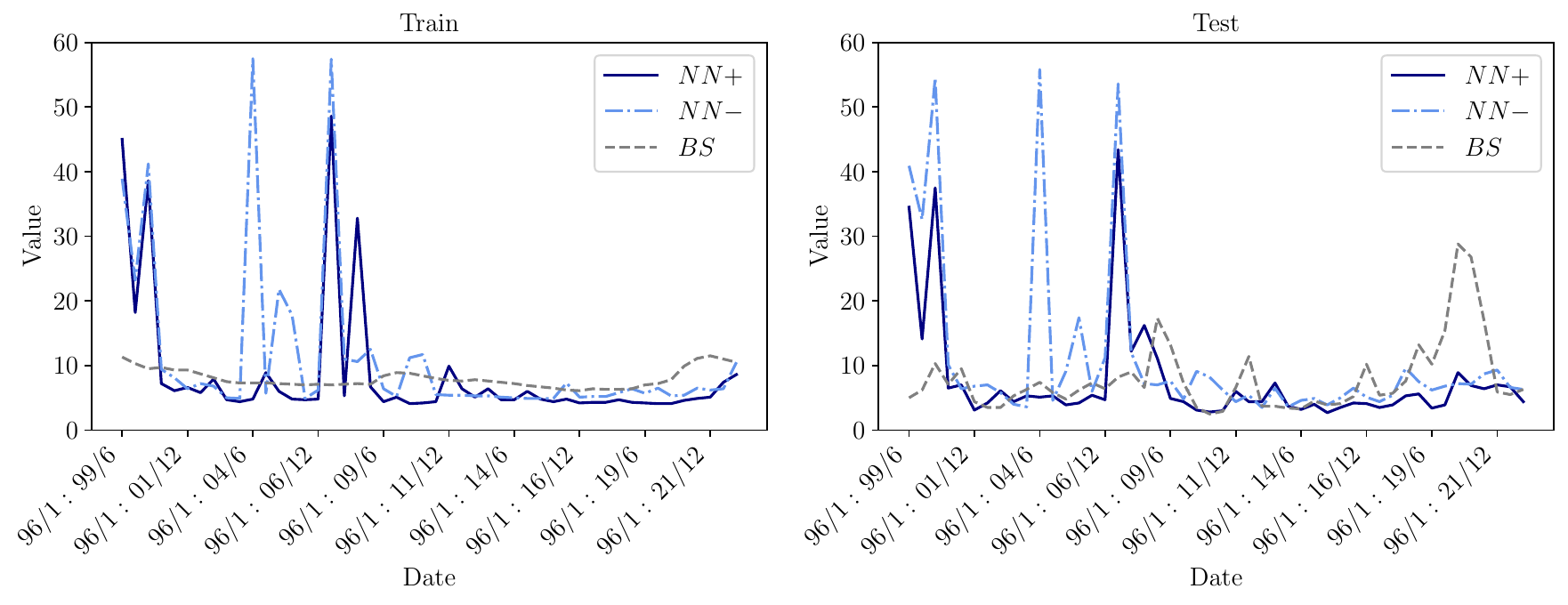}
    \end{subfigure}
    \\
    \begin{subfigure}[b]{0.78\textwidth}
        \includegraphics[width=\textwidth]{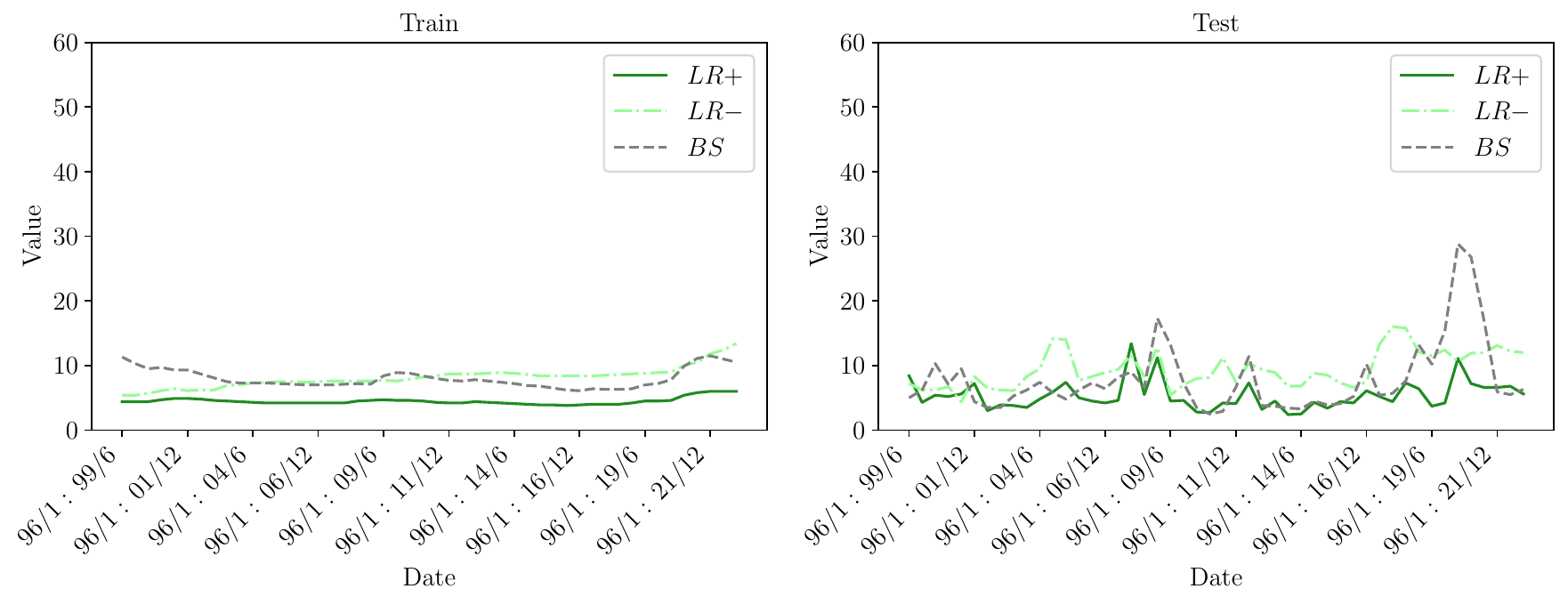}
    \end{subfigure}
    \\
    \begin{subfigure}[b]{0.78\textwidth}
        \includegraphics[width=\textwidth]{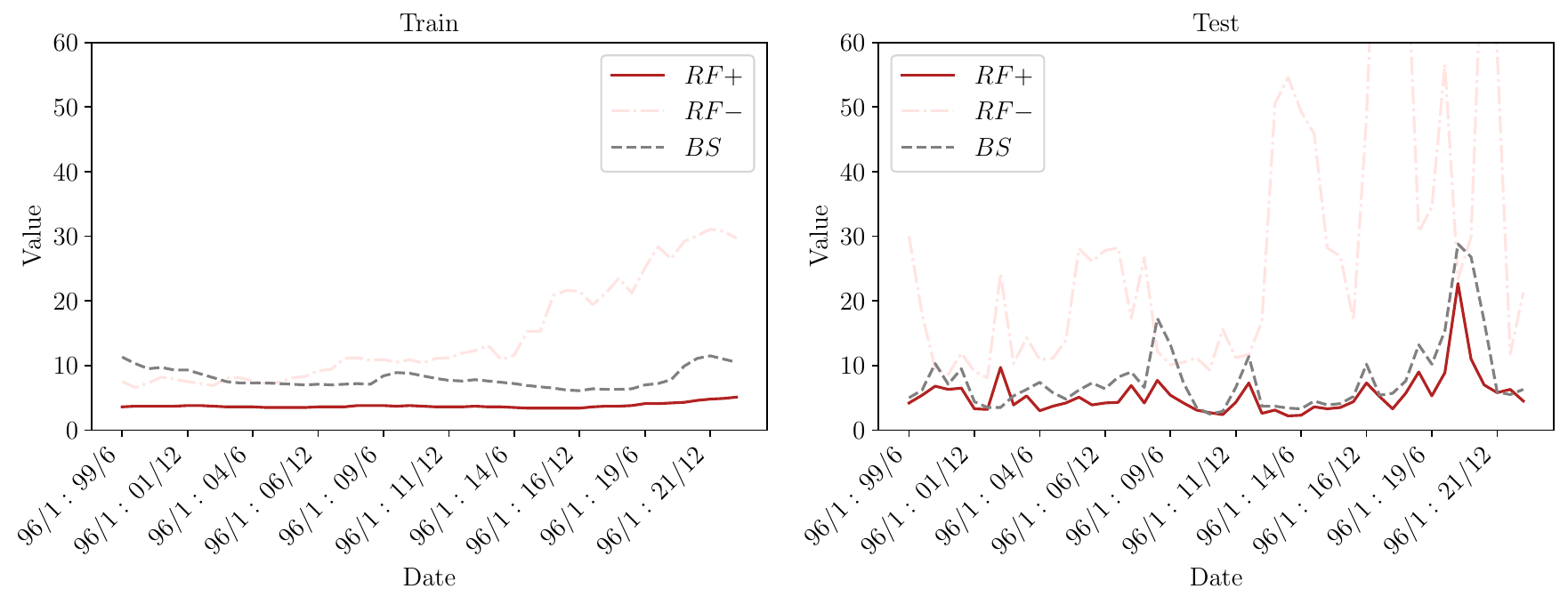}
    \end{subfigure}
    \caption{Evolution over time of the MAPE values for ITM options trained using an expanding window schema. The plots correspond to the values reported in Tables.~\ref{Tab:ITM_expanding_NN}, \ref{Tab:ITM_expanding_LR}, and \ref{Tab:ITM_expanding_RF} for the NN, LR, and RF models, respectively, from top to bottom.}
    \label{Fig:ITM_series_expanding}
\end{figure}

\begin{figure}[t]
    \centering
    \begin{subfigure}[b]{0.78\textwidth}
        \includegraphics[width=\textwidth]{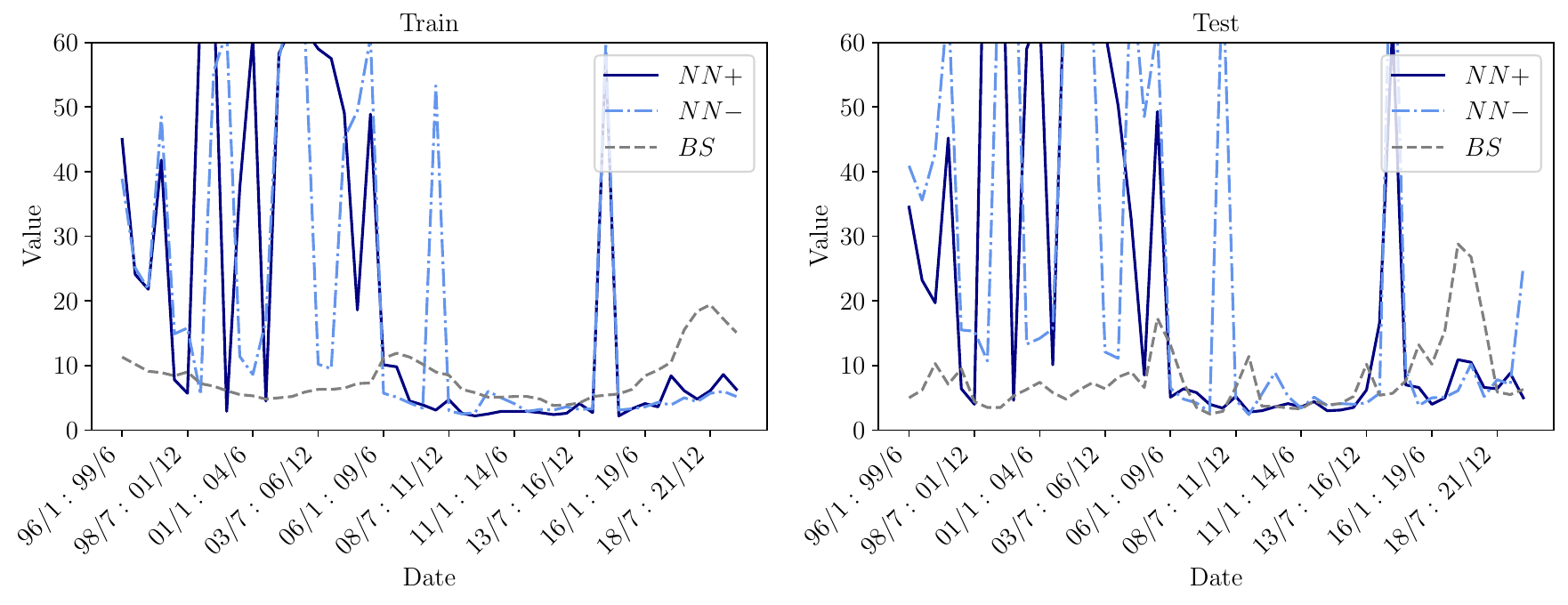}
    \end{subfigure}
    \\
    \begin{subfigure}[b]{0.78\textwidth}
        \includegraphics[width=\textwidth]{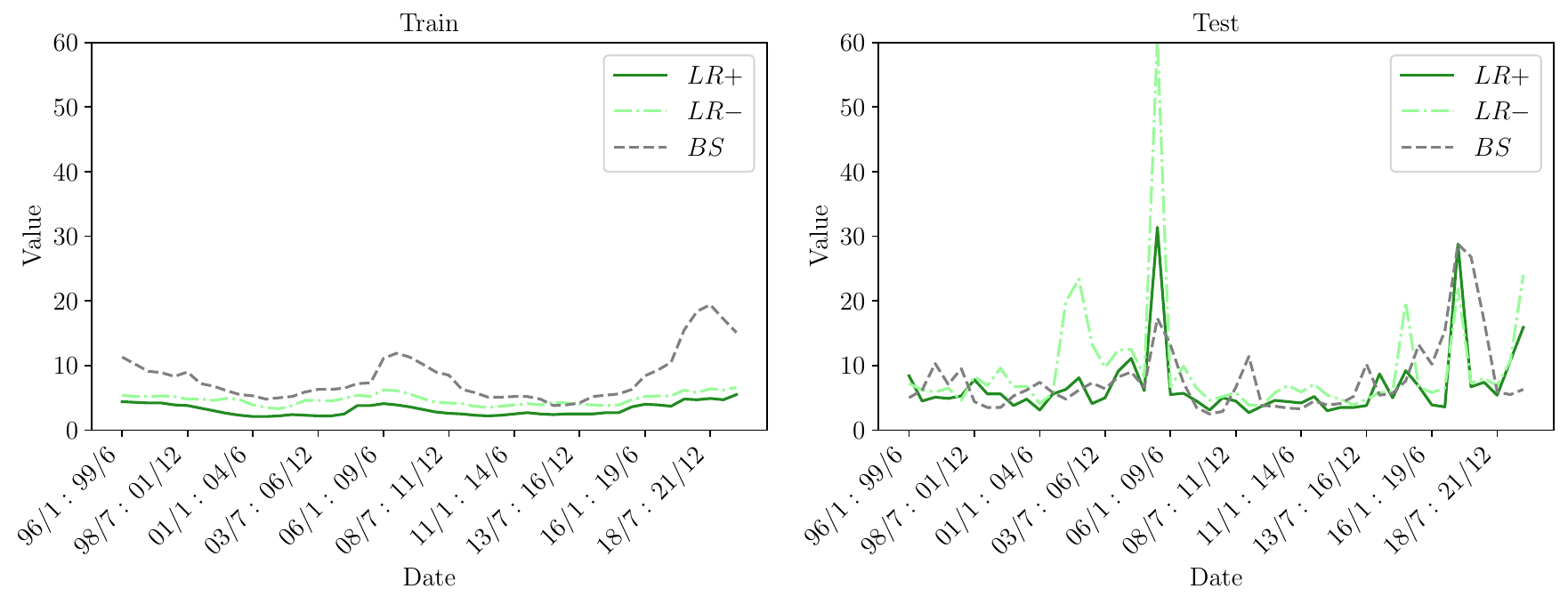}
    \end{subfigure}
    \\
    \begin{subfigure}[b]{0.78\textwidth}
        \includegraphics[width=\textwidth]{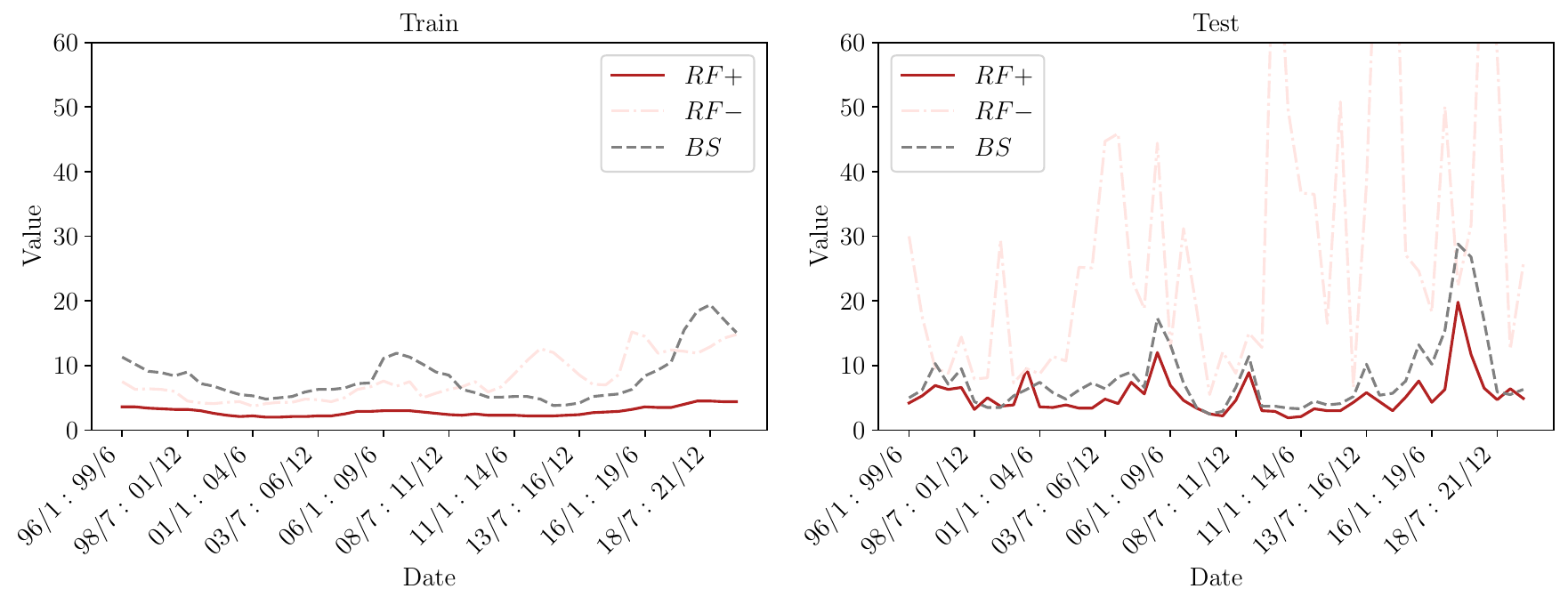}
    \end{subfigure}
    \caption{Evolution over time of the MAPE values for ITM options grouped by moneyness ranges trained using a rolling window schema. The plots correspond to the values reported in Tables.~\ref{Tab:ITM_rolling_NN}, \ref{Tab:ITM_rolling_LR}, and \ref{Tab:ITM_rolling_RF} for the NN, LR, and RF models, respectively, from top to bottom.}
    \label{Fig:ITM_series_rolling}
\end{figure}

The analysis of ITM options largely mirrors the findings for OTM contracts. Figures~\ref{Fig:ITM_series_expanding} and \ref{Fig:ITM_series_rolling} illustrate the MAPE dynamics over different out-of-sample periods for the three models.

\begin{figure}[t]
    \centering
    \includegraphics[width=\linewidth]{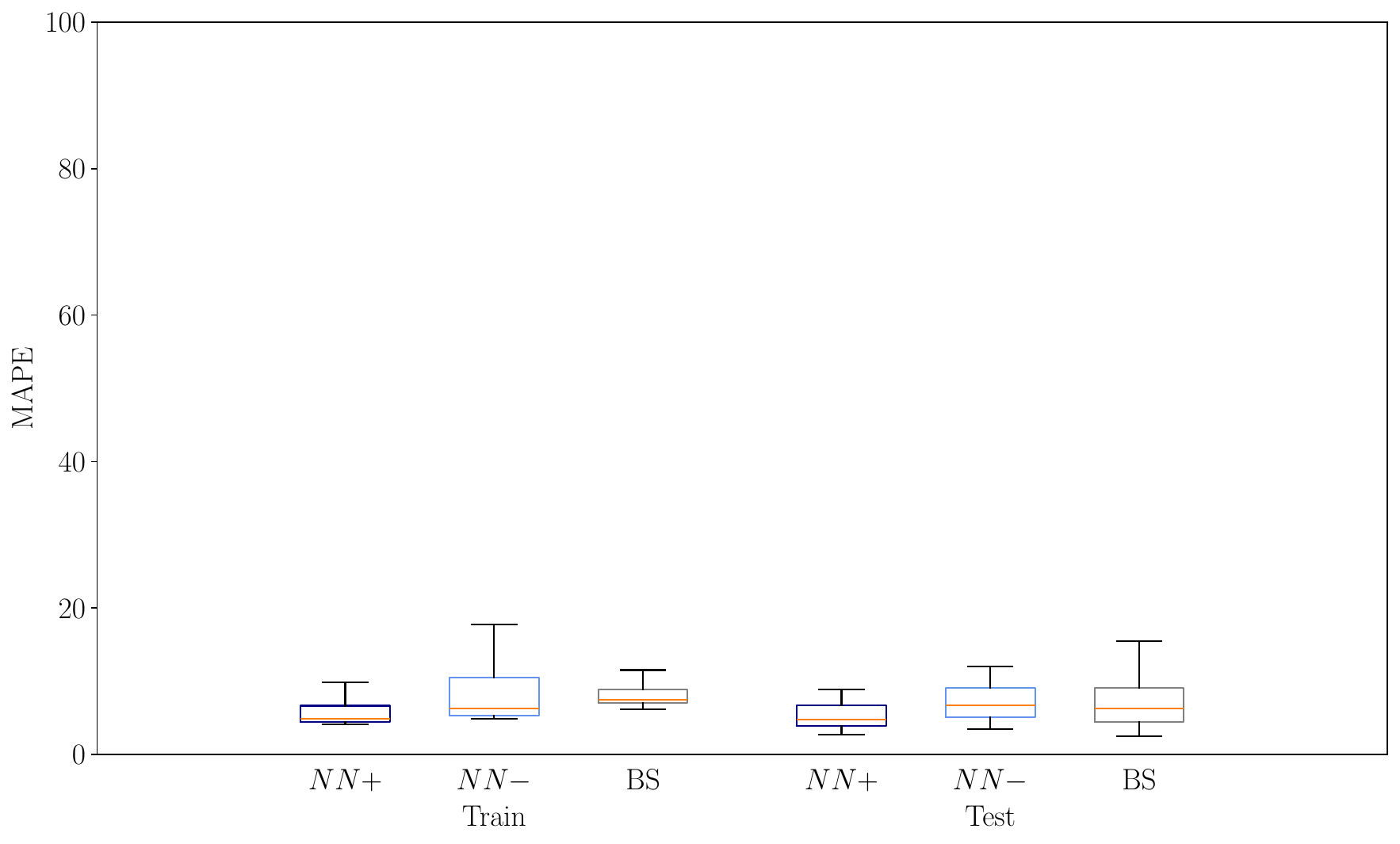}
    \includegraphics[width=\linewidth]{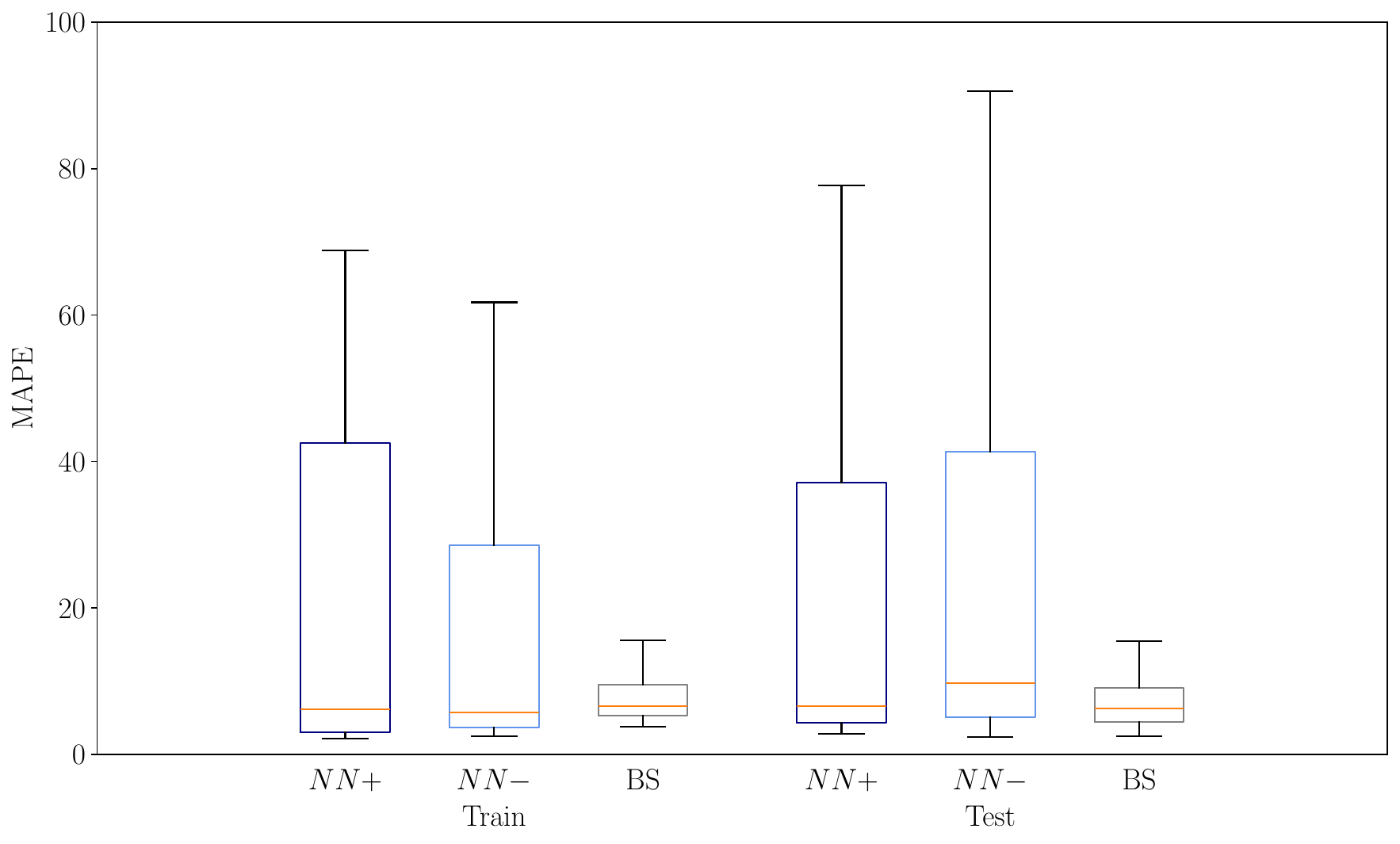}
    \caption{Boxplots of the mean absolute percentage error (MAPE) for ITM options under two data-splitting schemas. The top panel corresponds to the expanding window schema, while the bottom panel corresponds to the rolling window schema. In both panels, the data is grouped by subsets: the training set, the test set, the test set filtered by Black-Scholes price ($p_{\text{BS}} \geq 0.075$), and the test set filtered by $p_{\text{BS}} < 0.075$. Each group contains three boxplots, corresponding to the neural network trained with Black-Scholes information (NN+), the neural network trained without Black-Scholes information (NN-), and the Black-Scholes model (BS) itself. The exact numerical values corresponding to the expanding and rolling window results are presented in Tables.~\ref{Tab:ITM_expanding_NN} and \ref{Tab:ITM_rolling_NN}, respectively.}
    \label{Fig:ITM_combined_NN}
\end{figure}

\begin{figure}[t]
    \centering
    \includegraphics[width=\linewidth]{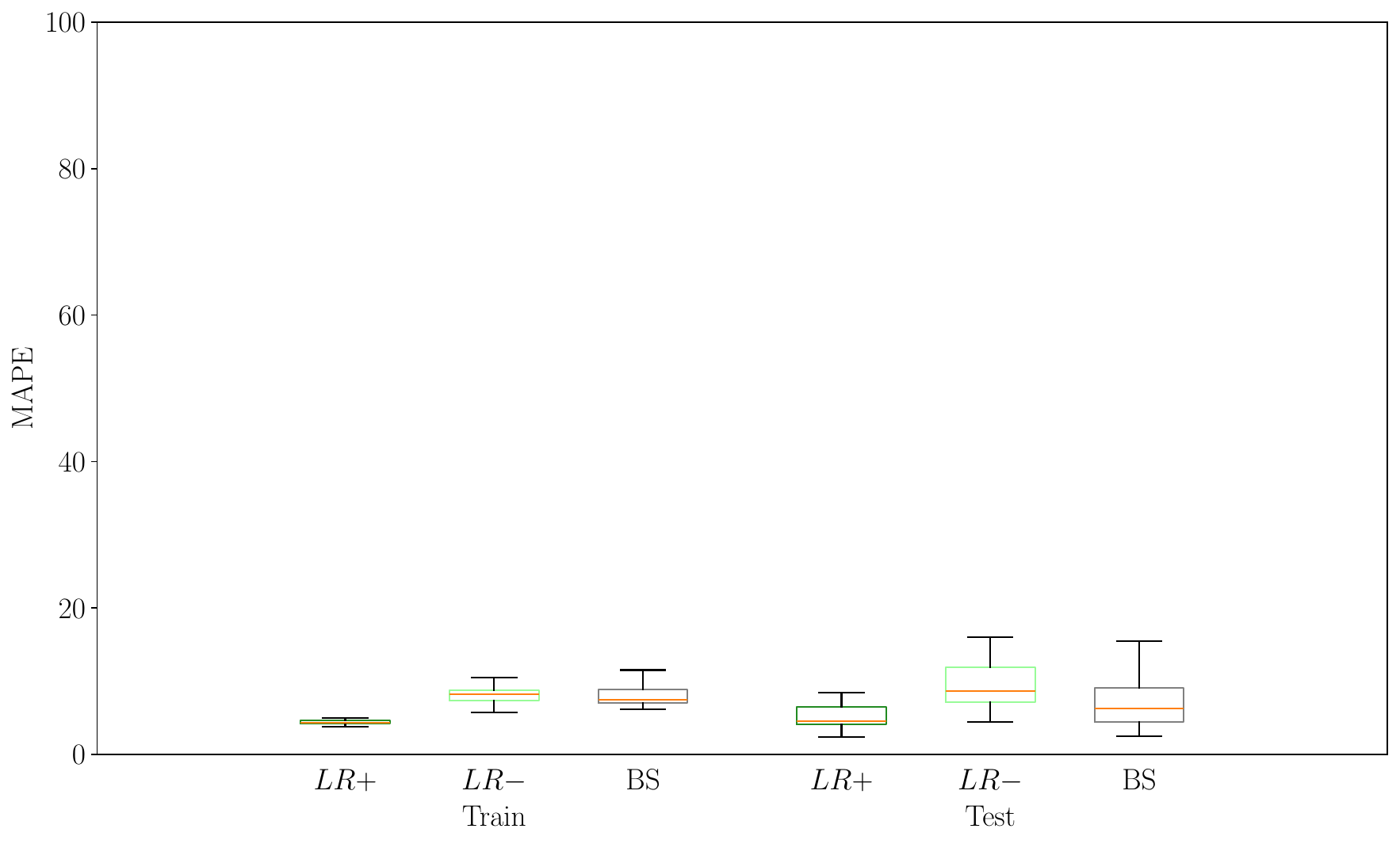}
    \includegraphics[width=\linewidth]{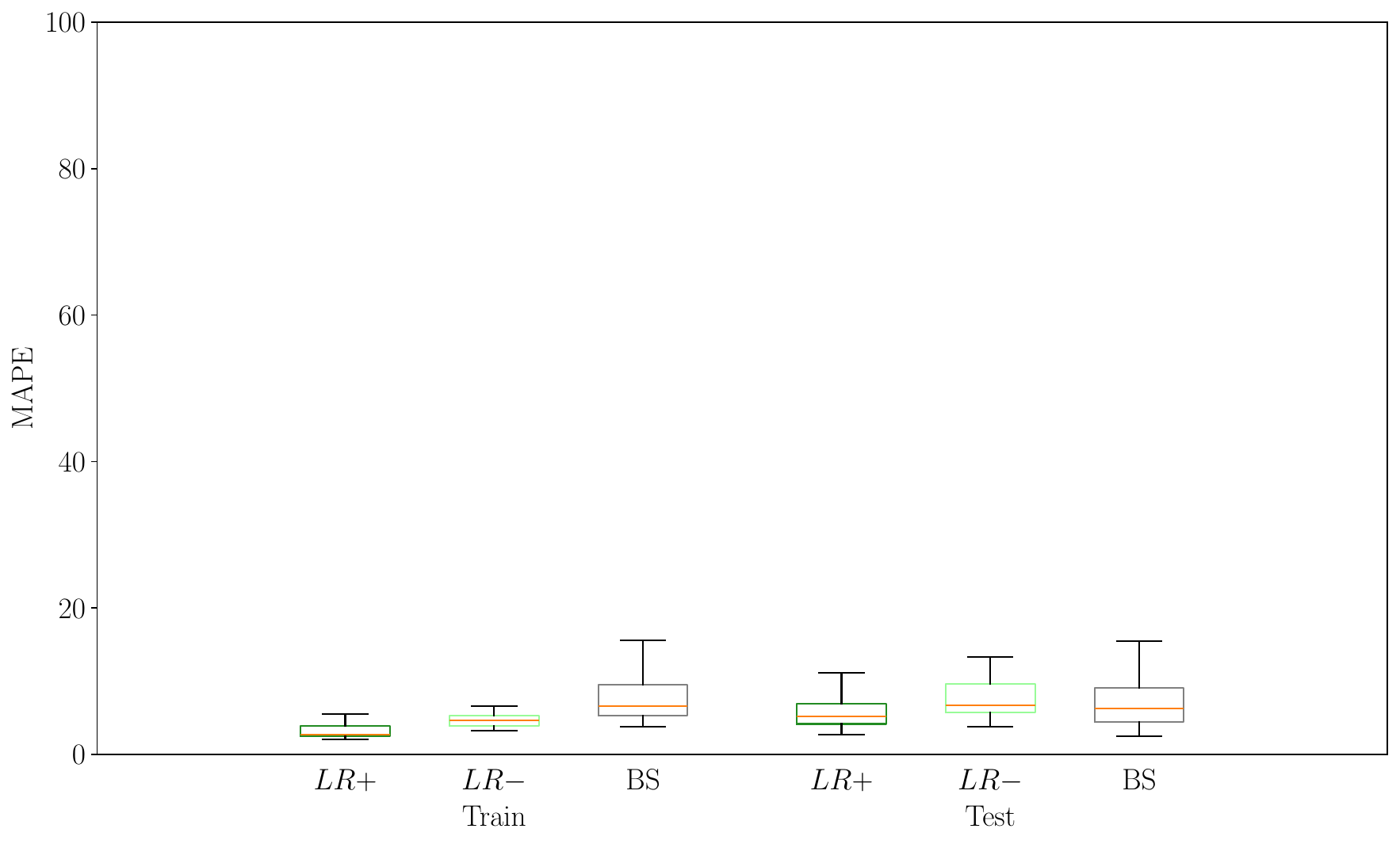}
    \caption{Boxplots of the mean absolute percentage error (MAPE) for ITM options under two data-splitting schemas. The top panel corresponds to the expanding window schema, while the bottom panel corresponds to the rolling window schema. In both panels, the data is grouped by subsets: the training set, the test set, the test set filtered by Black-Scholes price ($p_{\text{BS}} \geq 0.075$), and the test set filtered by $p_{\text{BS}} < 0.075$. Each group contains three boxplots, corresponding to the linear model trained with Black-Scholes information (LR+), the linear model trained without Black-Scholes information (LR-), and the Black-Scholes model (BS) itself. The exact numerical values corresponding to the expanding and rolling window results are presented in Tables.~\ref{Tab:ITM_expanding_LR} and \ref{Tab:ITM_rolling_LR}, respectively.}
    \label{Fig:ITM_combined_LR}
\end{figure}

\begin{figure}[t]
    \centering
    \includegraphics[width=\linewidth]{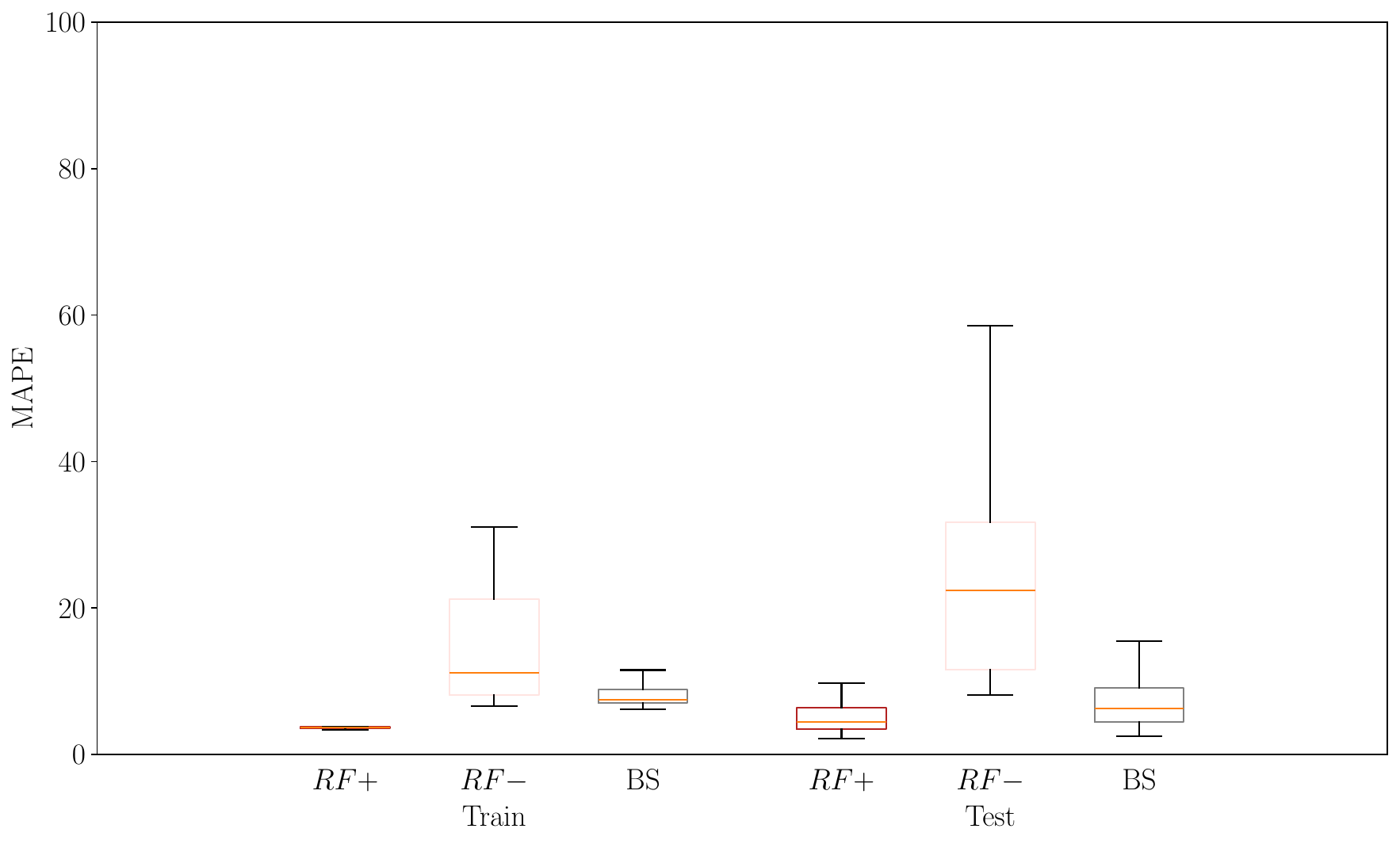}
    \includegraphics[width=\linewidth]{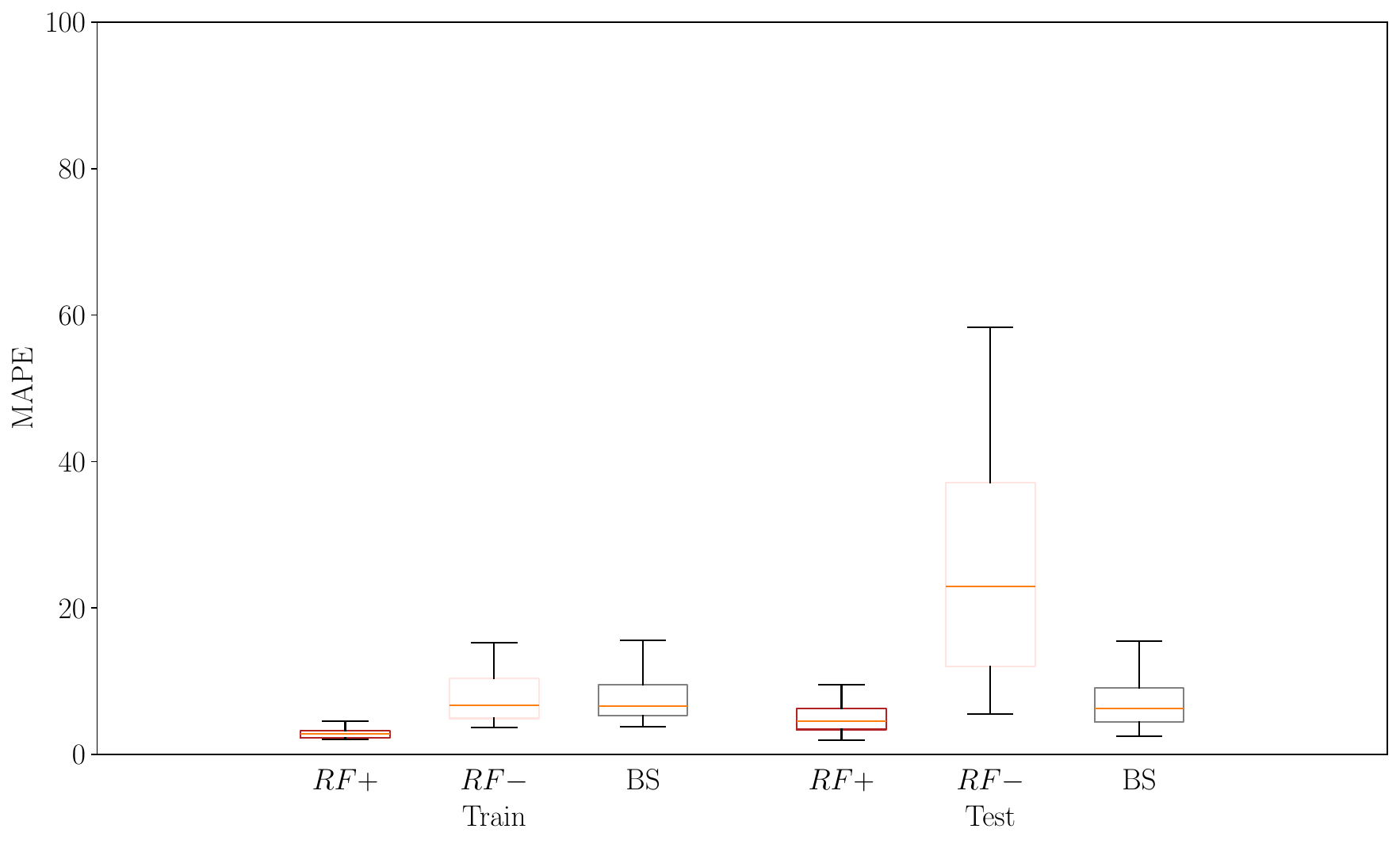}
    \caption{Boxplots of the mean absolute percentage error (MAPE) for ITM options under two data-splitting schemas. The top panel corresponds to the expanding window schema, while the bottom panel corresponds to the rolling window schema. In both panels, the data is grouped by subsets: the training set, the test set, the test set filtered by Black-Scholes price ($p_{\text{BS}} \geq 0.075$), and the test set filtered by $p_{\text{BS}} < 0.075$. Each group contains three boxplots, corresponding to the neural network trained with Black-Scholes information (RF+), the neural network trained without Black-Scholes information (RF-), and the Black-Scholes model (BS) itself. The exact numerical values corresponding to the expanding and rolling window results are presented in Tables.~\ref{Tab:ITM_expanding_RF} and \ref{Tab:ITM_rolling_RF}, respectively.}
    \label{Fig:ITM_combined_RF}
\end{figure}

Key observations from the OTM analysis remain consistent. NN exhibits instability when trained using a rolling window, while RF remains highly dependent on the inclusion of BS in the input space. This behavior is further evident when analyzing the MAPE distributions in Figures~\ref{Fig:ITM_combined_NN}, \ref{Fig:ITM_combined_LR}, and \ref{Fig:ITM_combined_RF}, where the variability in performance changes markedly when NN is trained using a rolling window, or when RF excludes BS.


\FloatBarrier

\FloatBarrier

\FloatBarrier

\section{Explaining Neural Networks Models with SHAP}\label{Sec:interpret}
To interpret the predictions of our NN models, we employ the \textit{SHapley Additive exPlanation (SHAP)} framework \cite{shapley201617,lundberg2017unified}.\footnote{For our analysis, we use the \texttt{shap} Python package}. SHAP provides a unified approach to explaining the contribution of each input feature to the model’s output, which makes it particularly suitable for interpreting complex nonlinear models like neural networks.

For a given input vector $x \in \mathbb{R}^K$ and a model $f$, the SHAP value $\phi_i(f,x)$ quantifies the effect of the $i$-th feature on the model’s prediction $f(x)$. This effect is determined by evaluating how the output changes when the feature is added to or removed from different subsets $S \subseteq \{1,\ldots, K\}$. The SHAP value is formally defined as:
\begin{equation}
\phi_i = \sum_{S \subseteq N \setminus \{i\}} \frac{|S|!(K - |S| - 1)!}{K!} \left[f_{S \cup \{i\}}(x) - f_S(x)\right]
\end{equation}
where $N = {1, \ldots, K}$ and the weighting ensures that the SHAP values sum to the total model output, i.e., $\sum_i \phi_i = f(x)$ and $f_S(x)$ denotes the model output when only features in $S$ are used. This allows us to interpret how each input contributes to the model output.

\begin{figure}[t]
    \begin{minipage}{\textwidth}
        \centering
        \begin{subfigure}{0.48\textwidth}
            \centering
            \includegraphics[width=\textwidth]{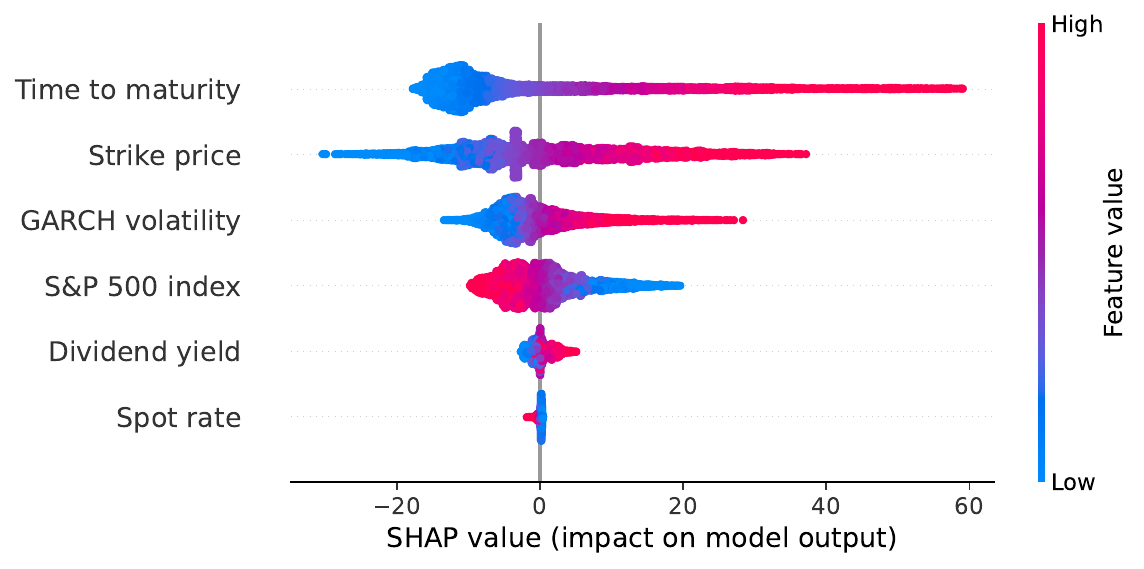}
            \caption{NN- (2009)}
        \end{subfigure}
        \begin{subfigure}{0.48\textwidth}
            \centering
            \includegraphics[width=\textwidth]{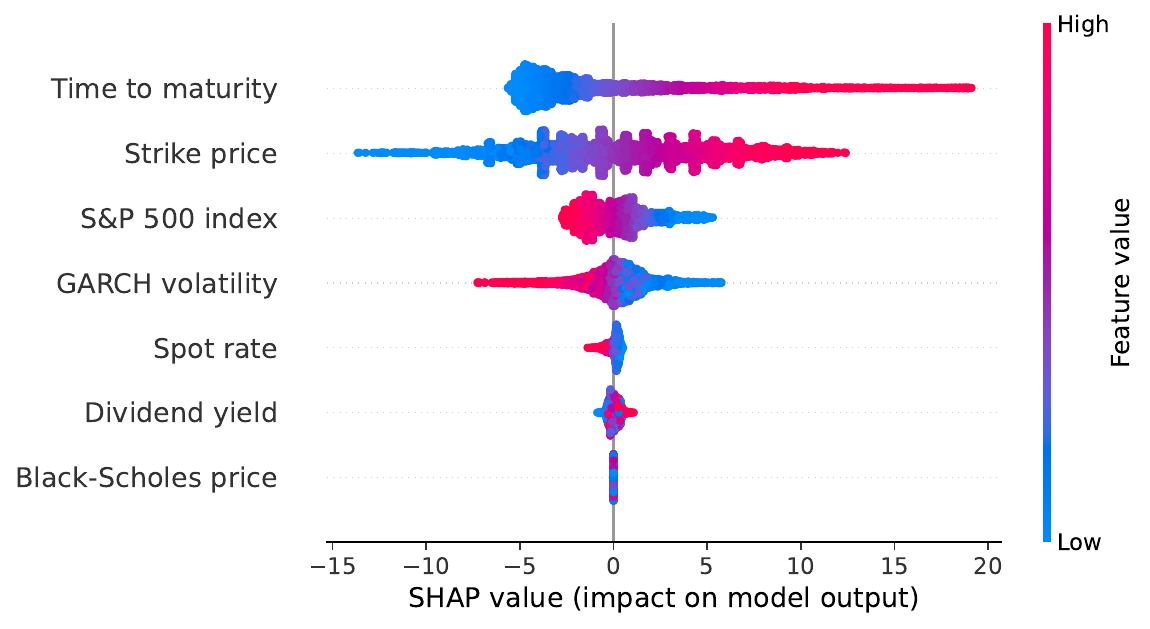}
            \caption{NN+ (2009)}
        \end{subfigure}

        \begin{subfigure}{0.48\textwidth}
            \centering
            \includegraphics[width=\textwidth]{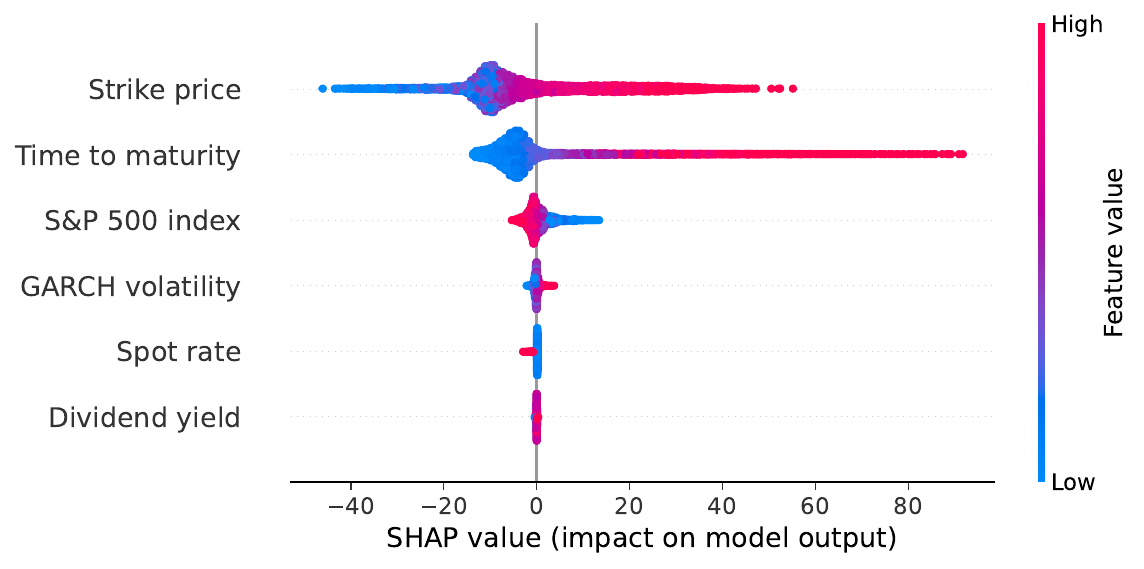}
            \caption{NN- (2015)}
        \end{subfigure}
        \begin{subfigure}{0.48\textwidth}
            \centering
            \includegraphics[width=\textwidth]{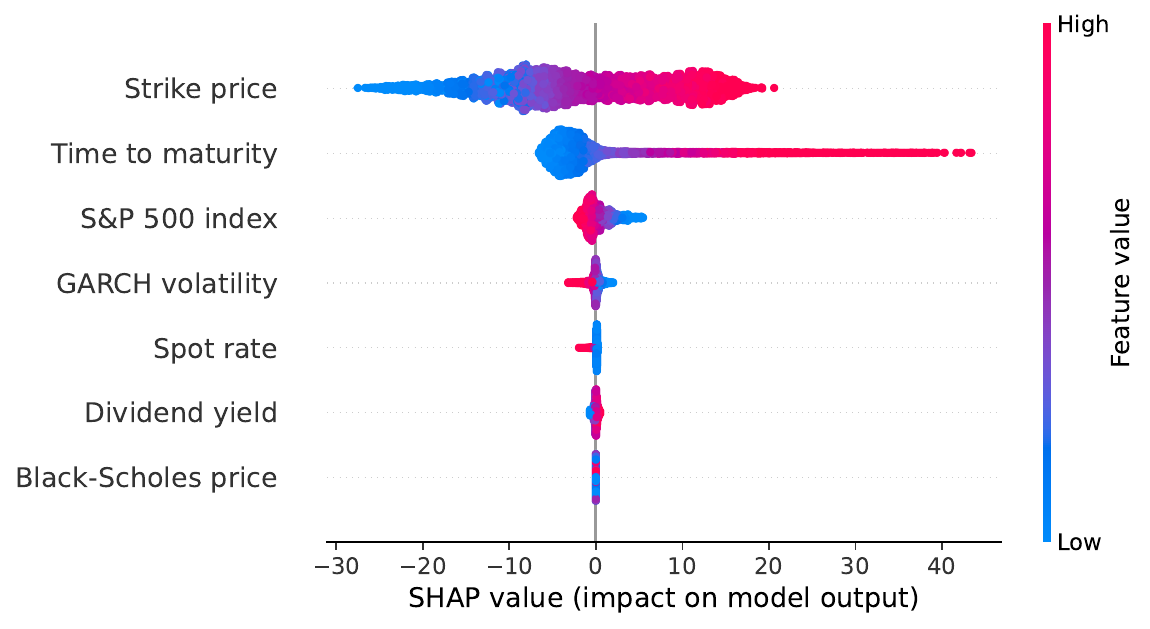}
            \caption{NN+ (2015)}
        \end{subfigure}

        \begin{subfigure}{0.48\textwidth}
            \centering
            \includegraphics[width=\textwidth]{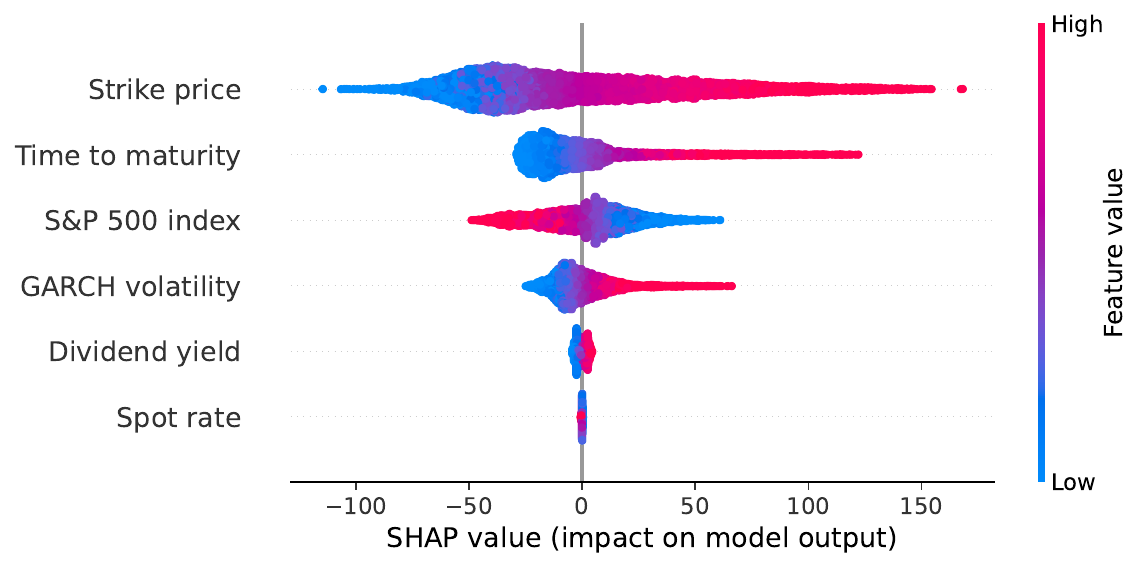}
            \caption{NN- (2021)}
        \end{subfigure}
        \begin{subfigure}{0.48\textwidth}
            \centering
            \includegraphics[width=\textwidth]{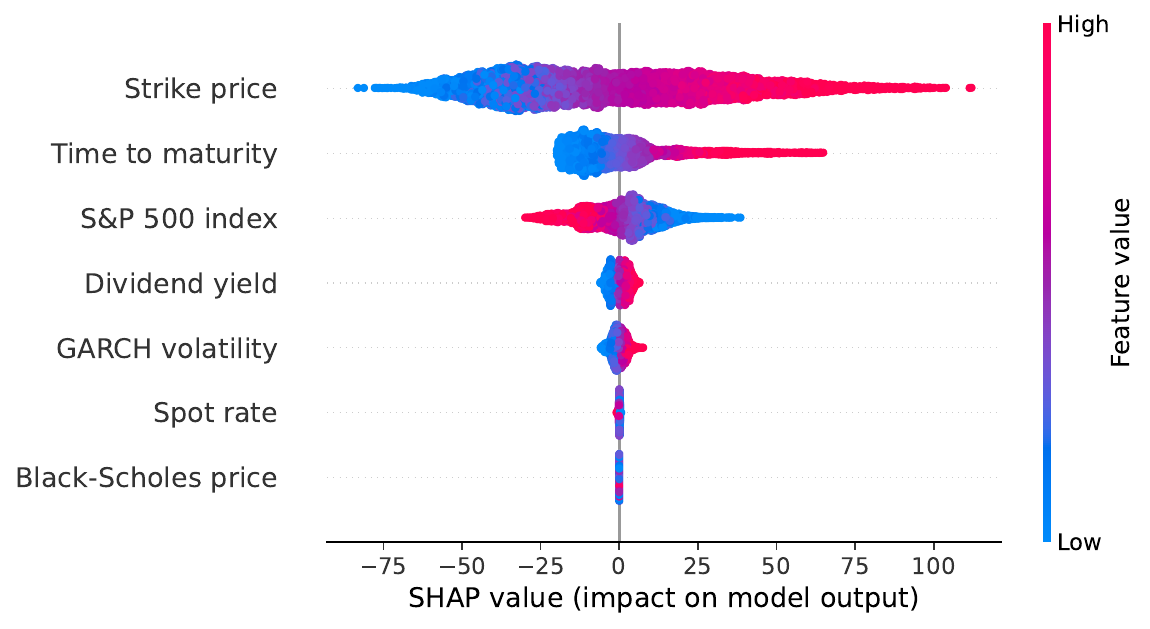}
            \caption{NN+ (2021)}
        \end{subfigure}
    \end{minipage}
        \centering
    \caption{SHAP Analysis for OTM Models. Each row in the grid represents a model tested on a specific timeframe, as indicated in the subcaptions. The left column (NN-) corresponds to neural networks that do not use the Black-Scholes price as an input, while the right column (NN+) represents neural networks trained with Black-Scholes as an additional feature.}

    \label{fig:shap_otm}
\end{figure}

\begin{figure}[t]
    \begin{minipage}{\textwidth}
        \centering
        \begin{subfigure}{0.48\textwidth}
            \centering
            \includegraphics[width=\textwidth]{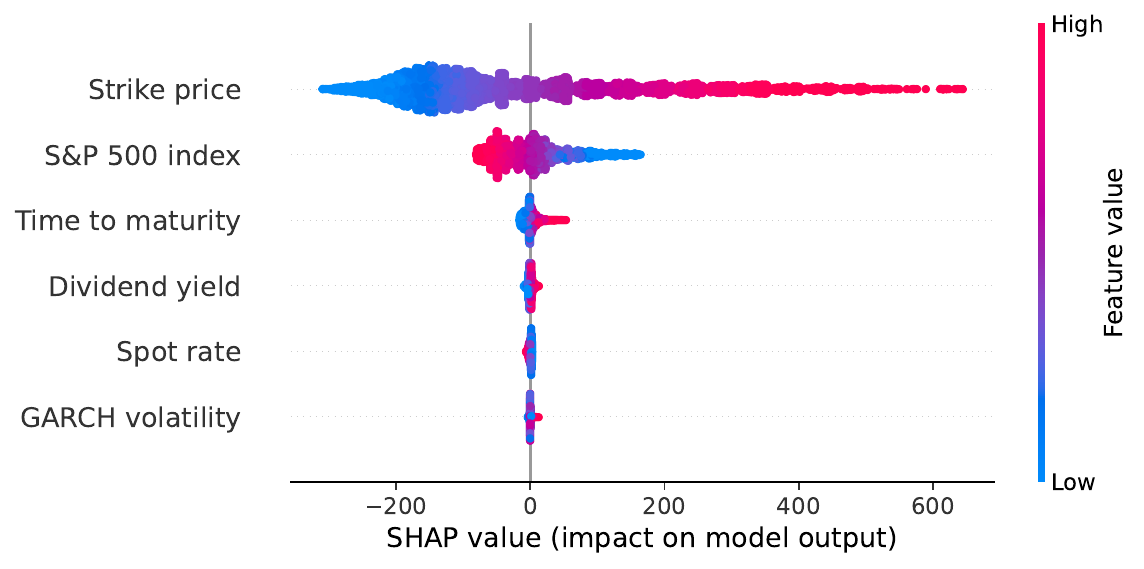}
            \caption{NN- (2009)}
        \end{subfigure}
        \begin{subfigure}{0.48\textwidth}
            \centering
            \includegraphics[width=\textwidth]{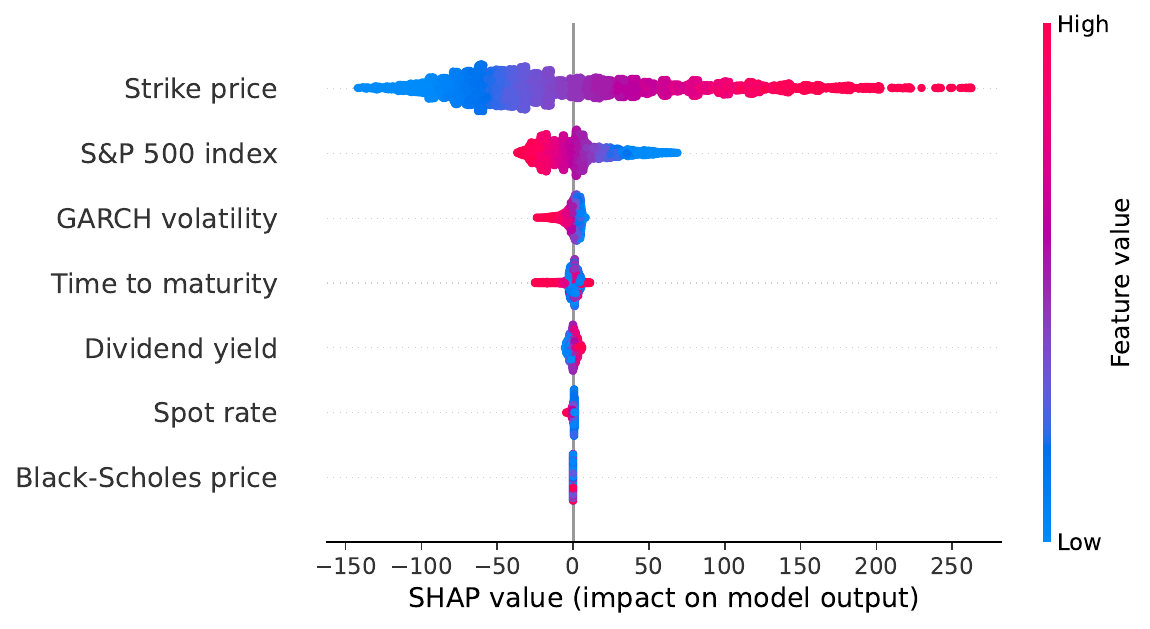}
            \caption{NN+ (2009)}
        \end{subfigure}

        \begin{subfigure}{0.48\textwidth}
            \centering
            \includegraphics[width=\textwidth]{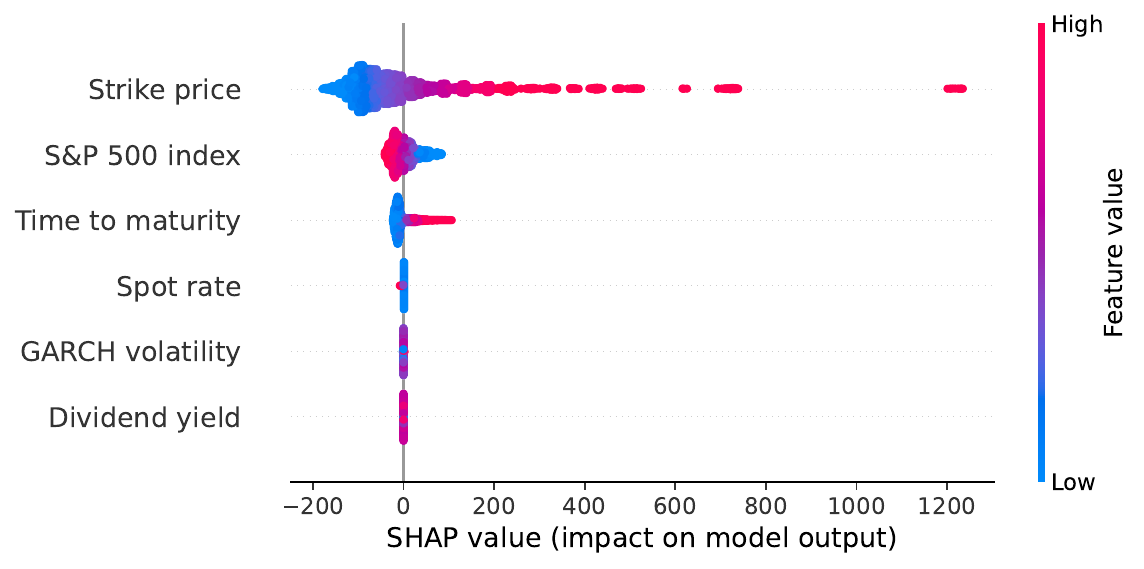}
            \caption{NN- (2015)}
        \end{subfigure}
        \begin{subfigure}{0.48\textwidth}
            \centering
            \includegraphics[width=\textwidth]{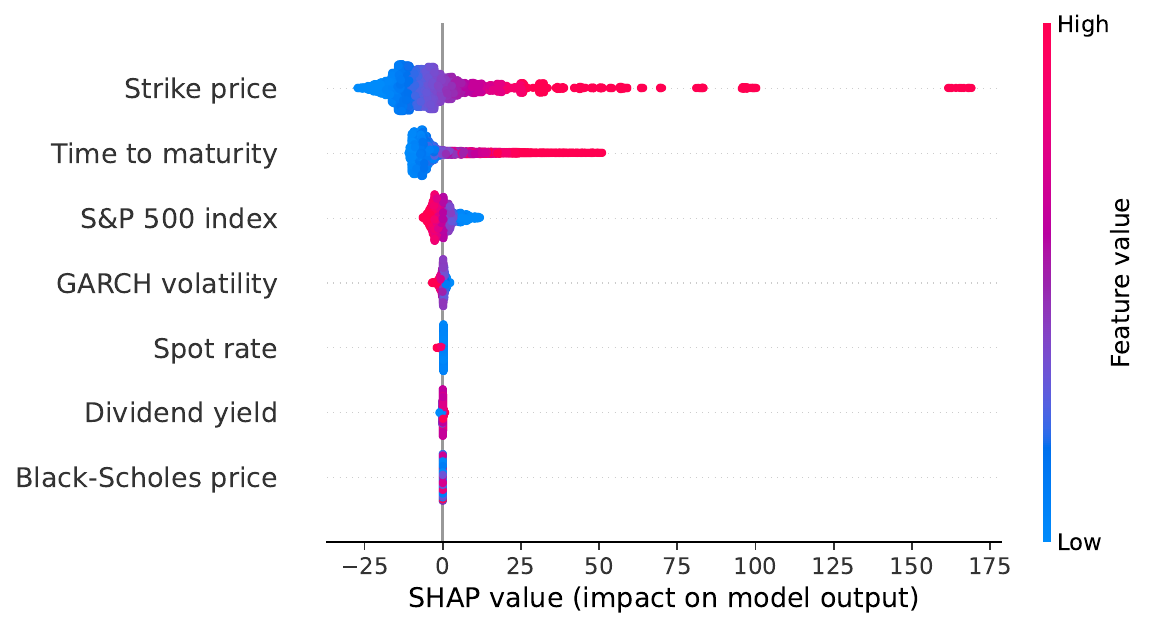}
            \caption{NN+ (2015)}
        \end{subfigure}

        \begin{subfigure}{0.48\textwidth}
            \centering
            \includegraphics[width=\textwidth]{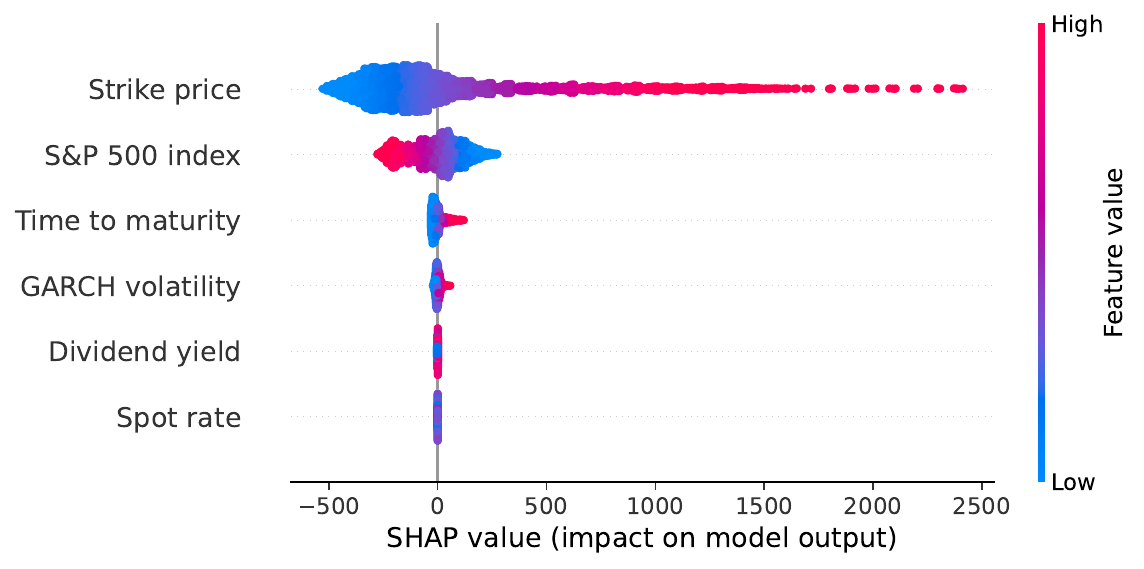}
            \caption{NN- (2021)}
        \end{subfigure}
        \begin{subfigure}{0.48\textwidth}
            \centering
            \includegraphics[width=\textwidth]{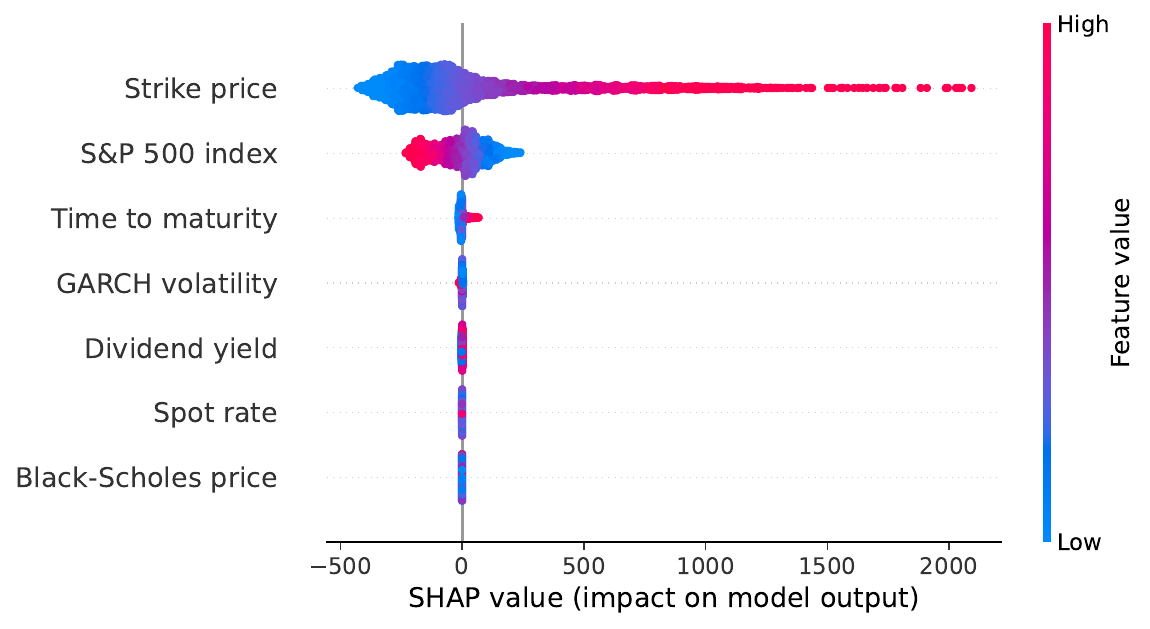}
            \caption{NN+ (2021)}
        \end{subfigure}
    \end{minipage}
        \centering
    \caption{SHAP Analysis for ITM Models. Each row in the grid represents a model tested on a specific timeframe, as indicated in the subcaptions. The left column (NN-) corresponds to neural networks that do not use the Black-Scholes price as an input, while the right column (NN+) represents neural networks trained with Black-Scholes as an additional feature.}
    \label{fig:shap_itm}
\end{figure}

Figures~\ref{fig:shap_otm} and \ref{fig:shap_itm} display SHAP value summary plots for the neural networks trained on OTM and ITM put options, respectively, highlighting how the importance of input features varies across model specifications and time periods. Each figure consists of a $3 \times 2$ grid of subplots. Each row corresponds to a different test period, arranged chronologically from top to bottom. We select two periods characterized by major financial shocks (2008 and 2020) and one relatively stable period (2014) to interpret how the models adjust their learned relationships under different market conditions. Within each row, the left column (NN-) represents neural networks trained without the Black-Scholes price as an input, whereas the right column (NN+) corresponds to neural networks trained with the Black-Scholes price as an additional feature. The year labels in the figures refer to the test sets used for SHAP value computation, not the training periods.

For each selected  NN model, we compute SHAP values using a randomly sampled batch of 10,000 observations from the respective test period. This design highlights how feature importance shifts across models and time periods.

Several key patterns emerge from the SHAP value plots in Fig.~\ref{fig:shap_otm}. First, including the Black-Scholes price in the input features does not appear to drastically alter the structure of the model’s predictions. In fact, for the 2009 and 2015 test periods, adding the Black-Scholes feature (NN+) leads to an overall rescaling of SHAP values and a lower magnitude for all other features. We can observe this effect in the x-axis range of the plots. Despite this rescaling, the relative importance of input features remains consistent across both NN- and NN+ models. Specifically, time to maturity and strike price are consistently among the most influential variables, which aligns well with classical option pricing theory and suggests that the neural network effectively learns key structural relationships in the option dataset.

GARCH volatility also emerges as a key driver. This feature plays a prominent role in the periods following major market disruptions, namely, the aftermath of the 2008 financial crisis (2009 test set) and the onset of the COVID-19 pandemic (2021 test set). In contrast, its importance diminishes significantly during the more stable 2015 period. This dynamic holds for both NN- and NN+ configurations. Furthermore, the direction of SHAP values aligns with theoretical expectations: features such as time to maturity and strike price show a positive contribution (red points) to the output, indicating that an increase in these values leads to a higher predicted option price. Conversely, the S\&P 500 index exhibits predominantly negative SHAP values (blue points), reflecting that lower index levels—typical of market downturns—increase the value of a put option.

Turning to the results in Fig.~\ref{fig:shap_itm}, we observe similar effects from the inclusion of the Black-Scholes price, with limited structural changes and a reduction in SHAP magnitude in some cases. However, the importance of time to maturity appears reduced in the ITM setting, especially during crisis periods. Instead, the model's predictions are predominantly driven by the strike price, which emerges as the most influential feature across all test periods. As in the OTM case, its effect on the output follows theoretical intuition. Lastly, the S\&P 500 index again displays more variability in its influence during periods of heightened uncertainty, such as 2009 and 2021, and less so during the relatively stable 2015 interval.

\begin{figure}[t]
    \centering
    \begin{subfigure}[t]{0.45\textwidth}
        \centering
        \includegraphics[width=\textwidth]{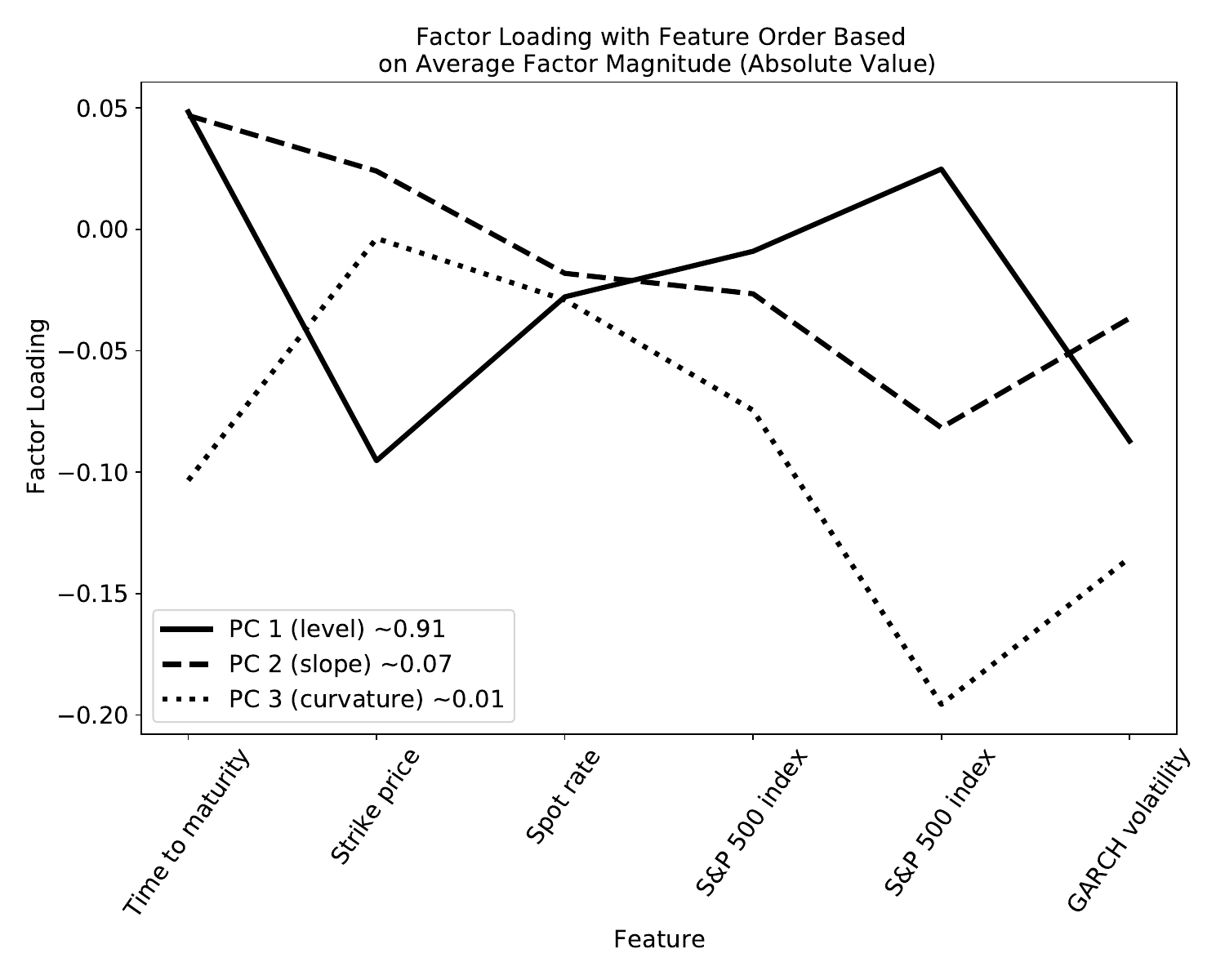}
        \caption{Top three Principal Component Analysis (PCA) averaged over time, as described by \cite{Litterman1991CommonFA}}
        \label{fig:PCA_load}
    \end{subfigure}
    \hfill
    \begin{subfigure}[t]{0.52\textwidth}
        \centering
        \includegraphics[width=\textwidth]{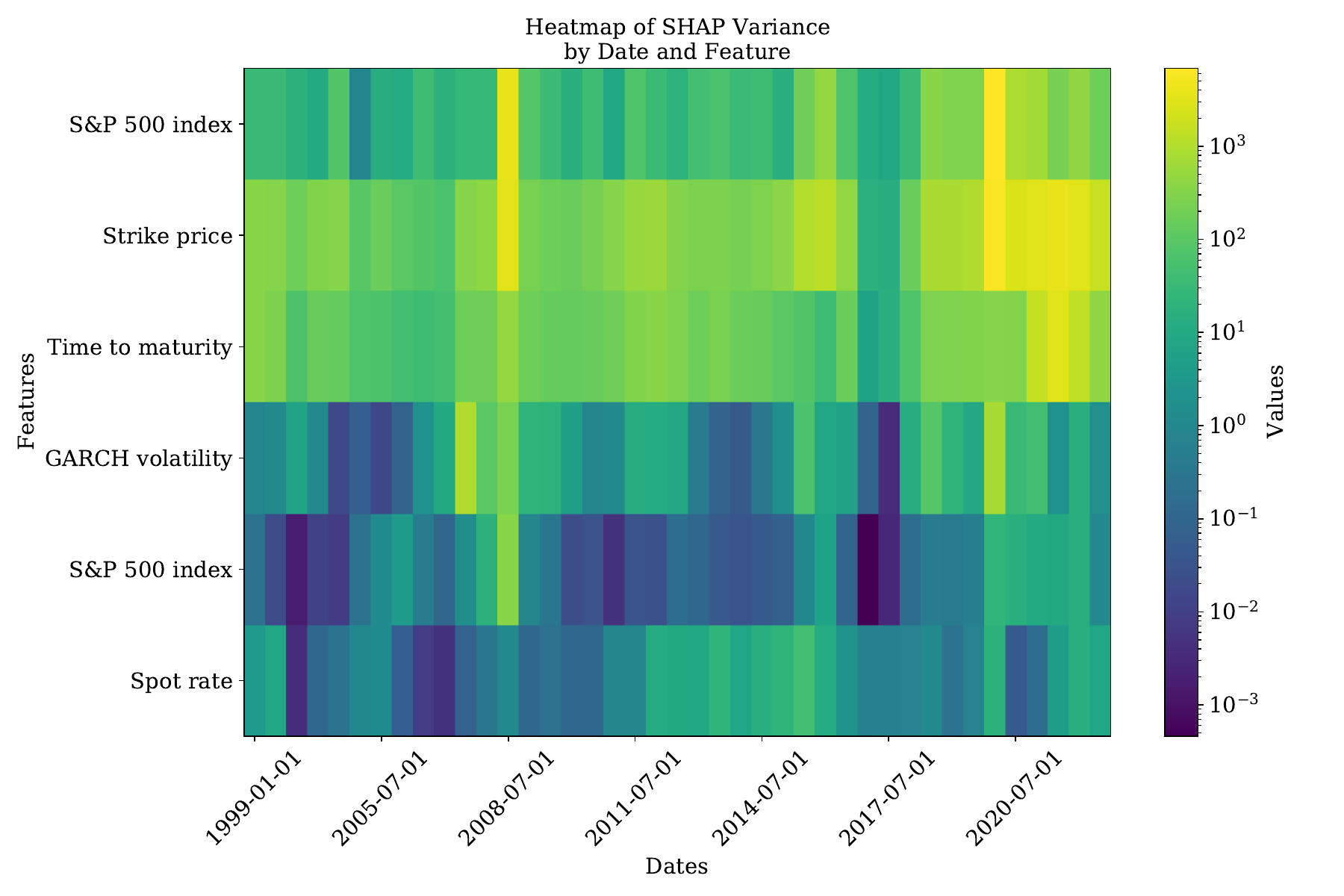}
        \caption{SHAP Analysis on 10,000 randomly selected points per test period as described by \cite{lundberg2017unified}}
        \label{fig:SHAP_heatmap}
    \end{subfigure}
    \caption{Principal Component and SHAP Analysis.}
    \label{fig:side_by_side}
\end{figure}

To further support and generalize the insights obtained from the SHAP analyses in Figures~\ref{fig:shap_otm} and \ref{fig:shap_itm}, we complement this investigation with a broader statistical perspective. In particular, we compare these results to a principal component analysis (PCA) conducted on the training data across different time periods. This exercise aims to uncover the main sources of variation in the input features of the option pricing problem. In Fig. \ref{fig:PCA_load}, we reference the work of \cite{Litterman1991CommonFA}, who found that bond returns were primarily determined by the first three principal components of the feature data, where the features of the analysis were bond yields at varying maturities. 
Unlike \cite{Litterman1991CommonFA}, our aim is to understand the correlations between characteristics of the option data set, as opposed to the asset returns. We conducted an analogous investigation of the options data used for our proposed models. In performing PCA on OTM options, we average the load value of each feature over the training time periods for the first, second, and third principal components. Then, we order the features on the $x$-axis based on the decreasing average explained variance\footnote{\cite{Litterman1991CommonFA} follows a natural ordering of the features by bond maturity}. 
We observe similarities between the results of this PCA on the options data and \cite{Litterman1991CommonFA} work on bond yields. 
In particular, similar behavior of the second and third principal components is observed, termed the \textit{slope} and \textit{curvature} components in \cite{Litterman1991CommonFA}. This PCA plot reveals parabolic and cubic shapes in the second and third components as referenced in the Litterman slope and curvature components, respectively.
An additional study we refer to, which evaluates the instrumented principal component analysis capability (IPCA), is developed by \citep{BUCHNER20221140}, who demonstrate that IPCA may be used to explain a significant portion of the variation in monthly returns of S\&P 500 options, and further explain the first three principle components in terms of level, slope, and curvature, as Litterman demonstrates. 

These findings align closely with the feature importances identified via SHAP in our neural network models. To make this connection explicit, we extend the SHAP analysis beyond individual models and visualize in Fig.~\ref{fig:SHAP_heatmap} the average feature importance scores across all trained networks for OTM options. The aggregated SHAP results show a similar pattern to the PCA findings, confirming that strike price, time to maturity, and the S\&P 500 index are the key features driving the model’s predictions. These results follow both the levels of the explained variance ratio recorded in the PCA results presented in Fig. \ref{fig:PCA_load} (the magnitude of the explained variance is reflected in the order of the characteristics on the horizontal axis of this graph) and intuition of what factors logically drive the price of options.  This consistency between statistical structure and model behavior strengthens our confidence in the robustness and interpretability of the learned relationships.

\section{Empirical Evaluation of No Arbitrage Conditions}
\label{Sec:no_arbitrage_check}

To validate the economic consistency of the neural network's predictions, we evaluated whether the model satisfies the basic no-arbitrage conditions such as monotonicity properties with respect to two key input variables: the strike price and the time to maturity of the option, as well as positivity of convexity with respect to the strike. From a theoretical standpoint, and holding all other variables constant, the price of a put option is expected to increase with a higher strike price and increase with a longer time to maturity. We implement an automated evaluation to empirically assess these relationships. There is an abundance of literature on constructions of arbitrage-free implied volatility surface, see \cite{fengler2009volatility}, for example. 

Our procedure perturbs one input variable at a time while keeping all others fixed and monitoring the resulting changes in the predicted option price. When perturbing the strike price, we apply a tolerance of \$$0.05$ to account for numerical approximation and the minimum pricing increment observed in the dataset. In contrast, when perturbing the time to maturity, we increment the original value by $5\%$ at each step. For each input, we verify whether the sequence of predicted prices follows a weakly monotonic pattern. When perturbing the inputs, the option’s moneyness may change, potentially causing it to switch between ITM and OTM classifications. To ensure consistency, we carefully select the appropriate trained model for each perturbed observation, using the model corresponding to the current moneyness classification at each step.

In addition to monotonicity, we also assess whether the predicted option prices exhibit convexity with respect to the strike price, a theoretical property implied by no-arbitrage bounds. We approximate second derivatives using discrete second-order differences, and consider a violation when this approximation is negative in at least two consecutive steps, after applying the same tolerance threshold of \$$0.05$ for step-to-step price variations.

We apply this evaluation to a random subsample of the data, totaling 123{,}587 options. We find that in 115{,}561 cases (93.51\%) the predictions respect monotonicity with respect to strike price, in 117{,}520 cases (95.09\%) they satisfy convexity with respect to strike price, and in 102{,}479 cases (82.92\%) they satisfy monotonicity with respect to time to maturity.

\begin{figure}[t]
    \centering
    \includegraphics[width=0.7\textwidth]{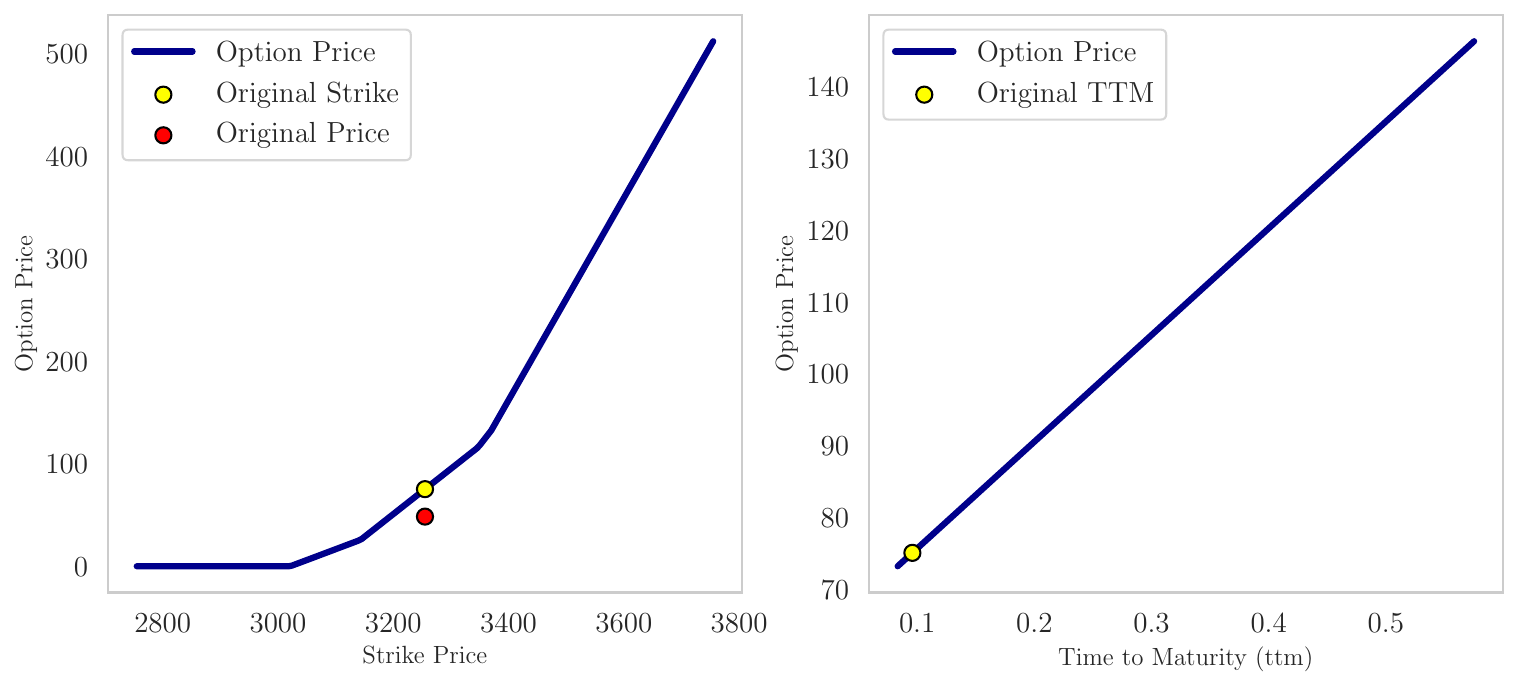}
    \includegraphics[width=0.7\textwidth]{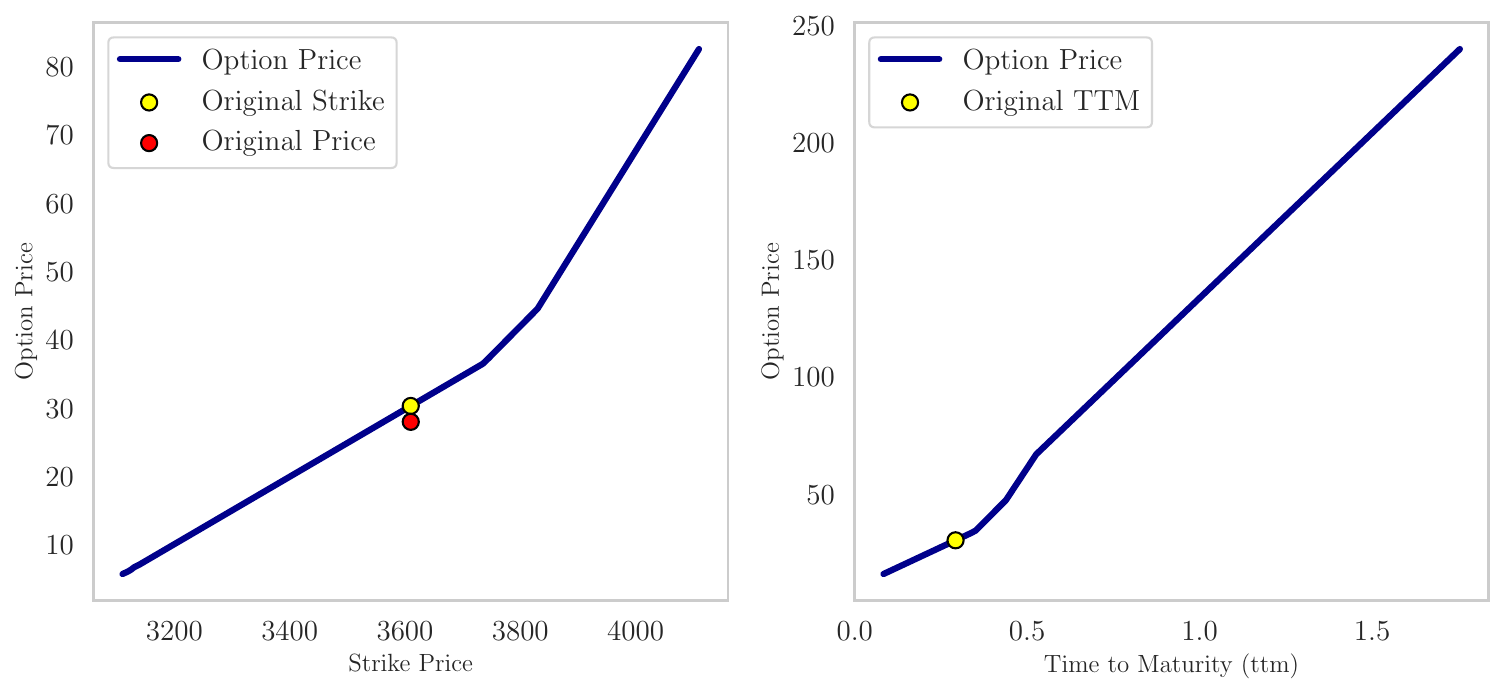}
    \caption{Option price variations for different samples.}
    \label{fig:option_variations}
\end{figure}

Fig.~\ref{fig:option_variations} provides two representative examples, each showing how the predicted option price changes when perturbing the strike and the time to maturity of a given option. The original configuration is marked for easier interpretation.

\begin{figure}[t]
    \centering
    \includegraphics[width=0.7\textwidth]{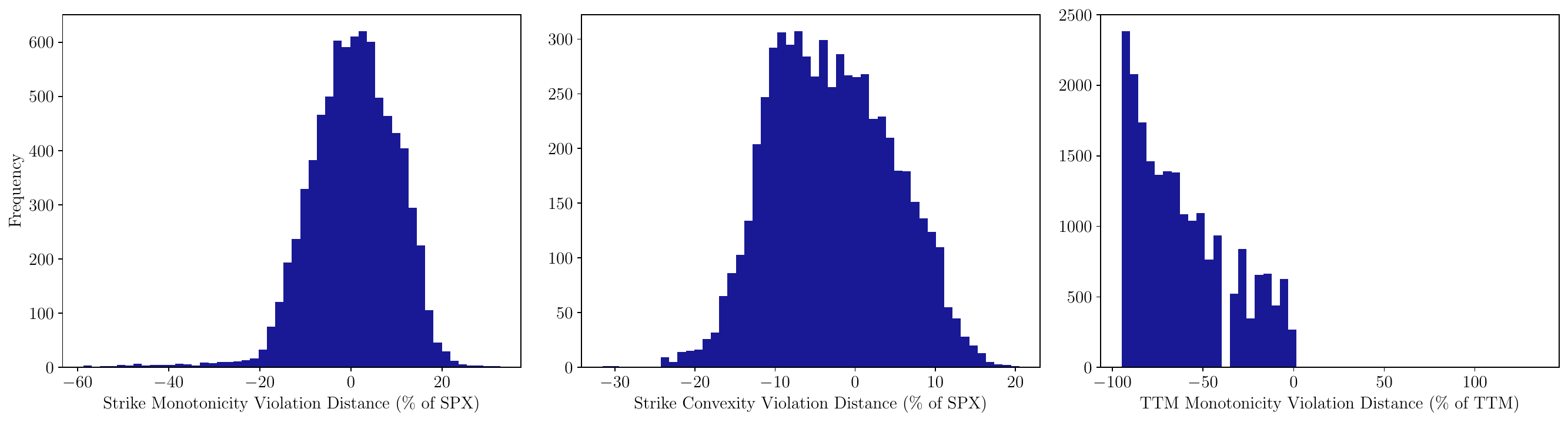}
    \caption{Distribution of monotonicity violation distances. Left: strike price violations measured in steps of \$5. Right: time-to-maturity (TTM) violations measured in steps of 5\% of the original TTM.}
    \label{fig:violation_distance}
\end{figure}

To better understand the nature of the violations, Fig.~\ref{fig:violation_distance} presents the distribution of violation distances for each of the three tests: monotonicity with respect to strike, convexity with respect to strike, and monotonicity with respect to time to maturity. The figure reveals several important patterns regarding the violations. For both monotonicity and convexity with respect to the strike price, which account for fewer than 7\% and 5\% of the analyzed cases, respectively, the distributions exhibit noticeable right tails. This indicates that most violations occur when the strike price is perturbed by a significant percentage relative to its original value. Moreover, many violations are observed at distant perturbation steps, suggesting that the model generally behaves well in the neighborhood of the original input configuration.

Turning to monotonicity with respect to time to maturity, the majority of violations occur when the TTM is reduced substantially from its original value. These violations often occur when moving several increments away from the initial TTM, making the option more short-term, but still within the boundaries enforced by our dataset filtering, which excludes options with less than one month to maturity. Similar to the strike price results, violations in TTM monotonicity tend to occur farther away from the original input point, reinforcing that the model's local behavior remains largely consistent with theoretical expectations.

\begin{figure}[t]
    \centering
    \begin{subfigure}[t]{0.32\textwidth}
        \centering
        \includegraphics[width=\textwidth]{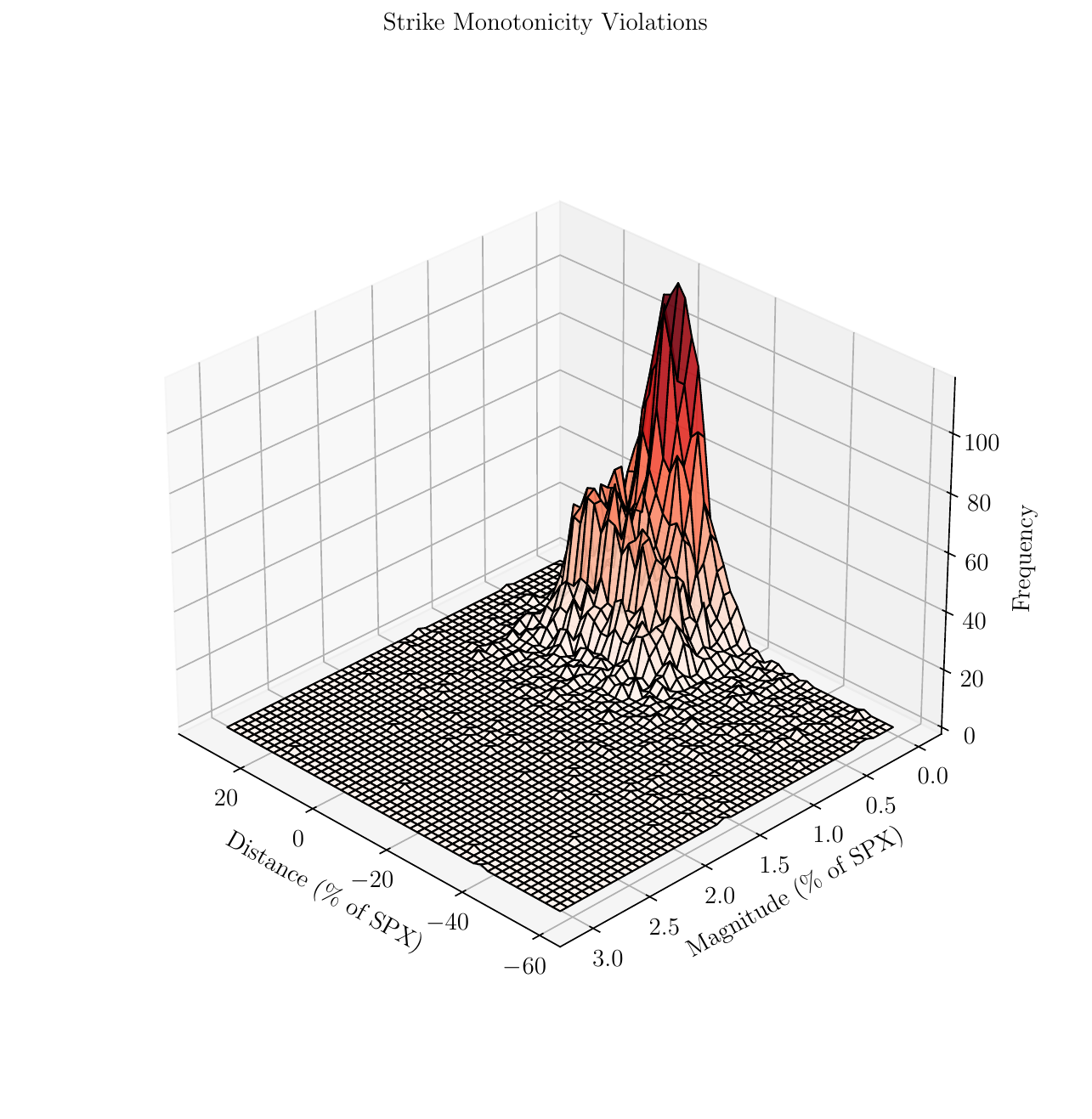}
    \end{subfigure}
    \begin{subfigure}[t]{0.32\textwidth}
        \centering
        \includegraphics[width=\textwidth]{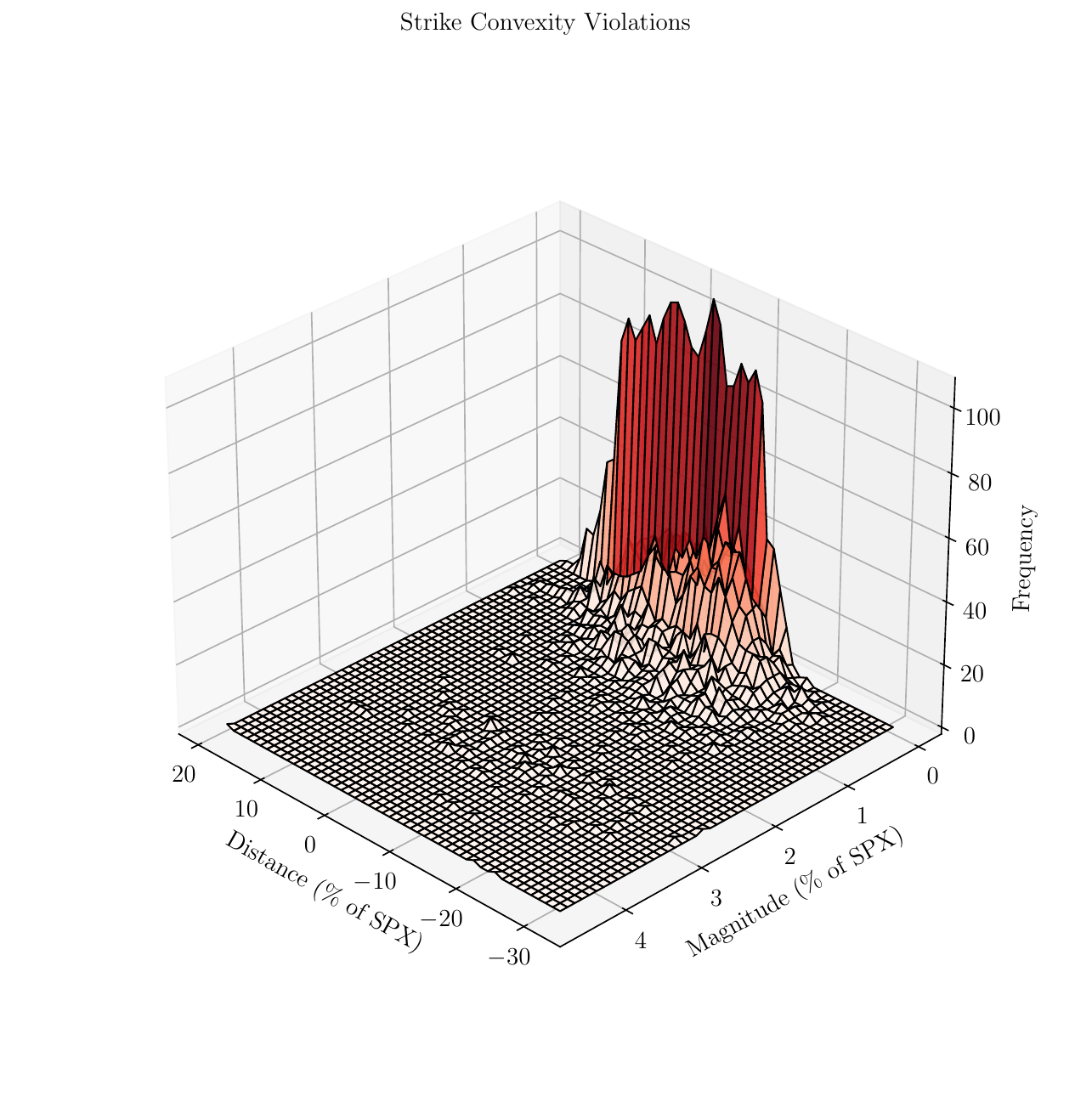}
    \end{subfigure}
    \begin{subfigure}[t]{0.32\textwidth}
        \centering
        \includegraphics[width=\textwidth]{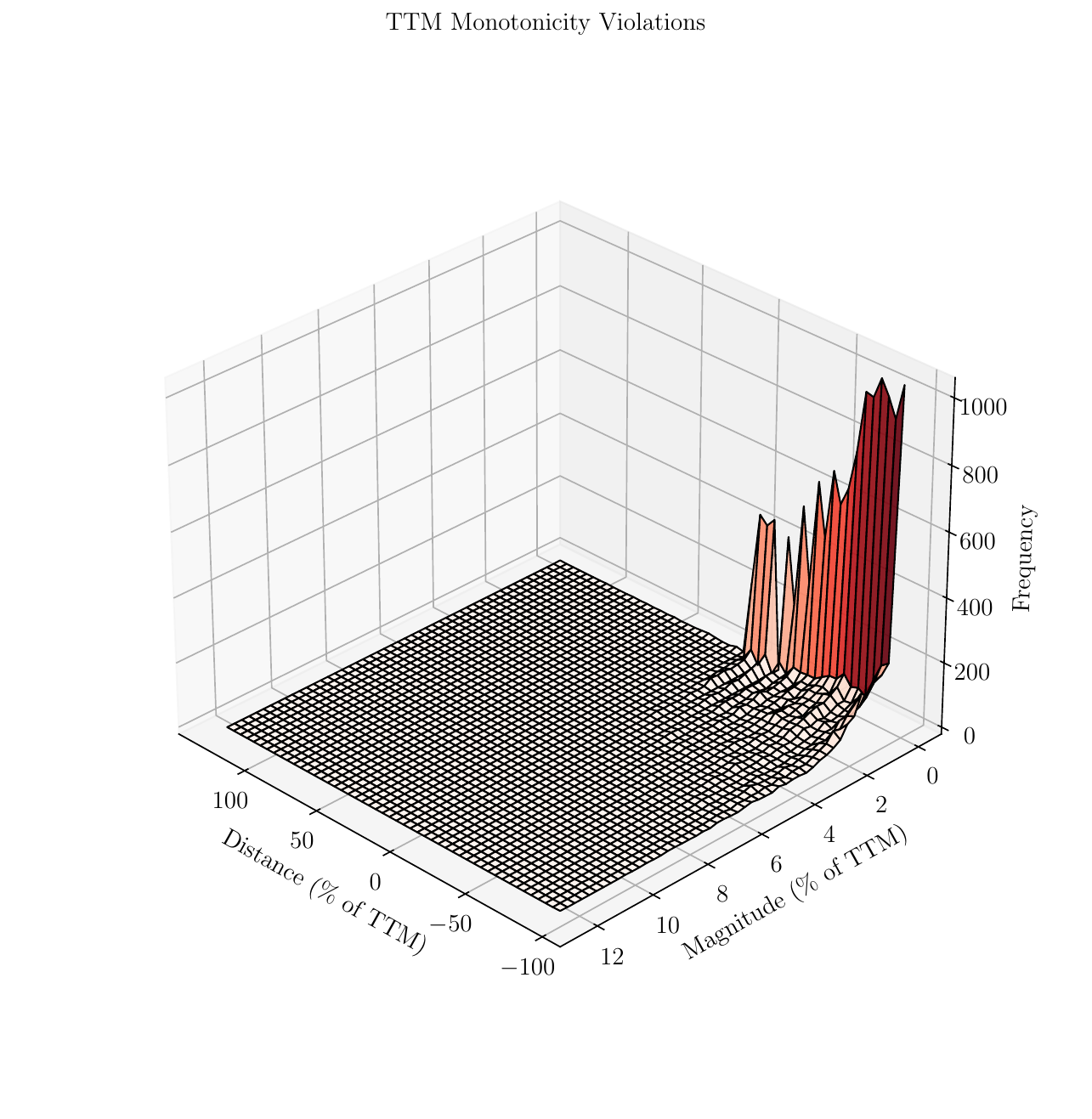}
    \end{subfigure}
    \caption{3D distribution of violation distance and magnitude. Each surface shows how often violations occur at various scaled distances and their corresponding severity.}
    \label{fig:violation_surfaces}
\end{figure}

To further investigate the nature of the violations, Fig.~\ref{fig:violation_surfaces} extends the previous analysis and introduces a third dimension representing the magnitude of the violation.. Specifically, after a violation is detected, we cumulate the magnitude of consecutive violations for the same perturbed option, where the magnitude is defined as the absolute difference between the predicted prices at consecutive perturbation steps, $|p_i - p_{i+1}|$. Each surface plot displays the frequency of violations as a function of both the scaled distance from the original input and the violation magnitude.

The results confirm that most violations are of negligible size. Most violations have negligible size, with the mass concentrated near zero, even when violations occur at distant perturbation steps. This observation further supports that the neural network predictions remain robust and economically consistent in the neighborhood of the original input configurations, with only minor deviations arising further away.

\section{Takeaways}\label{Sec:takeaways}
\vspace{5mm}

\subsection{Summary of Empirical Findings}

The analysis highlights key differences in model performance across training methodologies, moneyness segmentation, and feature configurations. NN consistently outperforms RF and LR on the out-of-sample set, particularly when pricing ITM options or OTM options with higher prices. However, NN exhibits greater instability, especially when trained using a rolling window. RF benefits significantly from including BS as a feature, demonstrating stable performance when BS is available but performing worse than the BS benchmark when it is removed. LR performs worst overall, struggling to generalize across both ITM and OTM options, with test set errors frequently exceeding the BS baseline.

Segmenting the dataset by BS price confirms that very low-priced OTM options contribute disproportionately to the overall pricing error. This explains much of the variability in MAPE across different training windows. Further moneyness segmentation shows MAPE increases as options move deeper OTM, MAPE increases significantly across all models. NN remains the most flexible model in these cases, maintaining competitive performance even without BS in the input space, while RF struggles without BS and LR remains unreliable.

The choice between expanding and rolling window methodologies also plays a critical role. Expanding windows incorporate all available data, offering greater model stability at the cost of adaptability to recent market shifts. Rolling windows, in contrast, improve adaptability but introduce instability—especially for NN. The expanding window approach incrementally incorporates all available data, while the rolling window approach restricts training to a fixed-length recent history. Although the rolling approach allows models to adapt to more recent trends, it introduces instability, particularly for NN. Empirical results indicate that expanding windows yield more stable performance across models but may reduce adaptability to market shifts by incorporating older data. A potential trade-off between the two approaches could involve  applying exponential decay to past data, placing more emphasis on recent data while gradually downweighting older observations rather than discarding them abruptly. This could mitigate the instability observed in NN models trained under the rolling window approach, where abrupt data removal leads to performance degradation.

Overall, these findings suggest that incorporating BS as a feature improves model robustness, but its absence affects models differently. NN demonstrates strong predictive capability but requires careful tuning under rolling window setups, while RF performs well only when BS is included. These findings inform model and feature selection decisions in real-world option pricing applications.

One possible difference between the rolling window and the expanding window is the increase in strikes and maturities of SPX options after the mid 2000s. To the extent that the pricing function is smooth between the option prices (i.e. outputs) and their characteristics (i.e. inputs), the additional strikes (maturities) that narrow the distance between strikes (maturities) would have minimal impact on the pricing function. 

\subsection{Business and Practical Implications}
The results suggest that NN provides a flexible approach to option pricing, effectively capturing patterns in the data even when the Black-Scholes price is excluded from the feature set. This adaptability is particularly relevant in real-world applications where market conditions deviate from theoretical assumptions. The explainability analysis further strengthens this case by highlighting that the NN model places consistent emphasis on economically meaningful features such as strike price, time to maturity, and the underlying index level. For practitioners, this adds transparency to what is often perceived as a “black box” model, making the model more acceptable in risk-sensitive settings..

The results also indicate that NN benefits from a broader training context, capturing more diverse data patterns when trained on an expanding window rather than being limited to only the most recent three years. This aligns with prior research in deep learning, where larger training sets have been shown to improve model generalization across domains such as computer vision \cite{sun2017revisiting} and natural language processing \cite{kaplan2020scaling}.

The strong dependence of RF on BS raises further concerns regarding its robustness. The findings suggest that when BS is excluded from the feature set, RF loses significant predictive power and often underperforms the BS benchmark itself. This dependency implies that RF may not be a reliable choice when market conditions deviate from the Black-Scholes framework, as it appears to function effectively only when BS is explicitly provided as input. Practitioners should consider this limitation, especially in environments where option pricing must remain robust under different volatility regimes or deviations from BS assumptions.

Although low-priced OTM options contribute disproportionately to the overall MAPE, they should not be disregarded in practical applications. Instead, selecting a model capable of pricing them more accurately could provide significant benefits, especially in risk management strategies where cost-effective downside protection is critical. One potential application is in tail-hedging strategies, which aim to protect portfolios from sudden market drawdowns while maintaining long-term investments. Deep OTM options, due to their low absolute prices, offer a cost-effective means of hedging against extreme downside risk. Identifying a pricing model that minimizes errors in this segment could enhance the effectiveness of such strategies, ensuring more reliable risk assessments and hedging decisions (see \cite{ilmanen2021tail} for a broader discussion on tail-hedging approaches).

Finally, the analysis of no-arbitrage violations provides an important practical lens for evaluating model reliability beyond accuracy metrics. Ensuring that pricing functions adhere to economic constraints such as monotonicity and convexity is crucial in institutional settings, where violations can lead to inconsistent valuations and potential trading inefficiencies. The observation that NN models can still produce such violations under certain training conditions highlights the need for additional regularization or post-hoc correction when deploying these models in production environments.



\section{Conclusions}\label{Sec:conclusion}

This paper presents a comprehensive evaluation of machine learning models for pricing index options, with a specific focus on their ability to simulate the evolution of option prices over time. We build on a GARCH(1,1) model of SPX log returns and a large dataset of S\&P500 options from 1996 to 2022 and explores both expanding and rolling training methodologies. Across a variety of model architectures, input feature sets, and moneyness regimes, we find that a a neural network with two hidden layers and four neurons each consistently outperforms both linear regression and random forest benchmarks, especially in pricing deep OTM options and during periods of elevated volatility.

Several practical insights emerge from our analysis. First, the inclusion of the Black-Scholes price as a feature markedly improves predictive accuracy for all models except the neural network, which remains competitive even when this feature is excluded. This robustness suggests that neural networks can serve as effective non-parametric alternatives to traditional models in settings where theoretical assumptions may not hold. Additionally, the choice of training window influences model stability and performance. While expanding windows offer greater consistency, rolling windows enable faster adaptation to market shifts but require careful tuning, particularly for neural networks. We also show that segmenting the data by moneyness and BS price levels can help mitigate distortions caused by low-priced options, which disproportionately contribute to overall pricing errors.

Importantly, we demonstrate that the neural network’s outputs respect key no-arbitrage requirements, including monotonicity and convexity with respect to strike and maturity. The SHAP analysis further confirms that the model's predictions align with financial intuition and statistical feature structure. Together, these results provide a flexible and explainable machine learning-based simulation framework for implied volatility surfaces, which can be applied in downstream tasks such as tail-risk hedging, stress testing, and scenario generation.

Future research could investigate hybrid modeling approaches that combine neural networks with no-arbitrage constraints more explicitly, or extend the current simulation framework to multi-asset settings. Moreover, incorporating option volume or liquidity measures as additional features may further enhance pricing accuracy and robustness in real-world trading environments. Finally, the framework developed in this paper could serve as a foundation for constructing and evaluating risk management goals, such as value-at-risk calculations and dynamic tail-hedge strategies, enabling practitioners to stress-test portfolios under simulated volatility scenarios with greater flexibility.

\begingroup \parindent 0pt \parskip 0.0ex \def\enotesize{\normalsize} \theendnotes \endgroup

%
%
%


\bibliography{biblio} 



\appendix
\section{Tables with Performance Results}\label{App:tables}

In each table, the Date column indicates the period covered, spanning from the beginning of the training set to the end of the test set. The last six months always serve as the test set. The format $ \text{YY/MM - YY/MM} $ is used, where the first date marks the start of the training set and the second denotes the end of the six-month test set.

\begin{table}[!ht]
    \centering
    \small
\begin{tabular}{p{1.75cm}|>{\centering\arraybackslash}p{0.6cm}>{\centering\arraybackslash}p{0.6cm}>{\centering\arraybackslash}p{0.6cm}|
                     >{\centering\arraybackslash}p{0.6cm}>{\centering\arraybackslash}p{0.6cm}>{\centering\arraybackslash}p{0.6cm}|
                     >{\centering\arraybackslash}p{0.6cm}>{\centering\arraybackslash}p{0.6cm}>{\centering\arraybackslash}p{0.6cm}|
                     >{\centering\arraybackslash}p{0.6cm}>{\centering\arraybackslash}p{0.6cm}>{\centering\arraybackslash}p{0.6cm}}
        \toprule
        & \multicolumn{3}{c|}{Train} 
        & \multicolumn{3}{c|}{Test} 
        & \multicolumn{3}{c|}{Test $p_{BS} \geq 0.075$} 
        & \multicolumn{3}{c}{Test $p_{BS} < 0.075$} \\
        \centering Date 
        & \textbf{$NN+$} & \textbf{$NN-$} & \textbf{$BS$} 
        & \textbf{$NN+$} & \textbf{$NN-$} & \textbf{$BS$} 
        & \textbf{$NN+$} & \textbf{$NN-$} & \textbf{$BS$} 
        & \textbf{$NN+$} & \textbf{$NN-$} & \textbf{$BS$} \\
        \midrule
96/1 : 99/6 & 61.0 & \textbf{30.1} & 66.9 & 71.3 & \textbf{25.5} & 55.6 & 31.6 & \textbf{18.8} & 51.7 & 525.3 & 103.0 & \textbf{99.5} \\
96/1 : 99/12 & 76.9 & \textbf{31.2} & 65.2 & 93.2 & \textbf{34.3} & 66.5 & 43.1 & \textbf{26.1} & 61.0 & 399.7 & \textbf{84.1} & 99.6 \\
96/1 : 00/6 & 168.5 & \textbf{36.6} & 65.3 & 315.5 & \textbf{50.7} & 50.8 & 117.7 & \textbf{35.6} & 45.2 & 2045.8 & 182.6 & \textbf{99.1} \\
96/1 : 00/12 & 51.7 & \textbf{45.3} & 63.9 & 56.2 & 41.8 & \textbf{41.6} & \textbf{37.4} & 39.2 & 37.6 & 321.5 & \textbf{77.6} & 98.2 \\
96/1 : 01/6 & 163.8 & 187.8 & \textbf{62.1} & 243.0 & 266.4 & \textbf{34.0} & 119.8 & 133.2 & \textbf{30.4} & 2446.7 & 2648.5 & \textbf{98.3} \\
96/1 : 01/12 & 169.3 & 186.4 & \textbf{60.1} & 172.7 & 191.9 & \textbf{44.0} & 82.4 & 88.7 & \textbf{39.0} & 1156.0 & 1315.7 & \textbf{99.2} \\
\bottomrule
\end{tabular}
    \caption{MAPE values for OTM options trained using an expanding window schema. Results are shown for models trained with Black-Scholes information (NN+), without it (NN-), and for the Black-Scholes model (BS).}
    \label{Tab:OTM_expanding_NN}
\end{table}
\FloatBarrier

\begin{table}[!ht]
    \centering
    \small
    \begin{tabular}{p{1.75cm}|>{\centering\arraybackslash}p{0.6cm}>{\centering\arraybackslash}p{0.6cm}>{\centering\arraybackslash}p{0.6cm}|
                     >{\centering\arraybackslash}p{0.6cm}>{\centering\arraybackslash}p{0.6cm}>{\centering\arraybackslash}p{0.6cm}|
                     >{\centering\arraybackslash}p{0.6cm}>{\centering\arraybackslash}p{0.6cm}>{\centering\arraybackslash}p{0.6cm}|
                     >{\centering\arraybackslash}p{0.6cm}>{\centering\arraybackslash}p{0.6cm}>{\centering\arraybackslash}p{0.6cm}}
        \toprule
        & \multicolumn{3}{c|}{Train} 
        & \multicolumn{3}{c|}{Test} 
        & \multicolumn{3}{c|}{Test $p_{BS} \geq 0.075$} 
        & \multicolumn{3}{c}{Test $p_{BS} < 0.075$} \\
        \centering Date 
        & \textbf{$NN+$} & \textbf{$NN-$} & \textbf{$BS$} 
        & \textbf{$NN+$} & \textbf{$NN-$} & \textbf{$BS$} 
        & \textbf{$NN+$} & \textbf{$NN-$} & \textbf{$BS$} 
        & \textbf{$NN+$} & \textbf{$NN-$} & \textbf{$BS$} \\
        \midrule
96/1 : 99/6 & 61.0 & \textbf{30.1} & 66.9 & 71.3 & \textbf{25.5} & 55.6 & 31.6 & \textbf{18.8} & 51.7 & 525.3 & 103.0 & \textbf{99.5} \\
96/7 : 99/12 & 75.9 & \textbf{29.8} & 64.1 & 90.7 & \textbf{33.0} & 66.5 & 41.4 & \textbf{25.5} & 61.0 & 392.9 & \textbf{79.2} & 99.6 \\
97/1 : 00/6 & 172.8 & 177.9 & \textbf{62.9} & 316.1 & 335.6 & \textbf{50.8} & 121.6 & 128.5 & \textbf{45.2} & 2017.6 & 2146.8 & \textbf{99.1} \\
97/7 : 00/12 & \textbf{39.8} & 182.5 & 59.2 & 69.9 & 382.2 & \textbf{41.6} & 43.6 & 164.9 & \textbf{37.6} & 441.2 & 3448.0 & \textbf{98.2} \\
98/1 : 01/6 & 138.6 & 180.4 & \textbf{55.9} & 85.3 & 196.0 & \textbf{34.0} & 43.4 & 113.0 & \textbf{30.4} & 834.3 & 1680.9 & \textbf{98.3} \\
98/7 : 01/12 & 116.8 & 177.4 & \textbf{52.6} & 209.7 & 199.2 & \textbf{44.0} & 91.4 & 87.4 & \textbf{39.0} & 1498.1 & 1417.0 & \textbf{99.2} \\
\bottomrule
\end{tabular}
    \caption{MAPE values for OTM options trained using a rolling window schema. Results are shown for models trained with Black-Scholes information (NN+), without it (NN-), and for the Black-Scholes model (BS).}
    \label{Tab:OTM_rolling_NN}
\end{table}
\FloatBarrier

\begin{table}[!ht]
    \centering
    \small
    \begin{tabular}{p{1.75cm}|>{\centering\arraybackslash}p{0.6cm}>{\centering\arraybackslash}p{0.6cm}>{\centering\arraybackslash}p{0.6cm}|
                     >{\centering\arraybackslash}p{0.6cm}>{\centering\arraybackslash}p{0.6cm}>{\centering\arraybackslash}p{0.6cm}|
                     >{\centering\arraybackslash}p{0.6cm}>{\centering\arraybackslash}p{0.6cm}>{\centering\arraybackslash}p{0.6cm}|
                     >{\centering\arraybackslash}p{0.6cm}>{\centering\arraybackslash}p{0.6cm}>{\centering\arraybackslash}p{0.6cm}}
        \toprule
        & \multicolumn{3}{c|}{Test $\frac{S}{K} \in (0, 1.1]$} 
        & \multicolumn{3}{c|}{Test $\frac{S}{K} \in (1.1, 1.2]$} 
        & \multicolumn{3}{c|}{Test $\frac{S}{K} \in (1.2, 1.3]$} 
        & \multicolumn{3}{c}{Test $\frac{S}{K} \in (1.3, 1.5]$} \\
        \centering Date 
        & \textbf{$NN+$} & \textbf{$NN-$} & \textbf{$BS$} 
        & \textbf{$NN+$} & \textbf{$NN-$} & \textbf{$BS$} 
        & \textbf{$NN+$} & \textbf{$NN-$} & \textbf{$BS$} 
        & \textbf{$NN+$} & \textbf{$NN-$} & \textbf{$BS$} \\
        \midrule
96/1 : 99/6 & 17.5 & \textbf{13.6} & 25.5 & 45.4 & \textbf{22.2} & 56.3 & 95.4 & \textbf{35.5} & 74.9 & 170.4 & \textbf{40.7} & 88.2 \\
96/1 : 99/12 & 37.0 & \textbf{22.5} & 36.3 & 70.5 & \textbf{28.1} & 73.4 & 137.2 & \textbf{40.3} & 88.1 & 191.5 & \textbf{59.6} & 96.4 \\
96/1 : 00/6 & 36.5 & \textbf{17.4} & 29.3 & 138.9 & \textbf{37.7} & 51.0 & 415.1 & 71.9 & \textbf{66.9} & 1190.1 & 129.9 & \textbf{84.0} \\
96/1 : 00/12 & 23.7 & 23.8 & \textbf{14.5} & 59.2 & 49.6 & \textbf{44.1} & 101.3 & \textbf{52.3} & 67.7 & 86.7 & \textbf{64.6} & 80.7 \\
96/1 : 01/6 & 32.3 & 44.1 & \textbf{17.6} & 129.8 & 129.8 & \textbf{32.1} & 354.6 & 374.9 & \textbf{49.1} & 879.1 & 978.9 & \textbf{63.5} \\
96/1 : 01/12 & 31.8 & 44.3 & \textbf{16.2} & 101.4 & 99.9 & \textbf{46.1} & 261.5 & 266.3 & \textbf{65.8} & 514.9 & 598.8 & \textbf{79.9} \\
\bottomrule
\end{tabular}
    \caption{Mean absolute percentage error (MAPE) values for OTM options grouped by moneyness ranges under the expanding window schema. The test set is divided into four moneyness ranges. Results are reported for models trained with Black-Scholes information (NN+), without it (NN-), and for the Black-Scholes model (BS).}
    \label{Tab:OTM_moneyness_expanding_NN}
\end{table}
\FloatBarrier

\begin{table}[!ht]
    \centering
    \small
    \begin{tabular}{p{1.75cm}|>{\centering\arraybackslash}p{0.6cm}>{\centering\arraybackslash}p{0.6cm}>{\centering\arraybackslash}p{0.6cm}|
                     >{\centering\arraybackslash}p{0.6cm}>{\centering\arraybackslash}p{0.6cm}>{\centering\arraybackslash}p{0.6cm}|
                     >{\centering\arraybackslash}p{0.6cm}>{\centering\arraybackslash}p{0.6cm}>{\centering\arraybackslash}p{0.6cm}|
                     >{\centering\arraybackslash}p{0.6cm}>{\centering\arraybackslash}p{0.6cm}>{\centering\arraybackslash}p{0.6cm}}
        \toprule
        & \multicolumn{3}{c|}{Test $\frac{S}{K} \in (0, 1.1]$} 
        & \multicolumn{3}{c|}{Test $\frac{S}{K} \in (1.1, 1.2]$} 
        & \multicolumn{3}{c|}{Test $\frac{S}{K} \in (1.2, 1.3]$} 
        & \multicolumn{3}{c}{Test $\frac{S}{K} \in (1.3, 1.5]$} \\
        \centering Date 
        & \textbf{$NN+$} & \textbf{$NN-$} & \textbf{$BS$} 
        & \textbf{$NN+$} & \textbf{$NN-$} & \textbf{$BS$} 
        & \textbf{$NN+$} & \textbf{$NN-$} & \textbf{$BS$} 
        & \textbf{$NN+$} & \textbf{$NN-$} & \textbf{$BS$} \\
        \midrule
96/1 : 99/6 & 17.5 & \textbf{13.6} & 25.5 & 45.4 & \textbf{22.2} & 56.3 & 95.4 & \textbf{35.5} & 74.9 & 170.4 & \textbf{40.7} & 88.2 \\
96/7 : 99/12 & 35.5 & \textbf{21.2} & 36.3 & 67.8 & \textbf{28.6} & 73.4 & 131.9 & \textbf{39.9} & 88.1 & 190.3 & \textbf{55.4} & 96.4 \\
97/1 : 00/6 & 40.2 & 42.6 & \textbf{29.3} & 148.7 & 145.2 & \textbf{51.0} & 427.6 & 434.9 & \textbf{66.9} & 1152.8 & 1268.7 & \textbf{84.0} \\
97/7 : 00/12 & 21.2 & 59.5 & \textbf{14.5} & 67.3 & 262.6 & \textbf{44.1} & 136.7 & 712.3 & \textbf{67.7} & 129.6 & 1105.1 & \textbf{80.7} \\
98/1 : 01/6 & 17.6 & 31.6 & \textbf{17.6} & 35.2 & 111.0 & \textbf{32.1} & 103.2 & 292.8 & \textbf{49.1} & 335.2 & 674.9 & \textbf{63.5} \\
98/7 : 01/12 & 16.7 & 50.8 & \textbf{16.2} & 94.6 & 87.7 & \textbf{46.1} & 279.0 & 242.5 & \textbf{65.8} & 759.9 & 671.7 & \textbf{79.9} \\
\bottomrule
\end{tabular}
    \caption{Mean absolute percentage error (MAPE) values for OTM options grouped by moneyness ranges under the rolling window schema. The test set is divided into four moneyness ranges. Results are reported for models trained with Black-Scholes information (NN+), without it (NN-), and for the Black-Scholes model (BS).}
    \label{Tab:OTM_moneyness_rolling_NN}
\end{table}
\FloatBarrier

\begin{table}[!ht]
    \centering
    \small
    \begin{tabular}{p{1.75cm}|>{\centering\arraybackslash}p{0.6cm}>{\centering\arraybackslash}p{0.6cm}>{\centering\arraybackslash}p{0.6cm}|
                     >{\centering\arraybackslash}p{0.6cm}>{\centering\arraybackslash}p{0.6cm}>{\centering\arraybackslash}p{0.6cm}}
        \toprule
        & \multicolumn{3}{c|}{Train} 
        & \multicolumn{3}{c}{Test}  \\
        \centering Date 
        & \textbf{$NN+$} & \textbf{$NN-$} & \textbf{$BS$} 
        & \textbf{$NN+$} & \textbf{$NN-$} & \textbf{$BS$} \\
        \midrule
96/1 : 99/6 & 45.0 & 38.9 & \textbf{11.3} & 34.5 & 40.9 & \textbf{5.0} \\
96/1 : 99/12 & 18.2 & 23.2 & \textbf{10.3} & 14.1 & 32.6 & \textbf{6.2} \\
96/1 : 00/6 & 38.6 & 41.2 & \textbf{9.5} & 37.5 & 54.5 & \textbf{10.3} \\
96/1 : 00/12 & \textbf{7.2} & 9.4 & 9.7 & \textbf{6.5} & 10.1 & 7.1 \\
96/1 : 01/6 & \textbf{6.1} & 8.1 & 9.3 & 7.0 & \textbf{6.4} & 9.5 \\
96/1 : 01/12 & 6.6 & \textbf{6.4} & 9.3 & \textbf{3.1} & 6.8 & 4.4 \\
\bottomrule
\end{tabular}
    \caption{MAPE values for ITM options trained using an expanding window schema. Results are shown for models trained with Black-Scholes information (NN+), without it (NN-), and for the Black-Scholes model (BS).}
    \label{Tab:ITM_expanding_NN}
\end{table}
\FloatBarrier

\begin{table}[!ht]
    \centering
    \small
    \begin{tabular}{p{1.75cm}|>{\centering\arraybackslash}p{0.6cm}>{\centering\arraybackslash}p{0.6cm}>{\centering\arraybackslash}p{0.6cm}|
                     >{\centering\arraybackslash}p{0.6cm}>{\centering\arraybackslash}p{0.6cm}>{\centering\arraybackslash}p{0.6cm}}
        \toprule
        & \multicolumn{3}{c|}{Train} 
        & \multicolumn{3}{c}{Test}  \\
        \centering Date 
        & \textbf{$NN+$} & \textbf{$NN-$} & \textbf{$BS$} 
        & \textbf{$NN+$} & \textbf{$NN-$} & \textbf{$BS$} \\
        \midrule
96/1 : 99/6 & 45.0 & 38.9 & \textbf{11.3} & 34.5 & 40.9 & \textbf{5.0} \\
96/7 : 99/12 & 24.1 & 25.1 & \textbf{10.2} & 23.2 & 35.6 & \textbf{6.2} \\
97/1 : 00/6 & 21.8 & 21.8 & \textbf{9.1} & 19.7 & 42.9 & \textbf{10.3} \\
97/7 : 00/12 & 41.8 & 48.5 & \textbf{8.9} & 45.2 & 66.9 & \textbf{7.1} \\
98/1 : 01/6 & \textbf{7.8} & 14.9 & 8.4 & \textbf{6.4} & 15.5 & 9.5 \\
98/7 : 01/12 & \textbf{5.7} & 15.8 & 9.0 & \textbf{4.0} & 15.3 & 4.4 \\
\bottomrule
\end{tabular}
        \caption{MAPE values for ITM options trained using a rolling window schema. Results are shown for models trained with Black-Scholes information (NN+), without it (NN-), and for the Black-Scholes model (BS).}
    \label{Tab:ITM_rolling_NN}
\end{table}
\FloatBarrier

\begin{table}[!ht]
    \centering
    \small
    \begin{tabular}{p{1.75cm}|>{\centering\arraybackslash}p{0.6cm}>{\centering\arraybackslash}p{0.6cm}>{\centering\arraybackslash}p{0.6cm}|
                     >{\centering\arraybackslash}p{0.6cm}>{\centering\arraybackslash}p{0.6cm}>{\centering\arraybackslash}p{0.6cm}|
                     >{\centering\arraybackslash}p{0.6cm}>{\centering\arraybackslash}p{0.6cm}>{\centering\arraybackslash}p{0.6cm}|
                     >{\centering\arraybackslash}p{0.6cm}>{\centering\arraybackslash}p{0.6cm}>{\centering\arraybackslash}p{0.6cm}}
        \toprule
        & \multicolumn{3}{c|}{Train} 
        & \multicolumn{3}{c|}{Test} 
        & \multicolumn{3}{c|}{Test $p_{BS} \geq 0.075$} 
        & \multicolumn{3}{c}{Test $p_{BS} < 0.075$} \\
        \centering Date 
        & \textbf{$LR+$} & \textbf{$LR-$} & \textbf{$BS$} 
        & \textbf{$LR+$} & \textbf{$LR-$} & \textbf{$BS$} 
        & \textbf{$LR+$} & \textbf{$LR-$} & \textbf{$BS$} 
        & \textbf{$LR+$} & \textbf{$LR-$} & \textbf{$BS$} \\
        \midrule

96/1 : 99/6 & \textbf{44.0} & 104.5 & 66.9 & \textbf{20.1} & 44.9 & 55.6 & \textbf{13.2} & 19.4 & 51.7 & 99.9 & 336.2 & \textbf{99.5} \\
96/1 : 99/12 & \textbf{39.8} & 105.0 & 65.2 & \textbf{57.0} & 137.7 & 66.5 & \textbf{23.9} & 32.0 & 61.0 & 259.4 & 785.3 & \textbf{99.6} \\
96/1 : 00/6 & \textbf{40.9} & 104.5 & 65.3 & 99.7 & 154.2 & \textbf{50.8} & \textbf{31.1} & 37.5 & 45.2 & 699.9 & 1175.7 & \textbf{99.1} \\
96/1 : 00/12 & \textbf{49.1} & 109.1 & 63.9 & 116.4 & 194.2 & \textbf{41.6} & 41.4 & 56.0 & \textbf{37.6} & 1174.9 & 2145.2 & \textbf{98.2} \\
96/1 : 01/6 & \textbf{54.8} & 113.3 & 62.1 & 127.9 & 145.5 & \textbf{34.0} & 90.4 & 102.7 & \textbf{30.4} & 799.2 & 910.0 & \textbf{98.3} \\
96/1 : 01/12 & \textbf{54.0} & 111.0 & 60.1 & 205.4 & 202.7 & \textbf{44.0} & 110.8 & 109.5 & \textbf{39.0} & 1235.6 & 1218.3 & \textbf{99.2} \\
\bottomrule
\end{tabular}
        \caption{MAPE values for OTM options trained using an expanding window schema. Results are shown for models trained with Black-Scholes information (LR+), without it (LR-), and for the Black-Scholes model (BS).}
    \label{Tab:OTM_expanding_LR}
\end{table}
\FloatBarrier

\begin{table}[!ht]
    \centering
    \small
    \begin{tabular}{p{1.75cm}|>{\centering\arraybackslash}p{0.6cm}>{\centering\arraybackslash}p{0.6cm}>{\centering\arraybackslash}p{0.6cm}|
                     >{\centering\arraybackslash}p{0.6cm}>{\centering\arraybackslash}p{0.6cm}>{\centering\arraybackslash}p{0.6cm}|
                     >{\centering\arraybackslash}p{0.6cm}>{\centering\arraybackslash}p{0.6cm}>{\centering\arraybackslash}p{0.6cm}|
                     >{\centering\arraybackslash}p{0.6cm}>{\centering\arraybackslash}p{0.6cm}>{\centering\arraybackslash}p{0.6cm}}
        \toprule
        & \multicolumn{3}{c|}{Train} 
        & \multicolumn{3}{c|}{Test} 
        & \multicolumn{3}{c|}{Test $p_{BS} \geq 0.075$} 
        & \multicolumn{3}{c}{Test $p_{BS} < 0.075$} \\
        \centering Date 
        & \textbf{$LR+$} & \textbf{$LR-$} & \textbf{$BS$} 
        & \textbf{$LR+$} & \textbf{$LR-$} & \textbf{$BS$} 
        & \textbf{$LR+$} & \textbf{$LR-$} & \textbf{$BS$} 
        & \textbf{$LR+$} & \textbf{$LR-$} & \textbf{$BS$} \\
        \midrule

96/1 : 99/6 & \textbf{44.0} & 104.5 & 66.9 & \textbf{20.1} & 44.9 & 55.6 & \textbf{13.2} & 19.4 & 51.7 & 99.9 & 336.2 & \textbf{99.5} \\
96/7 : 99/12 & \textbf{38.4} & 93.7 & 64.0 & \textbf{55.6} & 137.5 & 66.5 & \textbf{24.0} & 32.0 & 61.0 & 249.5 & 783.6 & \textbf{99.6} \\
97/1 : 00/6 & \textbf{40.2} & 82.8 & 62.9 & 124.5 & 171.3 & \textbf{50.8} & 41.0 & \textbf{39.6} & 45.2 & 854.6 & 1324.0 & \textbf{99.1} \\
97/7 : 00/12 & \textbf{42.2} & 75.2 & 59.2 & 112.8 & 174.4 & \textbf{41.6} & 40.7 & 54.8 & \textbf{37.6} & 1129.3 & 1862.2 & \textbf{98.2} \\
98/1 : 01/6 & \textbf{48.0} & 72.8 & 56.0 & 98.8 & 100.6 & \textbf{34.0} & 76.6 & 79.5 & \textbf{30.4} & 496.5 & 477.1 & \textbf{98.3} \\
98/7 : 01/12 & \textbf{41.5} & 64.9 & 52.6 & 300.4 & 302.2 & \textbf{44.0} & 150.7 & 151.3 & \textbf{39.0} & 1930.5 & 1945.5 & \textbf{99.2} \\
\bottomrule
\end{tabular}
            \caption{MAPE values for OTM options trained using a rolling window schema. Results are shown for models trained with Black-Scholes information (LR+), without it (LR-), and for the Black-Scholes model (BS).}
    \label{Tab:OTM_rolling_LR}
\end{table}
\FloatBarrier

\begin{table}[!ht]
    \centering
    \small
    \begin{tabular}{p{1.75cm}|>{\centering\arraybackslash}p{0.6cm}>{\centering\arraybackslash}p{0.6cm}>{\centering\arraybackslash}p{0.6cm}|
                     >{\centering\arraybackslash}p{0.6cm}>{\centering\arraybackslash}p{0.6cm}>{\centering\arraybackslash}p{0.6cm}|
                     >{\centering\arraybackslash}p{0.6cm}>{\centering\arraybackslash}p{0.6cm}>{\centering\arraybackslash}p{0.6cm}|
                     >{\centering\arraybackslash}p{0.6cm}>{\centering\arraybackslash}p{0.6cm}>{\centering\arraybackslash}p{0.6cm}}
        \toprule
        & \multicolumn{3}{c|}{Test $\frac{S}{K} \in (0, 1.1]$} 
        & \multicolumn{3}{c|}{Test $\frac{S}{K} \in (1.1, 1.2]$} 
        & \multicolumn{3}{c|}{Test $\frac{S}{K} \in (1.2, 1.3]$} 
        & \multicolumn{3}{c}{Test $\frac{S}{K} \in (1.3, 1.5]$} \\
        \centering Date 
        & \textbf{$LR+$} & \textbf{$LR-$} & \textbf{$BS$} 
        & \textbf{$LR+$} & \textbf{$LR-$} & \textbf{$BS$} 
        & \textbf{$LR+$} & \textbf{$LR-$} & \textbf{$BS$} 
        & \textbf{$LR+$} & \textbf{$LR-$} & \textbf{$BS$} \\
        \midrule
96/1 : 99/6 & \textbf{9.7} & 12.3 & 25.5 & \textbf{15.1} & 16.5 & 56.3 & \textbf{25.8} & 47.6 & 74.9 & \textbf{64.3} & 121.4 & 88.2 \\
96/1 : 99/12 & \textbf{21.5} & 22.6 & 36.3 & \textbf{47.9} & 51.6 & 73.4 & 110.0 & 188.6 & \textbf{88.1} & 279.7 & 454.8 & \textbf{96.4} \\
96/1 : 00/6 & \textbf{18.9} & 19.9 & 29.3 & 44.9 & \textbf{44.5} & 51.0 & 130.7 & 198.0 & \textbf{66.9} & 481.6 & 653.5 & \textbf{84.0} \\
96/1 : 00/12 & 24.7 & 20.6 & \textbf{14.5} & 96.5 & 111.8 & \textbf{44.1} & 267.6 & 431.0 & \textbf{67.7} & 528.4 & 714.8 & \textbf{80.7} \\
96/1 : 01/6 & 44.9 & 45.6 & \textbf{17.6} & 116.5 & 124.5 & \textbf{32.1} & 220.9 & 203.7 & \textbf{49.1} & 393.2 & 436.7 & \textbf{63.5} \\
96/1 : 01/12 & 42.1 & 44.4 & \textbf{16.2} & 105.1 & 104.4 & \textbf{46.1} & 221.7 & 251.0 & \textbf{65.8} & 456.5 & 447.3 & \textbf{79.9} \\
\bottomrule
\end{tabular}
    \caption{Mean absolute percentage error (MAPE) values for OTM options grouped by moneyness ranges under the expanding window schema. The test set is divided into four moneyness ranges. Results are reported for models trained with Black-Scholes information (LR+), without it (LR-), and for the Black-Scholes model (BS).}
    \label{Tab:OTM_moneyness_expanding_LR}
\end{table}
\FloatBarrier

\begin{table}[!ht]
    \centering
    \small
    \begin{tabular}{p{1.75cm}|>{\centering\arraybackslash}p{0.6cm}>{\centering\arraybackslash}p{0.6cm}>{\centering\arraybackslash}p{0.6cm}|
                     >{\centering\arraybackslash}p{0.6cm}>{\centering\arraybackslash}p{0.6cm}>{\centering\arraybackslash}p{0.6cm}|
                     >{\centering\arraybackslash}p{0.6cm}>{\centering\arraybackslash}p{0.6cm}>{\centering\arraybackslash}p{0.6cm}|
                     >{\centering\arraybackslash}p{0.6cm}>{\centering\arraybackslash}p{0.6cm}>{\centering\arraybackslash}p{0.6cm}}
        \toprule
        & \multicolumn{3}{c|}{Test $\frac{S}{K} \in (0, 1.1]$} 
        & \multicolumn{3}{c|}{Test $\frac{S}{K} \in (1.1, 1.2]$} 
        & \multicolumn{3}{c|}{Test $\frac{S}{K} \in (1.2, 1.3]$} 
        & \multicolumn{3}{c}{Test $\frac{S}{K} \in (1.3, 1.5]$} \\
        \centering Date 
        & \textbf{$LR+$} & \textbf{$LR-$} & \textbf{$BS$} 
        & \textbf{$LR+$} & \textbf{$LR-$} & \textbf{$BS$} 
        & \textbf{$LR+$} & \textbf{$LR-$} & \textbf{$BS$} 
        & \textbf{$LR+$} & \textbf{$LR-$} & \textbf{$BS$} \\
        \midrule
96/1 : 99/6 & \textbf{9.7} & 12.3 & 25.5 & \textbf{15.1} & 16.5 & 56.3 & \textbf{25.8} & 47.6 & 74.9 & \textbf{64.3} & 121.4 & 88.2 \\
96/7 : 99/12 & \textbf{21.7} & 23.0 & 36.3 & \textbf{46.5} & 51.3 & 73.4 & 102.8 & 183.8 & \textbf{88.1} & 257.2 & 431.6 & \textbf{96.4} \\
97/1 : 00/6 & 17.2 & \textbf{15.8} & 29.3 & 55.2 & \textbf{46.8} & 51.0 & 157.2 & 230.3 & \textbf{66.9} & 563.1 & 717.6 & \textbf{84.0} \\
97/7 : 00/12 & 24.1 & 21.8 & \textbf{14.5} & 79.1 & 99.2 & \textbf{44.1} & 209.6 & 372.2 & \textbf{67.7} & 445.4 & 587.0 & \textbf{80.7} \\
98/1 : 01/6 & 35.9 & 36.3 & \textbf{17.6} & 86.8 & 91.7 & \textbf{32.1} & 156.8 & 123.5 & \textbf{49.1} & 241.3 & 275.8 & \textbf{63.5} \\
98/7 : 01/12 & 54.7 & 54.9 & \textbf{16.2} & 156.9 & 154.3 & \textbf{46.1} & 369.5 & 404.0 & \textbf{65.8} & 850.6 & 834.1 & \textbf{79.9} \\
\bottomrule
\end{tabular}
    \caption{Mean absolute percentage error (MAPE) values for OTM options grouped by moneyness ranges under the rolling window schema. The test set is divided into four moneyness ranges. Results are reported for models trained with Black-Scholes information (LR+), without it (LR-), and for the Black-Scholes model (BS).}
    \label{Tab:OTM_moneyness_rolling_LR}
\end{table}
\FloatBarrier

\begin{table}[!ht]
    \centering
    \small
    \begin{tabular}{p{1.75cm}|>{\centering\arraybackslash}p{0.6cm}>{\centering\arraybackslash}p{0.6cm}>{\centering\arraybackslash}p{0.6cm}|
                     >{\centering\arraybackslash}p{0.6cm}>{\centering\arraybackslash}p{0.6cm}>{\centering\arraybackslash}p{0.6cm}}
        \toprule
        & \multicolumn{3}{c|}{Train} 
        & \multicolumn{3}{c}{Test}  \\
        \centering Date 
        & \textbf{$LR+$} & \textbf{$LR-$} & \textbf{$BS$} 
        & \textbf{$LR+$} & \textbf{$LR-$} & \textbf{$BS$} \\
        \midrule
96/1 : 99/6 & \textbf{4.4} & 5.4 & 11.3 & 8.4 & 7.2 & \textbf{5.0} \\
96/1 : 99/12 & \textbf{4.4} & 5.4 & 10.3 & \textbf{4.3} & 6.2 & 6.2 \\
96/1 : 00/6 & \textbf{4.4} & 5.7 & 9.5 & \textbf{5.4} & 6.2 & 10.3 \\
96/1 : 00/12 & \textbf{4.7} & 6.1 & 9.7 & \textbf{5.2} & 6.7 & 7.1 \\
96/1 : 01/6 & \textbf{4.9} & 6.4 & 9.3 & 5.6 & \textbf{4.4} & 9.5 \\
96/1 : 01/12 & \textbf{4.9} & 6.1 & 9.3 & 7.2 & 8.3 & \textbf{4.4} \\
\bottomrule
\end{tabular}
        \caption{MAPE values for ITM options trained using an expanding window schema. Results are shown for models trained with Black-Scholes information (LR+), without it (LR-), and for the Black-Scholes model (BS).}
    \label{Tab:ITM_expanding_LR}
\end{table}
\FloatBarrier

\begin{table}[!ht]
    \centering
    \small
    \begin{tabular}{p{1.75cm}|>{\centering\arraybackslash}p{0.6cm}>{\centering\arraybackslash}p{0.6cm}>{\centering\arraybackslash}p{0.6cm}|
                     >{\centering\arraybackslash}p{0.6cm}>{\centering\arraybackslash}p{0.6cm}>{\centering\arraybackslash}p{0.6cm}}
        \toprule
        & \multicolumn{3}{c|}{Train} 
        & \multicolumn{3}{c}{Test}  \\
        \centering Date 
        & \textbf{$LR+$} & \textbf{$LR-$} & \textbf{$BS$} 
        & \textbf{$LR+$} & \textbf{$LR-$} & \textbf{$BS$} \\
        \midrule
96/1 : 99/6 & \textbf{4.4} & 5.4 & 11.3 & 8.4 & 7.2 & \textbf{5.0} \\
96/7 : 99/12 & \textbf{4.3} & 5.2 & 10.2 & \textbf{4.5} & 6.2 & 6.2 \\
97/1 : 00/6 & \textbf{4.2} & 5.2 & 9.1 & \textbf{5.1} & 5.9 & 10.3 \\
97/7 : 00/12 & \textbf{4.2} & 5.3 & 8.9 & \textbf{4.9} & 6.5 & 7.1 \\
98/1 : 01/6 & \textbf{3.9} & 5.2 & 8.3 & 5.3 & \textbf{4.6} & 9.5 \\
98/7 : 01/12 & \textbf{3.8} & 4.8 & 9.0 & 7.8 & 8.3 & \textbf{4.4} \\
\bottomrule
\end{tabular}
        \caption{MAPE values for ITM options trained using a rolling window schema. Results are shown for models trained with Black-Scholes information (LR+), without it (LR-), and for the Black-Scholes model (BS).}
    \label{Tab:ITM_rolling_LR}
\end{table}
\FloatBarrier

\begin{table}[!ht]
    \centering
    \small
    \begin{tabular}{p{1.75cm}|>{\centering\arraybackslash}p{0.6cm}>{\centering\arraybackslash}p{0.6cm}>{\centering\arraybackslash}p{0.6cm}|
                     >{\centering\arraybackslash}p{0.6cm}>{\centering\arraybackslash}p{0.6cm}>{\centering\arraybackslash}p{0.6cm}|
                     >{\centering\arraybackslash}p{0.6cm}>{\centering\arraybackslash}p{0.6cm}>{\centering\arraybackslash}p{0.6cm}|
                     >{\centering\arraybackslash}p{0.6cm}>{\centering\arraybackslash}p{0.6cm}>{\centering\arraybackslash}p{0.6cm}}
        \toprule
        & \multicolumn{3}{c|}{Train} 
        & \multicolumn{3}{c|}{Test} 
        & \multicolumn{3}{c|}{Test $p_{BS} \geq 0.075$} 
        & \multicolumn{3}{c}{Test $p_{BS} < 0.075$} \\
        \centering Date 
        & \textbf{$RF+$} & \textbf{$RF-$} & \textbf{$BS$} 
        & \textbf{$RF+$} & \textbf{$RF-$} & \textbf{$BS$} 
        & \textbf{$RF+$} & \textbf{$RF-$} & \textbf{$BS$} 
        & \textbf{$RF+$} & \textbf{$RF-$} & \textbf{$BS$} \\
        \midrule
96/1 : 99/6 & \textbf{17.2} & 24.5 & 66.9 & \textbf{17.4} & 75.1 & 55.6 & \textbf{13.9} & 53.6 & 51.7 & \textbf{56.7} & 321.8 & 99.5 \\
96/1 : 99/12 & \textbf{16.4} & 26.4 & 65.2 & \textbf{26.3} & 56.9 & 66.5 & \textbf{24.2} & 39.1 & 61.0 & \textbf{39.2} & 165.9 & 99.6 \\
96/1 : 00/6 & \textbf{16.5} & 25.1 & 65.3 & \textbf{48.2} & 73.9 & 50.8 & \textbf{41.8} & 50.3 & 45.2 & 104.4 & 280.8 & \textbf{99.1} \\
96/1 : 00/12 & \textbf{17.2} & 25.3 & 63.9 & 51.5 & 64.9 & \textbf{41.6} & 45.6 & 54.8 & \textbf{37.6} & 135.1 & 208.3 & \textbf{98.2} \\
96/1 : 01/6 & \textbf{17.6} & 27.9 & 62.1 & 90.8 & 114.5 & \textbf{34.0} & 82.0 & 97.4 & \textbf{30.4} & 247.9 & 419.4 & \textbf{98.3} \\
96/1 : 01/12 & \textbf{19.5} & 28.6 & 60.1 & \textbf{21.8} & 43.2 & 44.0 & \textbf{17.0} & 34.5 & 39.0 & \textbf{73.2} & 137.3 & 99.2 \\
\bottomrule
\end{tabular}
    \caption{MAPE values for OTM options trained using an expanding window schema. Results are shown for models trained with Black-Scholes information (RF+), without it (RF-), and for the Black-Scholes model (BS).}
    \label{Tab:OTM_expanding_RF}
\end{table}
\FloatBarrier

\begin{table}[!ht]
    \centering
    \small
    \begin{tabular}{p{1.75cm}|>{\centering\arraybackslash}p{0.6cm}>{\centering\arraybackslash}p{0.6cm}>{\centering\arraybackslash}p{0.6cm}|
                     >{\centering\arraybackslash}p{0.6cm}>{\centering\arraybackslash}p{0.6cm}>{\centering\arraybackslash}p{0.6cm}|
                     >{\centering\arraybackslash}p{0.6cm}>{\centering\arraybackslash}p{0.6cm}>{\centering\arraybackslash}p{0.6cm}|
                     >{\centering\arraybackslash}p{0.6cm}>{\centering\arraybackslash}p{0.6cm}>{\centering\arraybackslash}p{0.6cm}}
        \toprule
        & \multicolumn{3}{c|}{Train} 
        & \multicolumn{3}{c|}{Test} 
        & \multicolumn{3}{c|}{Test $p_{BS} \geq 0.075$} 
        & \multicolumn{3}{c}{Test $p_{BS} < 0.075$} \\
        \centering Date 
        & \textbf{$RF+$} & \textbf{$RF-$} & \textbf{$BS$} 
        & \textbf{$RF+$} & \textbf{$RF-$} & \textbf{$BS$} 
        & \textbf{$RF+$} & \textbf{$RF-$} & \textbf{$BS$} 
        & \textbf{$RF+$} & \textbf{$RF-$} & \textbf{$BS$} \\
        \midrule
96/1 : 99/6 & \textbf{17.2} & 24.5 & 66.9 & \textbf{17.4} & 75.1 & 55.6 & \textbf{13.9} & 53.6 & 51.7 & \textbf{56.7} & 321.8 & 99.5 \\
96/7 : 99/12 & \textbf{15.9} & 26.7 & 64.1 & \textbf{26.7} & 55.5 & 66.5 & \textbf{24.5} & 39.0 & 61.0 & \textbf{40.3} & 156.7 & 99.6 \\
97/1 : 00/6 & \textbf{14.7} & 26.2 & 62.9 & \textbf{49.1} & 53.8 & 50.8 & 42.7 & \textbf{36.5} & 45.2 & 105.4 & 205.2 & \textbf{99.1} \\
97/7 : 00/12 & \textbf{13.7} & 23.8 & 59.2 & 54.2 & 62.1 & \textbf{41.6} & 47.7 & 51.7 & \textbf{37.6} & 146.0 & 209.3 & \textbf{98.2} \\
98/1 : 01/6 & \textbf{13.7} & 19.1 & 55.9 & 89.5 & 104.8 & \textbf{34.0} & 81.6 & 90.0 & \textbf{30.4} & 231.4 & 368.5 & \textbf{98.3} \\
98/7 : 01/12 & \textbf{16.4} & 17.4 & 52.6 & \textbf{17.7} & 40.7 & 44.0 & \textbf{14.5} & 32.3 & 39.0 & \textbf{52.4} & 132.4 & 99.2 \\
\bottomrule
\end{tabular}
    \caption{MAPE values for OTM options trained using a rolling window schema. Results are shown for models trained with Black-Scholes information (RF+), without it (RF-), and for the Black-Scholes model (BS).}
    \label{Tab:OTM_rolling_RF}
\end{table}
\FloatBarrier

\begin{table}[!ht]
    \centering
    \small
    \begin{tabular}{p{1.75cm}|>{\centering\arraybackslash}p{0.6cm}>{\centering\arraybackslash}p{0.6cm}>{\centering\arraybackslash}p{0.6cm}|
                     >{\centering\arraybackslash}p{0.6cm}>{\centering\arraybackslash}p{0.6cm}>{\centering\arraybackslash}p{0.6cm}|
                     >{\centering\arraybackslash}p{0.6cm}>{\centering\arraybackslash}p{0.6cm}>{\centering\arraybackslash}p{0.6cm}|
                     >{\centering\arraybackslash}p{0.6cm}>{\centering\arraybackslash}p{0.6cm}>{\centering\arraybackslash}p{0.6cm}}
        \toprule
        & \multicolumn{3}{c|}{Test $\frac{S}{K} \in (0, 1.1]$} 
        & \multicolumn{3}{c|}{Test $\frac{S}{K} \in (1.1, 1.2]$} 
        & \multicolumn{3}{c|}{Test $\frac{S}{K} \in (1.2, 1.3]$} 
        & \multicolumn{3}{c}{Test $\frac{S}{K} \in (1.3, 1.5]$} \\
        \centering Date 
        & \textbf{$RF+$} & \textbf{$RF-$} & \textbf{$BS$} 
        & \textbf{$RF+$} & \textbf{$RF-$} & \textbf{$BS$} 
        & \textbf{$RF+$} & \textbf{$RF-$} & \textbf{$BS$} 
        & \textbf{$RF+$} & \textbf{$RF-$} & \textbf{$BS$} \\
        \midrule
96/1 : 99/6 & \textbf{9.8} & 22.8 & 25.5 & \textbf{14.3} & 79.9 & 56.3 & \textbf{21.1} & 97.2 & 74.9 & \textbf{30.3} & 137.8 & 88.2 \\
96/1 : 99/12 & \textbf{17.7} & 30.6 & 36.3 & \textbf{25.7} & 50.1 & 73.4 & \textbf{29.8} & 70.5 & 88.1 & \textbf{40.6} & 104.3 & 96.4 \\
96/1 : 00/6 & \textbf{18.0} & 19.8 & 29.3 & 44.8 & \textbf{35.0} & 51.0 & 73.9 & 91.5 & \textbf{66.9} & 97.9 & 253.7 & \textbf{84.0} \\
96/1 : 00/12 & 23.1 & 30.0 & \textbf{14.5} & 50.9 & 65.3 & \textbf{44.1} & 73.4 & 97.8 & \textbf{67.7} & 105.4 & 121.8 & \textbf{80.7} \\
96/1 : 01/6 & 37.4 & 33.4 & \textbf{17.6} & 83.3 & 104.4 & \textbf{32.1} & 134.8 & 186.4 & \textbf{49.1} & 195.3 & 264.6 & \textbf{63.5} \\
96/1 : 01/12 & \textbf{12.4} & 22.9 & 16.2 & \textbf{18.6} & 40.1 & 46.1 & \textbf{26.6} & 52.3 & 65.8 & \textbf{42.7} & 83.7 & 79.9 \\
\bottomrule
\end{tabular}
    \caption{Mean absolute percentage error (MAPE) values for OTM options grouped by moneyness ranges under the expanding window schema. The test set is divided into four moneyness ranges. Results are reported for models trained with Black-Scholes information (RF), without it (RF-), and for the Black-Scholes model (BS).}
    \label{Tab:OTM_moneyness_expanding_RF}
\end{table}
\FloatBarrier

\begin{table}[!ht]
    \centering
    \small
    \begin{tabular}{p{1.75cm}|>{\centering\arraybackslash}p{0.6cm}>{\centering\arraybackslash}p{0.6cm}>{\centering\arraybackslash}p{0.6cm}|
                     >{\centering\arraybackslash}p{0.6cm}>{\centering\arraybackslash}p{0.6cm}>{\centering\arraybackslash}p{0.6cm}|
                     >{\centering\arraybackslash}p{0.6cm}>{\centering\arraybackslash}p{0.6cm}>{\centering\arraybackslash}p{0.6cm}|
                     >{\centering\arraybackslash}p{0.6cm}>{\centering\arraybackslash}p{0.6cm}>{\centering\arraybackslash}p{0.6cm}}
        \toprule
        & \multicolumn{3}{c|}{Test $\frac{S}{K} \in (0, 1.1]$} 
        & \multicolumn{3}{c|}{Test $\frac{S}{K} \in (1.1, 1.2]$} 
        & \multicolumn{3}{c|}{Test $\frac{S}{K} \in (1.2, 1.3]$} 
        & \multicolumn{3}{c}{Test $\frac{S}{K} \in (1.3, 1.5]$} \\
        \centering Date 
        & \textbf{$RF+$} & \textbf{$RF-$} & \textbf{$BS$} 
        & \textbf{$RF+$} & \textbf{$RF-$} & \textbf{$BS$} 
        & \textbf{$RF+$} & \textbf{$RF-$} & \textbf{$BS$} 
        & \textbf{$RF+$} & \textbf{$RF-$} & \textbf{$BS$} \\
        \midrule
96/1 : 99/6 & \textbf{9.8} & 22.8 & 25.5 & \textbf{14.3} & 79.9 & 56.3 & \textbf{21.1} & 97.2 & 74.9 & \textbf{30.3} & 137.8 & 88.2 \\
96/7 : 99/12 & \textbf{18.2} & 30.4 & 36.3 & \textbf{26.1} & 50.1 & 73.4 & \textbf{30.0} & 70.3 & 88.1 & \textbf{41.0} & 98.0 & 96.4 \\
97/1 : 00/6 & \textbf{18.3} & 19.4 & 29.3 & 45.3 & \textbf{32.6} & 51.0 & 75.8 & \textbf{56.9} & 66.9 & 99.5 & 171.9 & \textbf{84.0} \\
97/7 : 00/12 & 23.5 & 28.0 & \textbf{14.5} & 55.9 & 59.1 & \textbf{44.1} & 81.0 & 96.6 & \textbf{67.7} & 104.3 & 121.1 & \textbf{80.7} \\
98/1 : 01/6 & 37.8 & 32.8 & \textbf{17.6} & 83.5 & 98.9 & \textbf{32.1} & 129.2 & 165.0 & \textbf{49.1} & 191.8 & 236.6 & \textbf{63.5} \\
98/7 : 01/12 & \textbf{11.4} & 18.5 & 16.2 & \textbf{14.8} & 33.8 & 46.1 & \textbf{21.2} & 49.8 & 65.8 & \textbf{33.0} & 92.2 & 79.9 \\
\bottomrule
\end{tabular}
    \caption{Mean absolute percentage error (MAPE) values for OTM options grouped by moneyness ranges under the rolling window schema. The test set is divided into four moneyness ranges. Results are reported for models trained with Black-Scholes information (RF+), without it (RF-), and for the Black-Scholes model (BS).}
    \label{Tab:OTM_moneyness_rolling_RF}
\end{table}
\FloatBarrier

\begin{table}[!ht]
    \centering
    \small
    \begin{tabular}{p{1.75cm}|>{\centering\arraybackslash}p{0.6cm}>{\centering\arraybackslash}p{0.6cm}>{\centering\arraybackslash}p{0.6cm}|
                     >{\centering\arraybackslash}p{0.6cm}>{\centering\arraybackslash}p{0.6cm}>{\centering\arraybackslash}p{0.6cm}}
        \toprule
        & \multicolumn{3}{c|}{Train} 
        & \multicolumn{3}{c}{Test}  \\
        \centering Date 
        & \textbf{$RF+$} & \textbf{$RF-$} & \textbf{$BS$} 
        & \textbf{$RF+$} & \textbf{$RF-$} & \textbf{$BS$} \\
        \midrule
96/1 : 99/6 & \textbf{3.6} & 7.5 & 11.3 & \textbf{4.2} & 30.0 & 5.0 \\
96/1 : 99/12 & \textbf{3.7} & 6.6 & 10.3 & \textbf{5.4} & 17.9 & 6.2 \\
96/1 : 00/6 & \textbf{3.7} & 7.3 & 9.5 & \textbf{6.8} & 9.6 & 10.3 \\
96/1 : 00/12 & \textbf{3.7} & 8.2 & 9.7 & \textbf{6.3} & 8.6 & 7.1 \\
96/1 : 01/6 & \textbf{3.7} & 7.9 & 9.3 & \textbf{6.5} & 11.9 & 9.5 \\
96/1 : 01/12 & \textbf{3.8} & 7.5 & 9.3 & \textbf{3.3} & 9.1 & 4.4 \\
\bottomrule
\end{tabular}
    \caption{MAPE values for ITM options trained using an expanding window schema. Results are shown for models trained with Black-Scholes information (RF+), without it (RF-), and for the Black-Scholes model (BS).}
    \label{Tab:ITM_expanding_RF}
\end{table}
\FloatBarrier

\begin{table}[!ht]
    \centering
    \small
    \begin{tabular}{p{1.75cm}|>{\centering\arraybackslash}p{0.6cm}>{\centering\arraybackslash}p{0.6cm}>{\centering\arraybackslash}p{0.6cm}|
                     >{\centering\arraybackslash}p{0.6cm}>{\centering\arraybackslash}p{0.6cm}>{\centering\arraybackslash}p{0.6cm}}
        \toprule
        & \multicolumn{3}{c|}{Train} 
        & \multicolumn{3}{c}{Test}  \\
        \centering Date 
        & \textbf{$RF+$} & \textbf{$RF-$} & \textbf{$BS$} 
        & \textbf{$RF+$} & \textbf{$RF-$} & \textbf{$BS$} \\
        \midrule
96/1 : 99/6 & \textbf{3.6} & 7.5 & 11.3 & \textbf{4.2} & 30.0 & 5.0 \\
96/7 : 99/12 & \textbf{3.6} & 6.3 & 10.2 & \textbf{5.3} & 17.6 & 6.2 \\
97/1 : 00/6 & \textbf{3.4} & 6.4 & 9.1 & \textbf{6.9} & 9.6 & 10.3 \\
97/7 : 00/12 & \textbf{3.3} & 6.3 & 8.9 & \textbf{6.3} & 8.8 & 7.1 \\
98/1 : 01/6 & \textbf{3.2} & 6.0 & 8.4 & \textbf{6.6} & 14.4 & 9.5 \\
98/7 : 01/12 & \textbf{3.2} & 4.5 & 9.0 & \textbf{3.2} & 7.9 & 4.4 \\
\bottomrule
\end{tabular}
        \caption{MAPE values for ITM options trained using a rolling window schema. Results are shown for models trained with Black-Scholes information (RF+), without it (RF-), and for the Black-Scholes model (BS).}
    \label{Tab:ITM_rolling_RF}
\end{table}
\FloatBarrier




\end{document}